%% file: main.tex
\documentclass[journal,comsoc,twoside]{IEEEtran}
\PassOptionsToPackage{draft}{hyperref}
\PassOptionsToPackage{obeyspaces}{url}
\usepackage[figuresright]{rotating}
\usepackage{textcomp}
\usepackage{cite}
\usepackage{balance}
\usepackage{booktabs}
\usepackage{amssymb}
\usepackage{graphicx}
\usepackage{amsmath}
\usepackage{multicol, multirow}
\usepackage{rotating}
\usepackage{subcaption}
\usepackage{algorithm}
\usepackage{xspace}
\usepackage{dblfloatfix}
\usepackage{amssymb}% http://ctan.org/pkg/amssymb
\usepackage{pifont}% http://ctan.org/pkg/pifont
\usepackage{array}
\usepackage{cleveref}
\usepackage{amsmath,amssymb,amscd,mathtools}\usepackage{amsthm}
\usepackage{siunitx}
\usepackage[noend]{algpseudocode}
\usepackage{xcolor}
\usepackage{pgfplots}
\usepackage{tikz}
\definecolor{mediumturquoise}{rgb}{0.28, 0.82, 0.8}
\usepackage{array}
\usepackage{url}
\usepackage[numbers,sort&compress]{natbib}
\usepackage{pifont}% http://ctan.org/pkg/pifont
\newcommand{\cmark}{\ding{51}}
\newcommand{\xmark}{\ding{55}}
\usetikzlibrary{arrows, automata}
\usepackage[para, flushleft]{threeparttable}
\newcolumntype{P}[1]{>{\centering\arraybackslash}p{#1}}
\makeatletter
\def\BState{\State\hskip-\ALG@thistlm}
\makeatother
\usepackage{array}
\usepackage[acronym]{glossaries}
\newcolumntype{P}[1]{>{\centering\arraybackslash}p{#1}}

\input{glossary}

\makeglossaries{}

\begin{document}
\definecolor{mygray}{gray}{0.6}

\newcommand{\plus}{\raisebox{.4\height}{\scalebox{.6}{+}}}
\newcommand{\minus}{\raisebox{.4\height}{\scalebox{.6}{-}}}
\newcommand{\eleonora}[1]{\textcolor{black}{#1}}
\newcommand{\chhagan}[1]{\textcolor{mygray}{#1}}
\newcommand{\ankit}[1]{\textcolor{red}{#1}}
\newcommand{\hassan}[1]{\textcolor{brown}{#1}}

\newcommand{\review}[1]{\textcolor{black}{#1}}

\newcommand{\todobox}[3]{%
	\colorbox{#1}{\textcolor{white}{\sffamily\bfseries\scriptsize #2}}%
	~\textcolor{black}{#3} %
	\textcolor{#1}{$\triangleleft$}%
}
\newcommand{\todo}[1]{\todobox{red}{TODO}{#1}}
\newcommand{\comment}[1]{\todobox{green}{COMMENT}{#1}}

\newcommand{\questionbox}[3]{%
	\colorbox{#1}{\textcolor{white}{\sffamily\bfseries\scriptsize #2}}%
	~\textcolor{black}{#3} %
	\textcolor{#1}{$\triangleleft$}%
}
\newcommand{\question}[1]{\questionbox{orange}{QUESTION}{#1}}

\setlength{\abovedisplayskip}{4pt}
\setlength{\belowdisplayskip}{4pt}
%\title{Towards a \textcolor{red}{Secure} Coexistence of \\ICN and IP Architectures: \\State-of-the-art, and Open Issues}
\title{The Road Ahead for Networking: \\A Survey on ICN-IP Coexistence Solutions}
    %%%%%%%%%%%%%%%%%%%%%%%%%%%%%%%%%%%%%%%%%%%%%%%%%%
    %%HEADER
\author{Mauro Conti,~\IEEEmembership{Senior Member, IEEE}, Ankit Gangwal, Muhammad Hassan, Chhagan Lal, Eleonora Losiouk*
   \IEEEcompsocitemizethanks{
       \IEEEcompsocthanksitem \textit{* Corresponding author.}
       }
    \thanks{All authors are with the Department of Mathematics, University of Padua, 35121, Italy. email: \{conti, gangwal, hassan, chhagan, elosiouk\}@math.unipd.it}
    \thanks{Manuscript received Month DD, YYYY; revised Month DD, YYYY, Month DD, YYYY, and Month DD, YYYY; accepted Month DD, YYYY.}
    \thanks{Digital Object Identifier XX.XXXX/COMST.YYYY.XXXXXXX}   
}

%%%%%%%%%%%%%%%%%%%%%%%%%%%%%%%%%%%%%%%%%%%%%%%%%%


\maketitle
\begin{abstract}
In recent years, the usage model of the Internet has changed, pushing researchers towards the design of the \gls{ICN} paradigm as a possible replacement of the existing architecture. Even though both Academia and Industry have investigated the feasibility and effectiveness of \gls{ICN}, achieving the complete replacement of the \gls{IP} is a challenging task: (i)~the process involves multiple parties, such as \gls{ISPs}, that need to coordinate among each other; (ii)~it requires an indefinite amount of time to update hardware and software of network components; and (iii)~it is a high risk goal that might introduce unexpected complications. Thus, the process of replacing the current Internet will inevitably lead towards a period of coexistence between the old and the new architectures. Given the urgency of the problem, this transition phase will happen very soon and people should address it in a smooth way. 

Some research groups have already addressed the coexistence by designing their own architectures, but none of those is the final solution to move towards the future Internet considering the unaltered state of the networking. To design such architecture, the research community needs now a comprehensive overview of the existing solutions that have so far addressed the coexistence. The purpose of this paper is to reach this goal by providing the first comprehensive survey and classification of the coexistence architectures according to their features (i.e., deployment approach, deployment scenarios, addressed coexistence requirements and \review{additional} architecture or technology used) and evaluation parameters (i.e., challenges emerging during the deployment and the runtime behaviour of an architecture). We believe that this paper will finally fill the gap required for moving towards the design of the final coexistence architecture. 
\end{abstract}


% ACRONYMS
% ICN
% HTTP
% IP
% UDP 
% PIT
% SDN
% NFV
% CDN 
% DTN 

\input{introduction}

\input{background}
\input{classification_criteria}

\input{coexistence_architectures}
\input{discussion_conclusion}
\input{acks}
\bibliographystyle{IEEEtran}
\bibliography{main}
\input{bio}
\vfill
\end{document}

%% file: glossary.tex
\newacronym{3GPP}{3GPP}{Third Generation Partnership Project}
\newacronym{ACK}{ACK}{acknowledgement} 
\newacronym{API}{API}{Application Programming Interface}
\newacronym{APIs}{APIs}{Application Programming Interfaces}
\newacronym{BN}{BN}{Border-Node}
\newacronym{CA}{CA}{Congestion Avoidance}
\newacronym{CCN}{CCN}{Content-Centric Networking}
\newacronym{CDN}{CDN}{Content Delivery Network}
\newacronym{CL}{CL}{Convergence Layer}
\newacronym{cNAP}{cNAP}{consumer NAP}
\newacronym{COTS}{COTS}{Commercial Off-The-Shelf}
\newacronym{CSS}{CSS}{CONET Sub System}
\newacronym{DNS}{DNS}{Domain Name System}
\newacronym{DOCTOR}{DOCTOR}{DeplOyment and seCurisaTion of new functiOnalities in virtualized networking enviRonments}
\newacronym{DONA}{DONA}{Data-Oriented Networking Architecture}
\newacronym{DoS}{DoS}{Denial of Service}
\newacronym{D-switch}{D-switch}{Dual-Stack switch}
\newacronym{DTN}{DTN}{Delay-Tolerant Networking}
\newacronym{eGW}{eGW}{egress GateWay}
\newacronym{EN}{EN}{End-Node}
\newacronym{FF}{FF}{Forwarding Function}
\newacronym{FIB}{FIB}{Forwarding Information Base}
\newacronym{FIs}{FIs}{Forwarding Identifiers}
\newacronym{FN}{FN}{Forwarding Node}
\newacronym{FNs}{FNs}{Forwarding Nodes}
\newacronym{hICN}{hICN}{hybrid ICN}
\newacronym{HTTP}{HTTP}{HyperText Transfer Protocol}
\newacronym{ICN}{ICN}{Information-Centric Networking}
\newacronym{ICN BGW}{ICN BGW}{ICN Border GateWay}
\newacronym{iGW}{iGW}{ingress GateWay}
\newacronym{IN}{IN}{Internal-Node}
\newacronym{IoT}{IoT}{Internet of Things}
\newacronym{IP}{IP}{Internet Protocol}
\newacronym{IPN}{IPN}{Interplanetary Internet}
\newacronym{IPsec}{IPsec}{Internet Protocol Security}
\newacronym{ISPs}{ISPs}{Internet Service Providers}
\newacronym{IT}{IT}{Information Technology}
\newacronym{LAN}{LAN}{Local Area Network}
\newacronym{LTE}{LTE}{Long Term Evolution}
\newacronym{MAC}{MAC}{Media Access Control}
\newacronym{MANO}{MANO}{MANagement and Orchestration}
\newacronym{NAP}{NAP}{Network Attachment Point}
\newacronym{NAT}{NAT}{Network Address Translation}
\newacronym{NDO}{NDO}{Named Data Object}
\newacronym{NDN}{NDN}{Named-Data Networking}
\newacronym{NetInf}{NetInf}{Network of Information}
\newacronym{NSF}{NSF}{National Science Foundation}
\newacronym{NSN}{NSN}{Name-System-Node}
\newacronym{NFV}{NFV}{Network Functions Virtualization}
\newacronym{NRS}{NRS}{Name Resolution Service}
\newacronym{O-ICN}{O-ICN}{Overlay for Information-Centric Networking}
\newacronym{OICNSIM}{OICNSIM}{Overlay ICN simulator}
\newacronym{OSI}{OSI}{Open Systems Interconnection}
\newacronym{PIT}{PIT}{Pending Interest Table}
\newacronym{P2P}{P2P}{Point to Point}
\newacronym{PoA}{PoA}{Point of Attachment}
\newacronym{POINT}{POINT}{iP Over IcN- the betTer IP}
\newacronym{PoP}{PoP}{Points-of-Presence}
\newacronym{PURSUIT}{PURSUIT}{Publish-Subscribe Internet Technologies}
\newacronym{PARC}{PARC}{Palo Alto Research Center}
\newacronym{PSIRP}{PSIRP}{Publish-Subscribe Internet Routing Paradigm}
\newacronym{QoE}{QoE}{Quality of Experience}
\newacronym{RF}{RF}{Rendezvous Function}
\newacronym{RI}{RI}{Rendezvous Identifier}
\newacronym{RID}{RID}{Routing IDentifier}
\newacronym{RIFE}{RIFE}{architectuRe for an Internet For Everybody}
\newacronym{RINA}{RINA}{Recursive INternetwork Architecture}
\newacronym{RP}{RP}{Rendezvous Point}
\newacronym{SDN}{SDN}{Software-Defined Networking}
\newacronym{SI}{SI}{Scope Identifier}
\newacronym{SN}{SN}{Serving-Node}
\newacronym{FP7}{FP7}{Seventh Framework Program}
\newacronym{TCP}{TCP}{Transmission Control Protocol}
\newacronym{TF}{TF}{Topology Function}
\newacronym{TLS}{TLS}{Transport Layer Security}
\newacronym{TM}{TM}{Topology Manager}
\newacronym{TV}{TV}{Television}
\newacronym{UDP}{UDP}{User Datagram Protocol}
\newacronym{URL}{URL}{Uniform Resource Locator}
\newacronym{VNF}{VNF}{Virtualized Network Function}
\newacronym{VoD}{VoD}{Video on Demand}
\newacronym{VPN}{VPN}{Virtual Private Network}
\newacronym{WLAN}{WLAN}{Wireless Local Area Network}

%% file: introduction.tex
\section{Introduction}
\label{introduction}

The current Internet architecture was designed for a small research community over three decades ago with the purpose of interconnecting multiple heterogeneous networks. At that time, nobody foresaw the popularity and longevity that the Internet architecture started gaining in late `80s and early `90s and that led towards the connection of over 3 billion of mobile and desktop devices. Today, people exploit networking devices for a variety of purposes, that go from simple web browsing to video conferencing and content distribution, with the expectation of being always connected, regardless of their time and place. \review{The misalignment between the original design and the current usage highlighted the limitations of the \gls{IP}-based architecture and motivated researchers to explore new solutions to overcome them.} Among those limitations, the primary concern is the performance of the current Internet, which has to cope with the huge number of connected devices all over the world and with the new pattern of use of the network. According to this study~\cite{numberConnDevices}, currently there are around 23 billions of connected devices in the world, each one identified by a unique \gls{IP} address and consuming the network bandwidth. With such a huge number of devices, the first issue is the availability of unique \gls{IP} addresses to be assigned. Even though researchers originally chose to allocate 32 bits to compose an \gls{IP} address through the IPv4 protocol, they had to introduce the IPv6 protocol to extend the number of allocated bits from 32 to 128. \review{\gls{NAT}~\cite{RFC3022} is also another solution addressing the same problem, and it allows to assign the same public address to a set of devices belonging to the same private network. Thus, when using the private network each device has its own \gls{IP} address, chosen within a range of private \gls{IP} addresses, but, for an entry external to the network, all the devices have the same public \gls{IP} address. To enable the communication between the private network and the Internet a firewall is responsible for intercepting a request, forwarding it to the Internet with the public \gls{IP} and redirecting the incoming response to the appropriate device.}

Another problem is given by the type of network traffic: most of it is made of \gls{HTTP} requests, which means that users have changed the way they use Internet from a low-bandwidth interactive and store-and-forward approach towards a web and content dominated traffic. \review{To support this, Cisco Visual Networking Index~\cite{cisco2021} shows that in recent years video traffic delivery has suddenly become very popular on the Internet, with an Internet traffic that will be 194 exabytes per month by 2021, and multimedia traffic up to 82\%, from 70\% in 2015. Furthermore, due to the technological advancements in hardware devices and an increasing deployment of pervasive computing application, it is indicated that the number of communicating devices (including smart devices) will be three-times more than the world's population~\cite{zetta}. Moreover, it has also been reported~\cite{what2016} that 86\% of worldwide user traffic consists of only video data, which consists of \gls{VoD}, video streaming, \gls{P2P}, and \gls{TV}.} 

Finally, from a security and privacy point of view the current Internet is not even able to guarantee some essential requirements, such as origin authentication, data integrity or data confidentiality, because of its lack of security by design. This is the motivation for the introduction of solutions, such as \review{\gls{IPsec} suite~\cite{RFC4301} or \gls{TLS}~\cite{RFC8446}}, that work on top of the current Internet and are aimed at overcoming its limitations. 

For the above-mentioned reasons, researchers started designing new Internet architectures \review{(e.g., \gls{RINA}~\cite{rina}, \gls{ICN}~\cite{ccnx-1.0})}, that might replace the current one in the future. Among those, the most promising \review{architectures} adhere to the \gls{ICN} paradigm: a new network communication model in which the traditional host-centric paradigm has been moved to the new information-centric networking. While in the current Internet two endpoints can start communicating only if they know the respective \gls{IP} address, explicitly or by use of a \gls{DNS}, in \gls{ICN} they can send requests specifying only content names, without being aware of contents location in the network. This decoupling between request sending and content transferring introduces several benefits: reduction of latency and network load due to in-network caching \cite{Diallo2011LeveragingCF,TANG2019590,7467400,8057300}, inherent content integrity \cite{8539022} and better support for mobility due to name-based routing \cite{Anastasiades2014,8303694}. 

\review{The ongoing research shows that the inherent benefits of \gls{ICN} (e.g., fast, efficient, and secure data delivery, improved reliability) make \gls{ICN} a suitable networking model for various emerging technologies, such as \gls{IoT}~\cite{8478349, NOUR201995} and 5G~\cite{8303694, 8263145}. In the first scenario, \gls{ICN} can help with establishing the connectivity among smart devices in an IoT environment, as well as in a smart city, in a smart e-health, and in a smart grid context. Also, the management of the huge amount of data generated by \gls{IoT} devices (i.e., the \gls{IoT} big data) is challenging in the existing \gls{IP} architecture, while it is minimized by the in-network caching feature in \gls{ICN}. This feature allows to reduce the traffic load on data producers by caching data on intermediate routers. Additionally, the receiver-driven communication in \gls{ICN} allows \gls{IoT}-receivers to ask for data without revealing their location information, thus being privacy supporting. Similarly, there are various advantages coming up from an \gls{ICN}-based 5G architecture (i.e., 5G-\gls{ICN}): (i) 5G-\gls{ICN} provides a single protocol able to handle mobility and security, instead of using a diverse set of \gls{IP}-based \gls{3GPP} protocols (such as in the case of existing mobile networks, e.g., \gls{LTE}, 3G, 4G), (ii) it provides a unifying platform with the same layer-3 \gls{APIs} to integrate heterogeneous radios (e.g., Wifi, \gls{LTE}, 3G) and wired interfaces in the same network, (iii) it converges services like computing, storage, and networking over a single platform, which enhances the flexibility of enabling virtualized service logic and caching functions anywhere in the network.} 

\review{Due to the several advantages and the various potential next-generation applications, \gls{ICN} is gaining significant attention from both Industry and Academia~\cite{Kumar2019, 8624408}: the authors in~\cite{8027034} provide an in-depth study of the state-of-the-art techniques by focusing on security, privacy, and access control aspects of \gls{ICN} architectures; in~\cite{8240926}, the authors present a survey on \gls{ICN} cache management strategies, along with benefits and limitations; the authors in~\cite{8303694} focus on the state-of-the-art techniques proposed to achieve mobile \gls{ICN}. However, none of those survey articles discuss the research issues and challenges affecting an \gls{ICN}-\gls{IP} coexistence scenario, as we aim to do in this paper. Only in~\cite{RFC}, researchers from InterDigital Inc. and Huawei provided a comparison among the existing coexistence architectures, but they focused specifically on the different deployment approaches chosen by each solution.}

\textbf{Motivation.} The benefits of \gls{ICN} can occur only in a full-\gls{ICN} scenario, which implies a complete replacement of the current Internet. Despite its obvious need, this is a long and complex process, that requires the coordination among the different parties (i.e., \gls{ISPs}), time, costs for updating hardware and software of the network components and ability to face all the new possible challenges. Previous attempts to replace a widely used technology, protocol or architecture (e.g., IPv4/IPv6 protocol, 3G/4G technology, 4G/5G technology) have always faced a long period of coexistence between the old and the new solution. In the same way, the replacement of the current Internet will involve a transition phase during which \gls{IP} and \gls{ICN} architectures will coexist. More specifically, we envision that in a coexistence scenario there will be \gls{ICN} and \gls{IP} ``islands'' surrounded by an \gls{IP} or an \gls{ICN} ``ocean'', where an ``island'' will be a single device, a computer, an application or a server running either the \gls{ICN} or the \gls{IP} protocol, while an ``ocean'' will be a network containing components, that run different architectures.

Researchers working in this field have already addressed the coexistence of \gls{IP} and \gls{ICN} following two separate approaches. In the first, the research groups designed future Internet architectures facing the coexistence only during the deployment of their testbeds and without considering it as part of the initial design. On the contrary, in the second case, the design of the future Internet architectures specifically addressed the coexistence of \gls{IP} and \gls{ICN}. 

All the existing networking solutions that consider the coexistence are affected by a strong limitation: the lack of a comprehensive approach in addressing the coexistence. The purpose of those solutions is to improve a network performance indicator, without considering all the issues that arise in a coexistence scenarios, especially those regarding the security and privacy of the end users. To design the first complete coexistence architecture, it is necessary first to have a comprehensive overview of strengths and weaknesses of the existing solutions. 

\textbf{Contribution.} The purpose of this paper is to provide the first complete survey and classification of the existing coexistence solutions. Details of \gls{ICN} and of its working methodology are out of scope for this paper, since there are already several surveys addressing this aim~\cite{8303694, 8027034, 8240926}. Overall, the contributions of this paper are as follows: 
\begin{enumerate}
    \item We define a set of relevant features necessary for comprehensively analyze a coexistence architecture.
    \item We provide the first comprehensive classification of all the main coexistence solutions.
    \item We discuss the open issues and challenges affecting the existing coexistence architectures, by providing possible insights to design a more reliable future Internet architecture. 
\end{enumerate}

\textbf{Organization.} The paper is organized as follows: in Section~\ref{background}, we introduce the ICN concept, by comparing it with the current \gls{IP} architecture and by illustrating its main benefits; Section~\ref{classification_criteria} describes all the criteria we identified and used for the analysis and classification of the coexistence architectures; in Section~\ref{coexistence_architectures} we deeply illustrate each coexistence architecture and provide the motivation for our classification; in Section~\ref{discussion_conclusion}, we discuss the main strengths and limitations of the current coexistence architectures, providing insights for improving the design of the future Internet; finally, in Section~\ref{conclude} we conclude the paper.

%% file: background.tex
\section{Background}
\label{background}
The purpose of this section is to provide an overview of the \gls{ICN} paradigm (Section~\ref{ICN_intro}), a comparison of the main features of \gls{IP} and \gls{ICN} architectures (Section~\ref{features}), the benefits of \gls{ICN} (Section~\ref{icn_benefits}) and, finally, the emerging technologies (Section~\ref{emerging_technologies}). 

\subsection{\review{\gls{ICN} Overview}}
\label{ICN_intro}
%\comment{This paragraph does not add any new information. I would remove it. -- The fundamental idea behind ICN paradigm is that \textit{who} is communicating is far less important in communication than \textit{what} data is required. This major shift in communication paradigm has occurred due to the end-users usage requirements from today’s Internet, which indeed is more content-centric than location-centric, e.g., video sharing, social networking, and retrieving aggregated data. ICN style fundamentally decouples the content from its sources, through a clear location-identity split. The methodological assumption behind is that content should be named, directed, and matched independently of its location since it may be present anywhere in the network. Therefore, instead of specifying a source-to-destination host pair for communication, ICN names the piece of content, making content (data) as a first-class entity.}

\review{The \gls{ICN} concept was first implemented in 2001 in the TRIAD project~\cite{Cheriton00triad:a}, by introducing a new \textit{content layer} in the \gls{IP} communication model. This layer provided several content-based features, among which: hierarchical content caching, content replication and content discovery, multicast-based content distribution, and name-based routing. Moreover, the layer supported end-to-end communication based on content name and \gls{URL} by relying on \gls{IP} addresses only to reduce the role of transient routing tags. Although TRIAD routing mechanism used content names instead of \gls{IP} addresses, the \gls{TCP} and the \gls{IP} protocols were still the backbone of the proposed architecture. In 2006, UC Berkeley and ICSI proposed the \gls{DONA}~\cite{Koponen:2007:DNA:1282380.1282402}, which improved TRIAD by incorporating data authenticity and persistence as key objectives of the architecture, but still having a strong dependency on the underlying \gls{TCP}/\gls{IP}. In 2009, the \gls{PARC} revealed the \gls{CCN}~\cite{Jacobson} project. Soon after, the \gls{NSF} introduced its ``Future Internet Architecture'' program, which paved the way for \gls{NDN}~\cite{Zhang} - a branch of the \gls{CCN} project. Both \gls{CCN} and \gls{NDN} significantly moved the TRIAD and \gls{DONA} projects forward, by introducing a new network layer to definitely replace the existing \gls{TCP} and \gls{IP} ones. Thus, \gls{CCN} and \gls{NDN} are considered two key projects due to the considerable attention they brought to the \gls{ICN} paradigm from both Academia and Industry, influencing also the design of the \gls{ICN} architecture~\cite{ICNRG}. 
}

\subsection{\review{Comparison Between \gls{IP}-based and \gls{ICN}-based Internet Architectures}}
\label{features}
    Originally developed as part of the ARPANET project~\cite{mcquillan1980new} during the 1960s, the current Internet is now often referred as \gls{TCP}/\gls{IP} architecture due to its most well-known protocols (i.e., \gls{TCP} and \gls{IP}). On the contrary, the \gls{ICN} paradigm was first introduced in the TRIAD project~\cite{Cheriton00triad:a} in 2001 and, then, followed by several architectures adhering to its new communication model. \review{Since \gls{ICN} is a paradigm, we will consider here the five main architectures to describe the technical features of the future Internet, while we will provide a comprehensive description of all the architectures addressing the \gls{ICN}-\gls{IP} coexistence in Section~\ref{coexistence_architectures}}: (i)~the \gls{DONA} architecture~\cite{Koponen:2007:DNA:1282380.1282402}, (ii)~the \gls{CCN} architecture~\cite{Jacobson}, (iii)~the \gls{NDN} architecture~\cite{Zhang}, (iv)~the \gls{PURSUIT} architecture~\cite{Dimitrov:2010:PPP:1839379.1839409}, and (v)~the \gls{NetInf} architecture~\cite{Dannewitz:2013:NII:2459510.2459643}. 

\begin{figure}[h]
	\centering
	\includegraphics[width=0.4\textwidth]{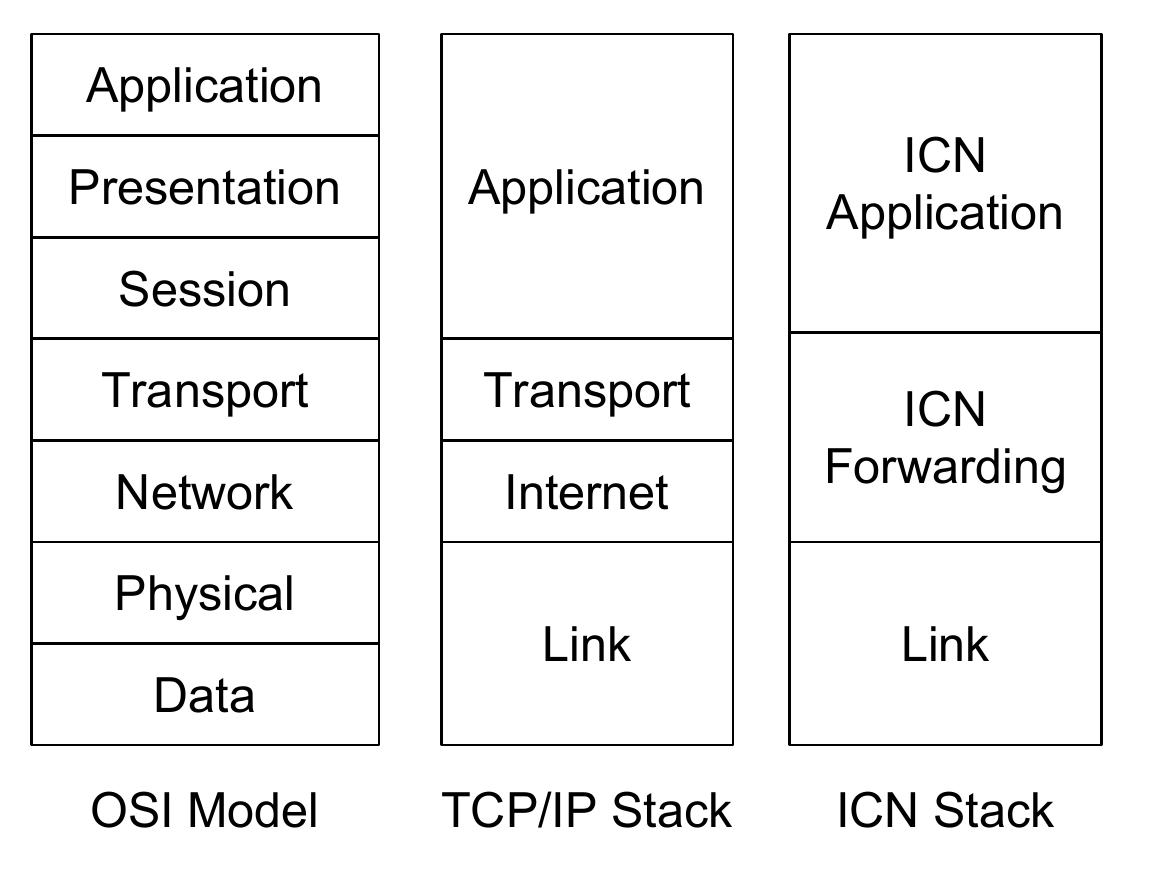}
	\caption{Adaptation of the \gls{OSI} seven layer model in the \gls{TCP}/\gls{IP} and \gls{ICN} protocol stacks.}
	\label{fig:stacks}
\end{figure}

\textbf{Protocol Stack.} Both \gls{TCP}/\gls{IP} and \gls{ICN} rely on a layered protocol stack, which is comparable to the \gls{OSI} Reference Model~\cite{zimmermann1980osi}, as shown in \review{Fig.}~\ref{fig:stacks}. 

The \gls{TCP}/\gls{IP} stack includes the following four layers\review{~\cite{TCPstack}}: 
\begin{itemize}
    \item \emph{Application} - it combines the functionality of the \emph{Application}, \emph{Presentation} and \emph{Session} layers of the \gls{OSI} model. It is responsible for sending and receiving data and it is specific for a particular type of application (e.g., \gls{DNS}, \gls{HTTP}).
    \item \emph{Transport} - it targets the \emph{Transport} layer of the \gls{OSI} model and it is responsible for the end-to-end data transfer and data streams. Its most important protocols are \gls{TCP}, which provides a reliable and connection-oriented service, and \gls{UDP}, which offers an unreliable and connection-less service.
    \item \emph{Internet} - equivalent to the \emph{Network} layer of the \gls{OSI} model, it provides addressing and routing functionalities to ensure the delivery of messages to their destination. \gls{IP} is the most important protocol, but it does not provide flow control or error handling.
    \item \emph{Link} - equivalent to the \emph{Data} and \emph{Physical} layers of the \gls{OSI} model, it manages the interaction among physical network components and it works as an interface with the network hardware.  
\end{itemize}

Since the \gls{ICN} stack is an evolution of the \gls{TCP}/\gls{IP} one\review{~\cite{icnstack1,icnstack2, point}}, each layer is described with respect to the corresponding one in the Internet stack. More specifically, the layers of the \gls{ICN} stack are the following ones: 
\begin{itemize}
    \item \emph{\gls{ICN} Application} - the protocols of this layer address content names instead of hosts locations. For example, the \gls{URL} inside an \gls{HTTP} request is replaced with the complete name of a content.
    \item \emph{\gls{ICN} Forwarding} - for any \gls{ICN}-compliant architecture this layer offers \review{routing functionalities for \gls{ICN} interest and data packets equivalent to} the  \gls{TCP}/\gls{IP} \emph{Network} layer in such a way that source and destination \gls{IP} addresses are removed from the network packets and only the addressed content name is declared. According to the specific architecture, this layer can also provide the features of the \gls{TCP}/\gls{IP} \emph{Transport} layer. In that case, the Interest/Data messages replace the \gls{TCP}/\gls{IP} segment/\gls{ACK} messages and the content requester becomes responsible for the message sending rate in place of the content source (producer or intermediate router).
    \item \emph{Link} - to be \gls{ICN}-compliant, this layer introduces a mapping between \gls{MAC} addresses and content names.
\end{itemize}

\textbf{Routing.} The purpose of the routing functionality is to route network packets from the source node till the destination node on one way and, then, from the destination to the source on the other. 

Each \gls{TCP}/\gls{IP} packet specifies both source and destination nodes by including their \gls{IP} addresses. An \gls{IP} address is the unique identifier of each network component and it contains both the address of the network and the address of the specific component within that network. In the current Internet, routers are the main responsible for the routing functionality. Equipped with at least two \gls{IP} interfaces (i.e., an incoming and an outgoing one), each router receives \gls{IP} packets in the incoming interface and checks whether there is a match, based on the longest prefix, in its \gls{FIB} internal data structure. The \gls{FIB} contains a mapping between a network prefix and a router's outgoing interface, together with the next-hop \gls{IP} address. If there is a match in the \gls{FIB} for the incoming packet, this is forwarded through the outgoing interface towards the next node in the network. 

In \review{ICN}, the routing functionality differs according to the specific design of each architecture, but they all have a common design choice: the packets sent by a requester contain only the full name of the content and no \gls{IP} addresses, neither the content requester's one nor the content source's one. In \gls{NDN} and \gls{CCN} architectures, contents are expressed through hierarchical names and routers use a longest-prefix match approach to find a possible entry in their \gls{FIB}, which returns the name-prefix/prefixes of the next node/nodes in the network. On the contrary, \gls{DONA} exploits a flat naming scheme to point to the contents available in the network and a name-based routing to redirect the packets until they reach the content source. A different approach is used by \gls{PURSUIT}, which relies on a publish/subscribe model. Publishers publish their contents in the network and subscribers ask for a specific content by using a flat name scheme, made of two components: the \gls{RI} and the \gls{SI}. The first element addresses the component responsible to find the match between publisher and subscriber for a specific content, while the second is used to identify the sub-network where the rendezvous is. Once the subscriber obtains the location of the publisher from the rendezvous node, it sends its packet to the \gls{TM} of the network where the content publisher is. The \gls{TM}, then, identifies the path from the publisher to the subscriber and adds a series of \gls{FIs} to the header of the packets. After that, the \gls{FNs} forward the packets only by using the \gls{FIs}, without any routing table. Finally, the \gls{NetInf} architecture adheres to both the approaches: name resolution, based on the publish/subscribe paradigm, and name-based routing. 

\textbf{Name Resolution.} In the \gls{TCP}/\gls{IP} architecture there is a dedicated network component responsible for the name resolution, which is the \gls{DNS}. This is a distributed service, which translates domain names, expressed in hierarchical \gls{URL}s, into the corresponding \gls{IP} addresses. The Internet is organized into separate \gls{DNS} zones, each one under the direct control of an authoritative \gls{DNS} server, and everytime a network device sends a request to its local \gls{DNS} server, this might reply with a value saved in its cache or, otherwise, forward the same request to a remote server. 

In \gls{ICN}, the name resolution differs according to the chosen forwarding approach. In case of name-based routing, the requester specifies a content by providing its full name, which is the same analyzed by the routers to find the next hop in the network. \review{On the other hand}, in the name resolution approach, used by \gls{PURSUIT} or \gls{NetInf}, there is always a dedicated node in the network, which is responsible for the mapping between publishers and subscribers. 

\textbf{Storing.} In the \gls{TCP}/\gls{IP} architecture, \review{routers do not have caching features, while in \gls{ICN}, caching is fundamental and almost any node is able to cache contents and to serve the corresponding requests.}

\textbf{Traffic Management.} \review{In the current Internet, the traffic management, in terms of connection management, flow control and congestion control, is guaranteed by the \gls{TCP} protocol. The establishment of a connection is regulated by the three-way handshake mechanism, through which the \gls{TCP} protocol checks for the availability of the remote server, before exchanging any data with it. Only at the end of the handshake, the real communication starts, together with the data exchange, and it is regulated by the introduction of sequence numbers in the message blocks that enable the destination node to properly order all the received messages. The flow control is provided by the \gls{ACK} messages received by the sender from the receiver every time a packet has been properly delivered. Thus, a sender never overflows the receiving host because the re-transmission of a packet is performed only after a timeout, which corresponds to either an \gls{ACK} not received by the sender or three \gls{ACK}s received. Finally, the congestion control refers to the prevention of the routers from becoming overflowed.}

In \gls{ICN}, some architectures, such as \gls{DONA}, still rely on the existing transport protocols so that all the forwarding mechanisms and transport functionalities are guaranteed. However, other \gls{ICN} solutions, such as \gls{NDN}, do not provide the \emph{Transport} layer functionalities and, instead, delegate them to the application itself or to the network packets. After a certain timeout, an application can transmit again a packet, which by design has a limited lifetime to prevent network congestion. Moreover, the availability of distributed caches, which means contents, all over the network should prevent losses due to congestion. 

\review{\subsection{Benefits of ICN-based architectures}
\label{icn_benefits}
The following ones are the key \gls{ICN} benefits, which better motivate why this architecture is a potential candidate for the future Internet. 
\subsubsection{Scalable and Cost-Efficient Content Distribution}
In a future world where the mobile video traffic will be dominant (e.g., video data will consume more than 80\% of the \gls{IP} traffic, wireless mobile devices will generate two-third of the Internet traffic~\cite{cisco}, Netflix and YouTube together amount nearly 50\% of Internet traffic), the current network operators will face challenges in meeting the bandwidth requirements from end users. Thus, the inherent \gls{ICN} support for caching at the network layer~\cite{Jacobson}, together with the receiver-driven mechanism, the inherent support for mobility and the multi-cast routing, make \gls{ICN} fit the new network use in a multimedia streaming context~\cite{dashovericn,6649319,dashoverccn,7169859,ndnavs,Carofiglio}.
}
\review{\subsubsection{Mobility and Multihoming}
\gls{ICN} also meets the requirements of the 5G network, such as global Internet access and user mobility over dense and heterogeneous networks by adapting to multiple radio access technologies (e.g., Wi-Fi and \gls{LTE}). As a matter of fact, \gls{ICN} supports the mobility at the network layer by decoupling time and space between request resolution and content transfer~\cite{8303694}. In particular, two fundamental \gls{ICN} features encourage seamless consumer mobility~\cite{Anastasiades2014,8303694}. The first is the receiver-driven communication model, where it is up to the consumer to request location-independent contents. The second is the connection-less request/response communication model between consumer and producer. Therefore, when a mobile consumer connects to a new \gls{PoA}, the above two features allow the consumer to re-issue interests for the data that he has not received from the previous \gls{PoA}. On the contrary, producer mobility is more challenging in \gls{ICN} because of no distinction between routing locator and content identifier. Previous work have already proposed new solutions for an efficient management of producer mobility in \gls{ICN}~\cite{7562050,Anastasiades2014}.}
\review{\subsubsection{Disruption Tolerance}
Achieving an end-to-end communication through \gls{TCP}/\gls{IP} transport sessions in challenged networks is often difficult due to the sparse connectivity, high-speed mobility, and disruptions of such networks. Since the application protocol sessions are bound to transport sessions, the communication fails as soon as the transport session fails. In the current Internet, several applications do not require seamless communication with end-to-end paths~\cite{Ott2004WhyST}. As the primary objective is to access data objects, \gls{ICN} is the perfect approach for \gls{DTN} architectures~\cite{Fall:2003:DNA:863955.863960,rfc4838} due to the in-network caching with hop-by-hop transport functionality, which provides a store-and-forward mechanism and enables a better performance and reliability.
}
\review{
\subsubsection{Security}
Unlike the \gls{TCP}/\gls{IP} architecture, the \gls{ICN} design comes with the security in mind. In particular, in \gls{ICN} the security follows a data-centric model, which focuses on the importance of guaranteeing content integrity and source authentication. For a content-centric architecture, where contents can be located and provided in any point of the network, and not only by the original content producer, the above-mentioned features are particularly significant. To achieve this aim, \gls{ICN} contents are always signed by the producer, thus allowing consumers to always verify content integrity and data-origin authentication~\cite{Compagno2018}.}

\subsection{Emerging Technologies}
\label{emerging_technologies}
Before thinking of redesigning the whole Internet architecture, researchers and companies have provided several solutions, which work on top of the current Internet, to overcome some of its limitations. Among those, the most successful attempts are the following emerging architectures: \gls{SDN}, \gls{NFV}, \gls{CDN} and \gls{DTN}. 

\subsubsection{Software-Defined Networking} 
SDN~\cite{farhady2015software} is an emerging networking paradigm that separates network control logic (i.e., the control plane) from the underlying switches and routers that forward the traffic (i.e., the data plane). By separating the control and data planes, the network switching/routing devices become simple forwarding devices and the control logic is incorporated in a logically centralized controller. This separation primarily helps in simplifying network (re)configuration, policy enforcement, and evolution~\cite{kreutz2015software}. The control plane and the data plane communicate via a well-defined programming interface, i.e., the forwarding elements of the data plane request for instructions from the controller as well as the controller has direct control over the data plane elements using \gls{APIs}. The most popular flavor of such \gls{APIs} is OpenFlow~\cite{mckeown2008openflow}. An OpenFlow switch has one or more flow tables for handling packet-rules. When a rule matches with the incoming traffic, the OpenFlow switch performs certain actions (forwarding, modifying, dropping, etc.) on the traffic flow. The rules installed by the controller decide the role of an OpenFlow switch, i.e., it can behave as a switch, router, firewall, or middlebox (such as traffic-shaper, load-balancer).

\subsubsection{Network Functions Virtualization} Diversity and dominance of proprietary appliances made service deployment, as well as testing, complex. \gls{NFV}~\cite{li2015software} was designed as a technology to leverage \gls{IT} virtualization by exporting network functions from the underlying dedicated hardware equipment to general software running on \gls{COTS} devices. Using \gls{NFV}, the key network functions can be performed at various network locations, e.g., network nodes, data-centers, network edge, as required. \gls{NFV} is different from \gls{SDN}, and it only deals with the virtualization of network functions.

\subsubsection{Content Delivery Network}
%CDN architectures~\cite{1250586,6674399,8046000} are one of the most essential components of the existing Internet \cite{cisco,6688724,7948965} and are aimed at improving bandwidth availability, accessibility and precise content delivery through content replication. Although Internet straightforward foundation has scaled substantially well, still its end-to-end,  \textit{“best effort”}  layout   is premised on the communication paradigm which grounds on passive traffic management \cite{1250586}.  Therefore, in order to offer improved support for delivering commercial content  which   was  earlier provided by  “best effort” design of Internet transport services~\cite{STOCKER20171003}, CDNs now appeared as  an overlay network on the Internet~\cite{Clark2005TheGO, Medagliani2017OverlayRF}.
The initial implementation of the Internet was designed to manage the traffic in a passive, end-to-end, and ``best effort'' approach~\cite{1250586}. With the explosion of user data and commercial content over the Internet, the ``best effort'' approach for traffic management became inefficient and unscalable. To handle this situation, \gls{CDN}~\cite{1250586, 6674399, 8046000} was designed~\cite{cisco,6688724,7948965}. Nowadays, \gls{CDN} appears as an integral and essential overlay network for the Internet~\cite{STOCKER20171003, Clark2005TheGO, Medagliani2017OverlayRF} since it primarily aims to improve bandwidth availability, accessibility, and precise content delivery through content replication.
\par
\gls{CDN} architecture consists of several cache servers that are strategically located across the Internet. Typically, \gls{CDN} holds a hierarchy of servers with multiple \gls{PoP} that stores copies of identical content to satisfy user's demand from most appropriate/closest site~\cite{NivenJenkins2012ContentDN}. It also has back-end servers for intra-\gls{CDN} content distribution. \gls{CDN} categorically distributes web contents to the cache servers, which are positioned close to the users. As a result, \gls{CDN} offers fast, efficient, and reliable web services to the users.
\par
There are two fundamental approaches for the deployment of \gls{CDN}: (i)~overlay model, where content is replicated to thousand of servers worldwide, and (ii)~network model, where routing configurations recognize the application services and forward them based on the predefined policies.
\par
Even though \gls{CDN}s improve content delivery, their performance is limited by the underlaying \gls{ISPs}. Usually, \gls{CDN}s do not manage independent packet data services, rather they rely on the \gls{ISPs} to make packet routing decisions. Moreover, both \gls{ISPs} and {CDN} collectively provide end-to-end \gls{QoE}\footnote{QoE is an all-inclusive model, which defines the quality perceived by a user when retrieving content or applications over the Internet.} for content delivery. Thus, coordination between \gls{ISPs} and \gls{CDN} providers causes a massive impact on the overall \gls{QoE}~\cite{ STOCKER20171003}.

\subsubsection{\glsentrylong{DTN}} \review{In the late 1990s, the widespread use of wireless protocols, together with an increasing interest in vehicular communication, encouraged researchers to design the \gls{IPN} architecture. This was the first attempt to address the need of long distance communications that were inevitably affected by packets loss/corruption and delays. \gls{DTN}~\cite{5770277} was first introduced as an adaptation of the \gls{IPN} for terrestrial networks~\cite{DTNstory}: it is an overlay architecture that operates above the protocol stack of \textit{ad-hoc} wireless networks and enables gateway functionality to interconnect them. To provide communication among networks having excessive delays due to highly repetitive link disruptions, \gls{DTN} adopts the ``store-carry-forward'' routing scheme~\cite{storecarryforward}: the main idea of this scheme is to have multiple nodes distributed over the network, each one able to receive a copy of the same message and then send it back to the destination node. This way, the delivery performance is improved and the destination node can receive the message from any location inside the network.}

%% file: classification_criteria.tex
%%%%%%%%%%%%%%%%%%%%%%%%%%%%%%%%%%%%%%%%%%%%%%%%%%%%%%%%%%%%%%%%%%%%%%%%%%%%%%%%%%%%%%%%%%%%%%%%%%%%%%%%%%%%%%%%%%%%%%%
\begin{figure*}[!b]
	\centering
	\includegraphics[width=1.5\columnwidth]{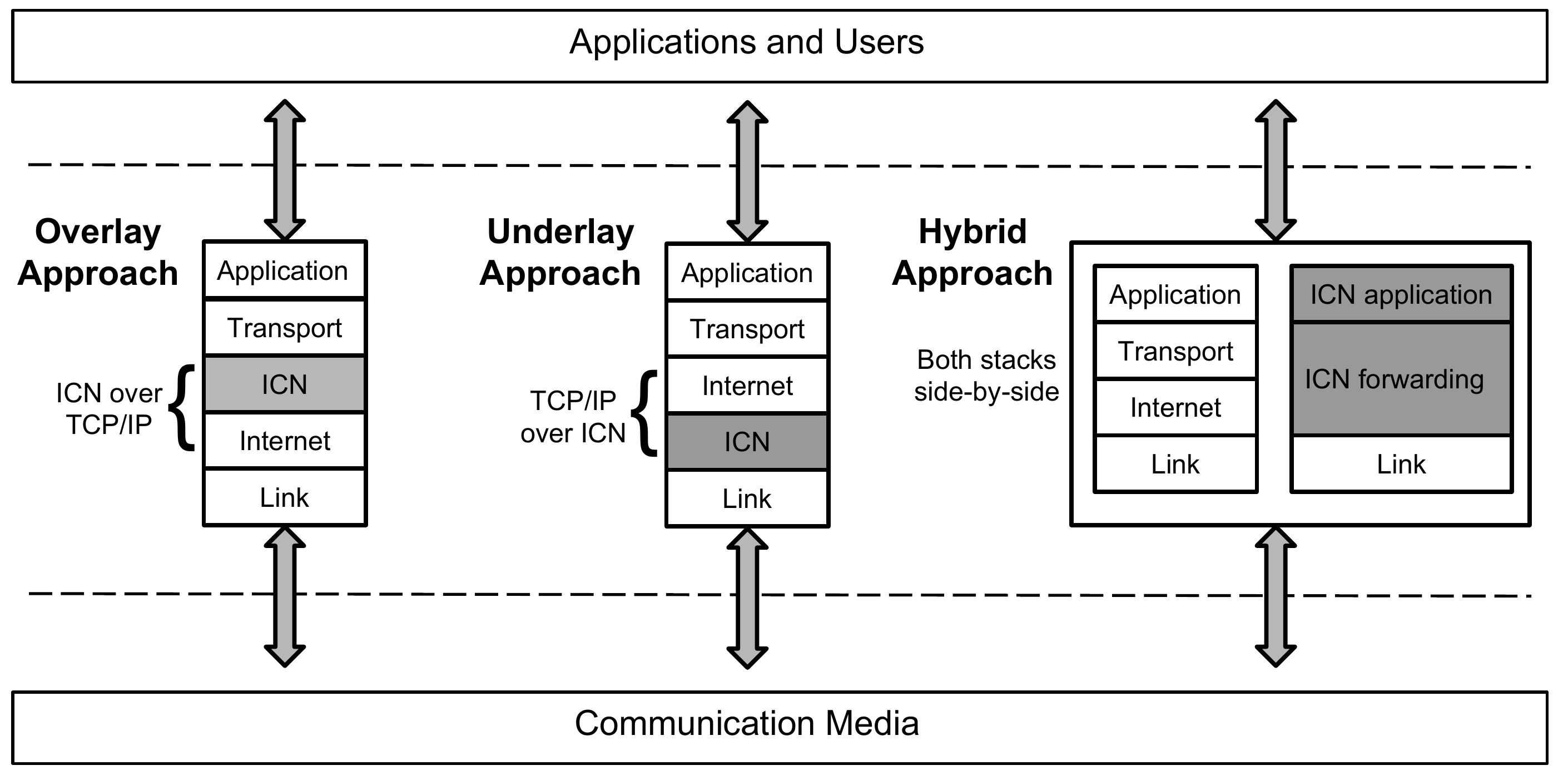}
	\caption{Deployment approaches of \gls{ICN} into the \gls{TCP}/\gls{IP} architecture.}
	\label{fig:Category}
\end{figure*}

\section{Coexistence Architectures: Features and Evaluation Parameters}
\label{classification_criteria}
In order to classify the existing architectures, we identified the necessary features and evaluation parameters to have a complete overview of each coexistence solution. The former come with the design of a coexistence architecture, while the latter refer to the challenges introduced during its deployment in a real scenario. The features are as follows: \emph{deployment approaches}, \emph{deployment scenarios}, \emph{addressed coexistence requirements} and \emph{\review{additional architecture or technology used}}. \review{On the other side}, the evaluation parameters are: \emph{traffic management}, \emph{access control}, \emph{scalability}, \emph{dynamic network management} and \emph{latency}. 
In the remaining part of this section, we will describe features (Section~\ref{c_features}) and evaluation parameters (Section~\ref{evaluation_parameters}) used for analyzing each coexistence architecture. 

%%%%%%%%%%%%%%%%%%%%%%%%%%%%%%%%%%%%%%%%%%%%%%%%%%%%%%%%%%%%%%%%%%%%%%%%%%%%%%%%%%%%%%%%%%%%%%%%%%%%%%%%%%%%%%%%%%%%%%%
\subsection{Features}
\label{c_features}
%%%%%%%%%%%%%%%%%%%%%%%%%%%%%%%%%%%%%%%%%
\subsubsection{Deployment Approaches}
The deployment of \gls{ICN} into the \gls{TCP}/\gls{IP} architecture inevitably \review{raises} the following question: \emph{How to introduce the \gls{ICN} protocol into the \gls{TCP}/\gls{IP} protocol?} To achieve this aim, researchers identified three possible approaches, shown in \review{Fig.}~\ref{fig:Category}: \emph{overlay} in case of \gls{ICN} running on top of the \gls{IP} protocol, \emph{underlay} in case of \gls{ICN} running under the \gls{IP} protocol and \emph{hybrid} in case of a coexistence of both \gls{IP} and \gls{ICN} protocols\review{~\cite{RFC}}. In the \emph{overlay} deployment approach, the aim is to enable the communication among several \gls{ICN} ``islands" in an \gls{IP} ``ocean" and is achieved through a tunnel over the Internet protocol. On the contrary, the \emph{underlay} solution involves the introduction of proxies and protocol conversion gateways near to either \gls{ICN} or \gls{IP} ``islands" to properly deliver and receive outgoing and incoming requests. As an example, an \gls{HTTP} request sent to an \gls{ICN}  ``island" is intercepted by a gateway, which is responsible for translating it into an \gls{ICN} Interest. Then, the resulting \gls{ICN} data packet is translated again into an \gls{HTTP} reply sent back to the requester. Finally, the \emph{hybrid} approach claims the coexistence of both \gls{ICN} and \gls{IP}, by adopting dual stack nodes able to handle the semantics of both \gls{IP} and \gls{ICN} packets. Given the diversity of the two protocols, from a semantic and format point of view, a dual stack node can use various options to infer content names from an \gls{IP} packet, such as performing deep packet inspection in the payload or looking into the content name in the \gls{IP} option header. 

        %\todo{traditional IP ones will be the default IP layers; ICN modified ones will be the new ones; in the hybrid the ICN stack is fully red; place a device to make it clear that it will be within the device; red parts should be bold and black; in the bottom "communication media" + in the top "users" + "apps"}

%\begin{figure*}
%	\centering
%	\includegraphics[width=0.9\columnwidth]{images/deployment_approaches.pdf}
%	\caption{Deployment approaches of ICN into the TCP/IP architecture.}
%	\label{fig:Category}
%\end{figure}

%%%%%%%%%%%%%%%%%%%%%%%%%%%%%%%%%%%%%%%%%
\subsubsection{Deployment Scenarios}
\begin{figure}[h]
	\centering
	\includegraphics[width=0.55\textwidth]{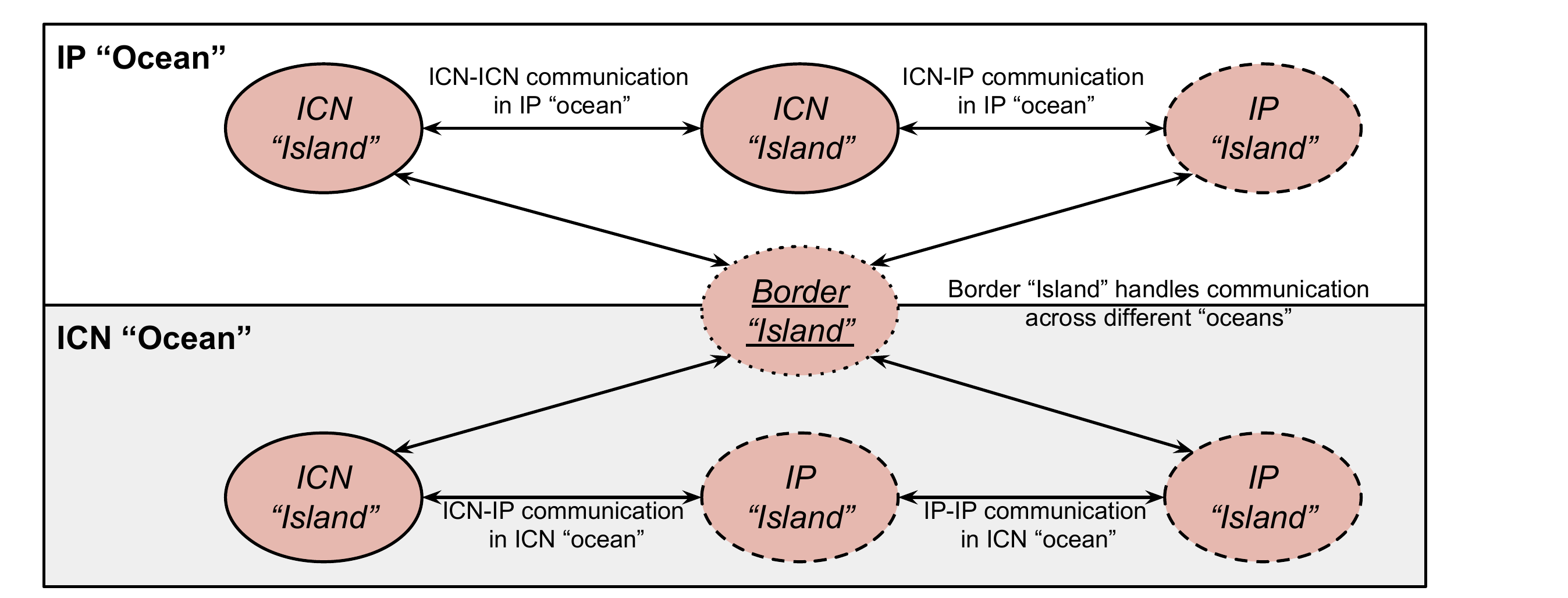}
	\caption{Deployment scenarios for a coexistence architecture.}
	\label{fig:coexistence_overview}
\end{figure}

The purpose of this feature is to analyze all the possible scenarios in which a coexistence architecture can be deployed among the others we identified and that are illustrated in \review{Fig.}~\ref{fig:coexistence_overview}. 

Each deployment scenario involves two ``islands", which run either the same networking architecture or two separate ones, surrounded by an \gls{ICN} or an \gls{IP} ``ocean". 
The possible different deployment scenarios are as follows: 
\begin{itemize}
    \item \emph{ICN-ICN communication in IP ``ocean''}.
    \item \emph{ICN-IP communication in IP ``ocean''}.
    \item \emph{ICN-IP communication in ICN ``ocean''}.
    \item \emph{IP-IP communication in ICN ``ocean''}.
    \item \emph{Border Island} - communication between different ``islands'' in separate ``oceans''. 
\end{itemize}

\subsubsection{Addressed Coexistence Requirements}
In a coexistence scenario, the heterogeneity of the different networks might generate conflicts that prevent each individual architecture from guaranteeing its main features and properties. For example, since most of the \gls{ICN} architectures do not preserve the native transport functionalities provided by the \gls{TCP} protocol of the current Internet, one of their most significant limitations is the traffic management. In a coexistence scenario, there would be a conflict between an \gls{IP} ``island'' implementing its own logic for managing the traffic network and an \gls{ICN} ``island'', which does not support the same features. 

Examining previous works~\cite{ren2015deployment}, we consider the following requirements as the necessary ones to be supported in a coexistence scenario:
\begin{itemize}
    \item \emph{Forwarding} - the network forwarding devices should be able to handle packets with diverse routing identifiers (e.g., the variable-lengths of content names lead to dissimilar size of prefix-set and thus, different forwarding table look-ups).
    \item \emph{Storage} - the network devices should support in-network caching to serve the content request and reduce bandwidth consumption. Nevertheless, the storage capacity of network devices also affects the size of the index table for the cached content and the time required to match the content name in the index table.
    \item \emph{Security} - the network devices should preserve the security policies enforced in one (source) network to another (destination) network such as authenticating the digital signatures of content objects for content-based security or privacy policies.
    \item \emph{Management} - the network devices should support management-related operations such as traffic-shaping/engineering, load-balancing, and explicit path steering.
\end{itemize}

%\textbf{Interoperability.} The network devices should support the interoperability among the different ICN architectures proposed, which are substantially different from each other (e.g., the edge routers must translate packet headers from overlay-to-underlay and vice versa).

%%%%%%%%%%%%%%%%%%%%%%%%%%%%%%%%%%%%%%%%%
\subsubsection{\review{Additional architecture or Technology Used}} 
%\todo{DNS based technology missing, .e.g., iDNS. Similarly the paper description  as well.}
\gls{ICN} and \gls{IP} are not the only architectures that can coexist, and even the coexistence could be improved using other technologies. More specifically, \gls{ICN} well fits with several different technologies that are already deployed in the current Internet infrastructure. Among those, there are \gls{SDN}, \gls{NFV} or \gls{CDN}. The purpose of this feature is to collect all the architectures that the coexistence solutions involve. 

%%%%%%%%%%%%%%%%%%%%%%%%%%%%%%%%%%%%%%%%%%%%%%%%%%%%%%%%%%%%%%%%%%%%%%%%%%%%%%%%%%%%%%%%%%%%%%%%%%%%%%%%%%%%%%%%%%%%%%%
\subsection{Evaluation Parameters}
\label{evaluation_parameters}
As evaluation parameters, we considered the following challenges arising during the deployment of a coexistence architecture in a real scenario: 
\begin{itemize}
    \item \emph{Access control} - in a networking context, access control uses a set of protocols to define, implement, and maintain policies that describe how the network nodes can be accessed by users/devices. Typically, it includes:
    \begin{itemize}
        \item Authorization, authentication, and accounting of network connections.
        \item Identity and access management.
        \item Mitigation of non-zero-day attacks.
        \item Policy lifecycle management.
        \item Role-based controls of user, device, application.
        \item Security posture check.
    \end{itemize}
    \item \emph{Scalability} - it ensures that the overall performance of a network will be not affected by the size of the network. In other words, scalability describes the ability of a network to grow and manage increasing demand.
    \item \emph{Dynamic network management} - it is the process of administering and managing dynamic changes in computer networks, such as topology changes and handovers for seamless host mobility.
    \item \emph{Latency} - it is defined as the amount of time a message takes to traverse a system. In a computer network, it is typically measured as the time required for a packet to be returned to its sender. The major factors for the network latency include propagation delays and delays due to routers, as well as storage devices.
    \item \emph{Traffic management} - \review{for a detailed description of the traffic management, we refer to Section~\ref{features}}.
\end{itemize}

%% file: coexistence_architectures.tex
\section{Classification of the Coexistence Architectures}
\label{coexistence_architectures}
The purpose of this section is to illustrate the classification of the coexistence architectures according to the features and the evaluation parameters described in Section~\ref{classification_criteria}. The summary of our findings is listed in Table~\ref{table:comparison}.

\begin{sidewaystable*}[!htbp]
\captionsetup{font=Large}
\centering
	\resizebox{1.02\columnwidth}{!}
	{
		\begin{threeparttable}
        \centering
        \caption{\fontsize{14}{14}\selectfont Classification of the coexistence architectures (\cmark~Addressed ~ \xmark~Not addressed).}
        \label{table:comparison}

\begin{tabular}{|c|c|c|c|c|c|c|c|c|c|c|c|c|c|c|c|c|}
\hline
\multicolumn{3}{|c|}{\begin{tabular}[c]{@{}c@{}}\textbf{Parameter}\end{tabular}} & \rotatebox{90}{\textbf{PURSUIT~\cite{6231280}}} & \rotatebox{90}{\textbf{NetInf~\cite{Dannewitz:2013:NII:2459510.2459643}}} & \textbf{\rotatebox{90}{NDN~\cite{NDNProject} \& CCN~\cite{Jacobson}~}} & \rotatebox{90}{\textbf{O-ICN~\cite{7084921}}} & \rotatebox{90}{\textbf{CONET~\cite{detti2011conet}}} & \textbf{\rotatebox{90}{GreenICN~\cite{vahlenkamp2013enabling}}} & \textbf{\rotatebox{90}{coCONET~\cite{veltri2012supporting}}} & \rotatebox{90}{\textbf{DOCTOR~\cite{doctor}}} & \rotatebox{90}{\textbf{POINT~\cite{point}}} & \rotatebox{90}{\textbf{RIFE~\cite{rife}}} & \rotatebox{90}{\textbf{CableLabs~\cite{cableLabs}}} & \rotatebox{90}{\textbf{NDN-LAN~\cite{NDNLAN}}} & \rotatebox{90}{\textbf{\review{hICN}~\cite{hICN}}} & \textbf{\rotatebox{90}{OFELIA~\cite{melazzi2012openflow}}} \\ \hline

\multicolumn{3}{|c|}{\begin{tabular}[c]{@{}c@{}}Duration of the project/\\ Year of publication\end{tabular}} & \begin{tabular}[c]{@{}c@{}}2010\\ to\\ 2013\end{tabular} & \begin{tabular}[c]{@{}c@{}}2010\\ to\\ 2013\end{tabular} & \begin{tabular}[c]{@{}c@{}}2010\\ till\\ today\end{tabular} & 2015 & \begin{tabular}[c]{@{}c@{}}2010\\ to\\ 2013\end{tabular} & 2013 & 2012 & \begin{tabular}[c]{@{}c@{}}2014\\ to\\ 2017\end{tabular} & \begin{tabular}[c]{@{}c@{}}2015\\ to\\ 2017\end{tabular} & \begin{tabular}[c]{@{}c@{}}2015\\ to\\ 2018\end{tabular} & 2016 & 2017 & 2018 & 2012 \\ \hline
\multirow{13}{*}{Features} & \multirow{3}{*}{\begin{tabular}[c]{@{}c@{}}Deployment\\ approaches\end{tabular}} & Overlay & \cmark & \cmark & \cmark & \cmark & \cmark & \cmark & \cmark &  &  &  &  &  &  &  \\ \cline{3-17} 
 &  & Underlay &  &  &  &  &  &  &  & \cmark & \cmark & \cmark & \cmark &  &  &  \\ \cline{3-17} 
 &  & Hybrid &  &  &  &  & \cmark &  &  &  &  &  &  & \cmark & \cmark & \cmark \\ \cline{2-17} 
 & \multirow{5}{*}{\begin{tabular}[c]{@{}c@{}}Deployment\\ scenarios\end{tabular}} & \begin{tabular}[c]{@{}c@{}}ICN-ICN\\ communication\\ in IP ``ocean''\end{tabular} & \cmark & \cmark & \cmark & \cmark & \cmark & \cmark & \cmark & \cmark &  &  & \cmark & \cmark & \cmark &  \\ \cline{3-17} 
 &  & \begin{tabular}[c]{@{}c@{}}ICN-IP\\ communication\\ in IP ``ocean''\end{tabular} &  &  &  &  &  & \cmark & \cmark & \cmark &  &  & \cmark & \cmark & \cmark &  \\ \cline{3-17} 
 &  & \begin{tabular}[c]{@{}c@{}}ICN-IP\\ communication\\ in ICN ``ccean''\end{tabular} &  &  &  &  &  &  &  & \cmark &  &  & \cmark & \cmark & \cmark &  \\ \cline{3-17} 
 &  & \begin{tabular}[c]{@{}c@{}}IP-IP\\ communication\\ in ICN ``ocean''\end{tabular} &  &  &  &  &  &  &  & \cmark &  &  & \cmark & \cmark & \cmark &  \\ \cline{3-17} 
 &  & Border Island &  &  &  & \cmark & \cmark &  &  & \cmark & \cmark & \cmark &  &  & \cmark & \cmark \\ \cline{2-17} 
 & \multirow{4}{*}{\begin{tabular}[c]{@{}c@{}}Addressed\\ coexistence\\ requirements\end{tabular}} & Forwarding & \cmark & \cmark & \cmark & \cmark & \cmark & \cmark & \cmark & \cmark & \cmark & \cmark & \cmark & \cmark & \cmark & \cmark \\ \cline{3-17} 
 &  & Storage & \cmark & \cmark & \cmark & \cmark & \cmark &  & \cmark & \cmark & \cmark & \cmark & \cmark & \cmark & \cmark & \cmark \\ \cline{3-17} 
 &  & Security & \cmark & \cmark & \cmark &  &  &  & \cmark & \cmark & \cmark & \cmark &  &  & \cmark & \cmark \\ \cline{3-17} 
 &  & Management &  &  &  &  & \cmark & \cmark & \cmark & \cmark &  &  &  &  & \cmark & \cmark \\ \cline{2-17} 
 & \multicolumn{2}{c|}{\review{Additional architecture or technology used}} & \begin{tabular}[c]{@{}c@{}}PSIRP\\ LAN\end{tabular} &  &  & \begin{tabular}[c]{@{}c@{}}SAIL\\ SDN\end{tabular} &  & SDN & SDN & \begin{tabular}[c]{@{}c@{}}NFV\\ SDN\end{tabular} & \begin{tabular}[c]{@{}c@{}}PURSUIT\\ SDN\end{tabular} & \begin{tabular}[c]{@{}c@{}}PURSUIT\\ DTN\end{tabular} & CDN & LAN & DNS & \begin{tabular}[c]{@{}c@{}}CONET\\ SDN\end{tabular} \\ \hline
\multicolumn{2}{|c|}{\multirow{6}{*}{\begin{tabular}[c]{@{}c@{}}Evaluation\\ parameters\end{tabular}}} & \begin{tabular}[c]{@{}c@{}}Traffic\\ management\end{tabular} & \xmark & \xmark & \xmark & \xmark &  &  &  &  &  &  & \xmark & \xmark &  &  \\ \cline{3-17} 
\multicolumn{2}{|c|}{} & \begin{tabular}[c]{@{}c@{}}Access\\ control\end{tabular} &  & \xmark &  &  &  &  &  &  &  &  &  &  &  &  \\ \cline{3-17} 
\multicolumn{2}{|c|}{} & Scalability &  &  &  & \xmark &  & \xmark &  &  & \xmark &  &  & \xmark & \xmark &  \\ \cline{3-17} 
\multicolumn{2}{|c|}{} & \begin{tabular}[c]{@{}c@{}}Dynamic\\ network\\ management\end{tabular} &  &  &  & \xmark &  &  &  &  & \xmark &  &  & \xmark &  &  \\ \cline{3-17} 
\multicolumn{2}{|c|}{} & Latency &  &  &  &  &  &  &  & \xmark & \xmark &  &  & \xmark & \xmark &  \\ \cline{3-17} 
\multicolumn{2}{|c|}{} & Other &  & \begin{tabular}[c]{@{}c@{}}NetInf\\ transport\\ functions\\ Interaction.\end{tabular} &  &  & \begin{tabular}[c]{@{}c@{}}New IP\\ option\\ overhead.\end{tabular} & \begin{tabular}[c]{@{}c@{}}SDN controller\\ must manage\\ every \gls{ICN}\\ request and\\ rewrite several\\ headers fields\\ for every\\ response packet.\end{tabular} & \begin{tabular}[c]{@{}c@{}}ICN capable\\ OpenFlow-\\ compliant\\ network.\end{tabular} &  &  &  & \begin{tabular}[c]{@{}c@{}}Optimization\\ of CCN router,\\ cache/content\\ implementation,\\ protocol\\ translation\\ between CCN\\ and HTTP.\end{tabular} &  &  & \begin{tabular}[c]{@{}c@{}}OpenFlow-\\ compliant\\ networking\\ elements.\end{tabular} \\ \hline
\end{tabular}

\end{threeparttable}
}
\end{sidewaystable*}

\subsection{\gls{PURSUIT}}
\begin{figure}[H]
	\centering
	\includegraphics[width=0.9\columnwidth]{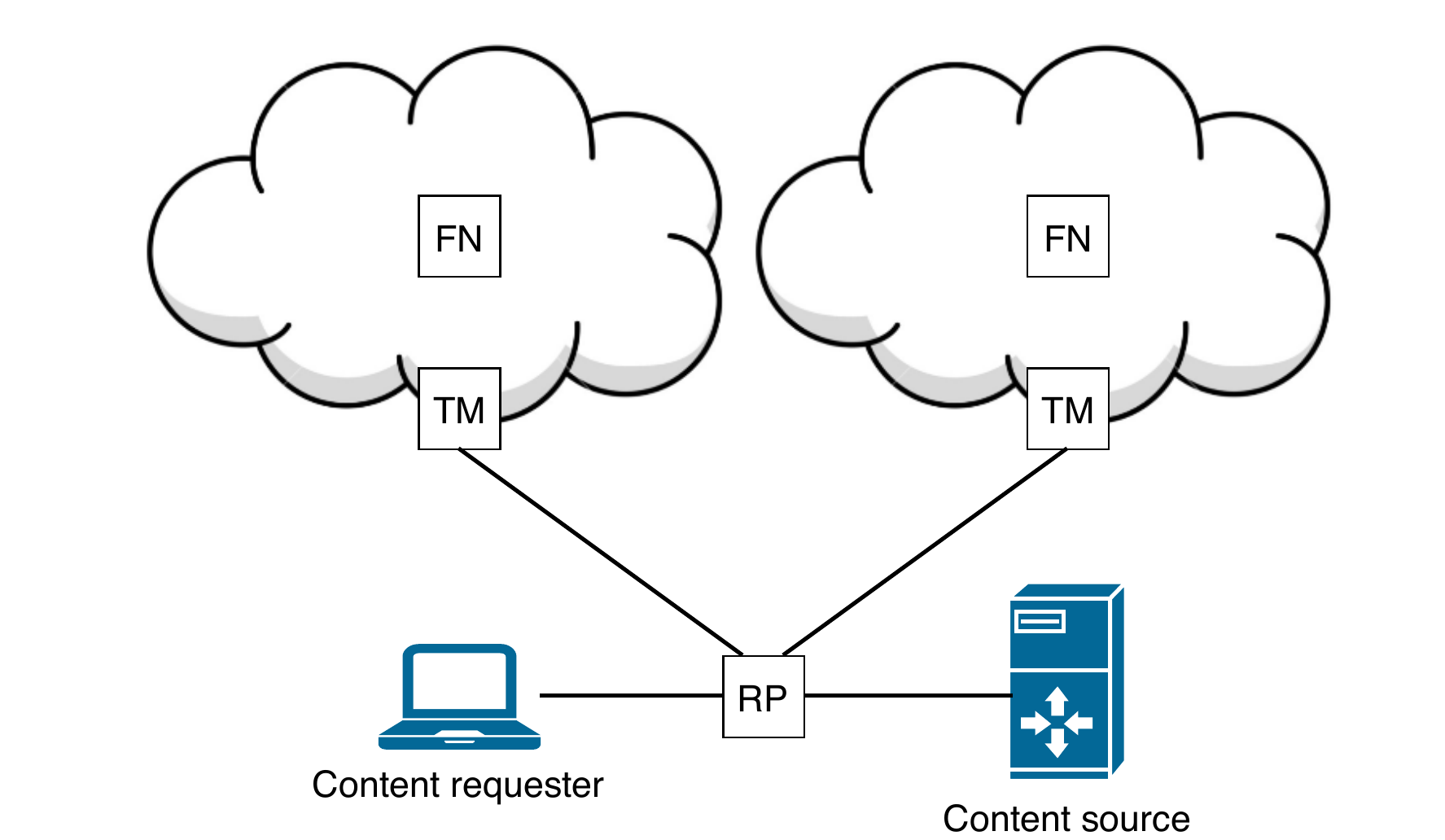}
	\caption{Simplified view of the \gls{PURSUIT} architecture.}
	\label{fig:pursuit}
\end{figure}

PURSUIT \cite{6231280} \review{was} a European project financed by the \gls{FP7}, started in September 2010 and ended in February 2013. \gls{PURSUIT} is an evolution of the \gls{FP7} project \gls{PSIRP}~\cite{Dimitrov:2010:PPP:1839379.1839409}, proposing an \gls{ICN} model based on a source node, that publishes an information, and on a client node, that subscribes to the content it desires. If the information is available, it will be delivered to the client. \gls{PURSUIT} aims at improving \gls{PSIRP}, meanwhile evaluating its performance, scalability, and  coexistence with the current Internet network. \review{Fig.}~\ref{fig:pursuit} shows a simplified form of the architecture proposed in \gls{PURSUIT} project.

%The architecture proposed in PURSUIT, which is shown in Figure~\ref{fig:pursuit} in a simplified form, relies on the definition of a new data format and on the introduction of three new network components.
\gls{PURSUIT} architecture relies on the definition of a new data format and on the introduction of three new network components. \gls{PURSUIT} addresses the data as information items, which consist of pair of identifiers, i.e., \gls{RI} and \gls{SI}. The former represents the real piece of information, while the latter specifies the group which the information belongs to. The three additional network functions addressed by \gls{PURSUIT} are: \gls{RF}, \gls{TF}, and \gls{FF}. The \gls{RF} plays a fundamental role in \gls{PURSUIT} since it maps subscribers to publishers and supports names resolution. Moreover, it also initializes the delivery of information item to the client. The \emph{\gls{RP}} performs the \gls{RF} and relies on a hierarchical distributed hash table internal data structure. The \emph{\gls{TM}} implements the \gls{TF} by deploying a routing protocol to collect the topology of its domain and by exchanging routing information with other domains for global routing. The \emph{\gls{FN}} implements the \gls{FF} and it is also responsible for redirecting the information item to the client. In particular, the forwarding mechanism is label-based and uses a bloom filter \cite{Jokela2009LIPSINLS} to speed up the information delivery. In addition, the \emph{FN} offers also a caching facility.

\begin{figure}[H]
	\centering
	\includegraphics[width=0.4\textwidth]{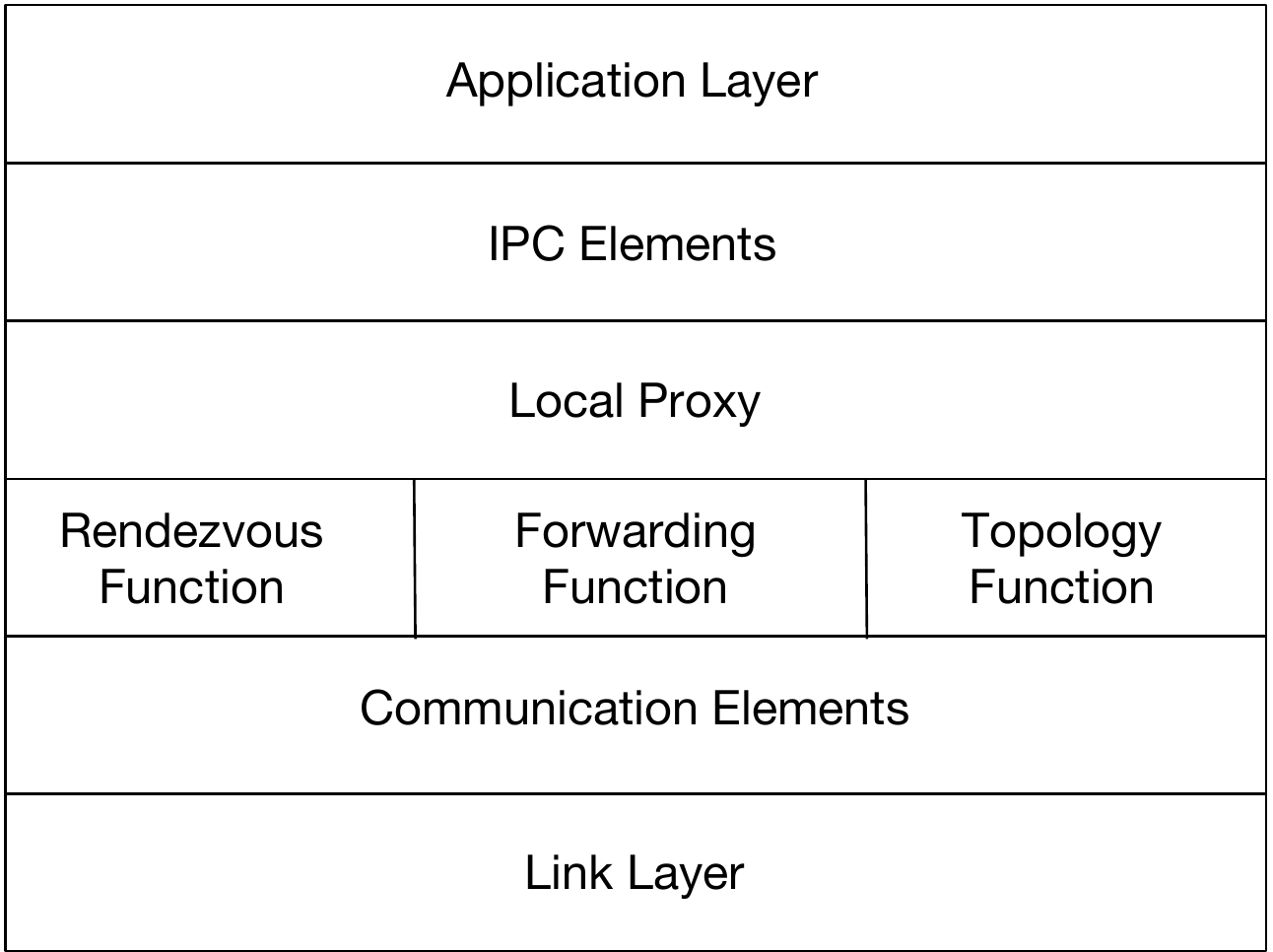}
	\caption{Internal architecture of a \gls{PURSUIT} node.}
	\label{fig:pursuit-node}
\end{figure}

As shown in \review{Fig.} \ref{fig:pursuit-node}, the \gls{PURSUIT} node internal architecture encompasses several components%and it exploits the Click Router framework \cite{Kohler:2000:CMR:354871.354874}
, enabling the publish/subscribe communication model among the different stack layers. The \emph{IPC Elements} implement a non-blocking inter-process mechanism, allowing users-space applications to issue publish/subscribe requests and communicate through the proposed prototype. The functionality of the \emph{Local Proxy} element is to maintain a local record for all the pending subscriptions and, after receiving a request, dispatch it to the appropriate functions (i.e., \emph{\gls{RF}}, \emph{\gls{FF}}, \emph{\gls{TF}}). Finally, the \emph{Communication Elements} are responsible for transmitting publications to the network.  
The design implementation of PURSUIT is based on Click elements \cite{Kohler:2000:CMR:354871.354874}: it creates Ethernet frames and forwards them to the appropriate network interface. In addition, it provides the ability to utilize raw \gls{IP} data packets as an alternative mechanism. This enables the prototype to be tested in Internet-wide scenarios.

\textbf{Deployment Approach.}
Trossen et al. \cite{6231280} implemented a Layer-2 \gls{VPN}-based \emph{overlay} solution of \gls{PURSUIT} among multiple nodes located in Europe, US and Asia. The prototype is established and verified on three different testbeds for experimental purposes, functioning as an overlay on LAN environment. 
To showcase a specimen of native \gls{ICN} application, multimedia streaming services were hosted as a demonstration, showing a lossless transmission and comparable performance.

\textbf{Deployment Scenarios.}
The \emph{ICN-ICN communication in IP ``ocean''} is the most suitable scenario for deploying PURSUIT, as it is also confirmed by the \emph{overlay} approach adopted in the testbed.
 %Figure \ref{fig:pursuit-node} describes the node architecture of overlay deployment  where all core functions of runs over the top of existing TCP/IP or UDP.  
  
%Although the original design of PURSUIT implies a \emph{clean-slate} deployment approach based on the publish-subscribe paradigm, the research group working on this project designed an experimental testbed working as an \emph{overlay} solution on the second layer of the OSI model. 

\textbf{Addressed Coexistence Requirements.}
\gls{PURSUIT} guarantees the following three coexistence requirements: 
\begin{itemize} 
    \item Forwarding - this is specifically provided by the \emph{FN}, a software-based forwarder used for \gls{ICN} messages exchange.
    \item Storage - the \emph{FN}, which has the responsibility of redirecting information to the client, provides caching facility to furnish storage of information.
    \item Security - the security measures provided by \gls{PURSUIT} refer to the access of information. Besides gathering information into groups, \gls{PURSUIT} supports the information categorization into scopes, used for the definition of access privileges and policy implementations.
    %\item Interoperability: TM implements   the  routing  protocol  which collects the topology of respective domain and manages global routing by exchanging routing information with other domains.
    \end{itemize}
%\question{Why interoperability? Topology  Function organizes  the  routing  protocol  which collects the topology of respective domain and manages global routing by exchanging routing information with other domains. It is operated by Topology Manager (TM)}

\textbf{\review{Additional architecture or Technology Used.}}
\gls{PURSUIT} is an evolution of the \gls{PSIRP} project and its testbed has been realized as an \emph{overlay} solution over a \gls{LAN} environment. 

\textbf{Evaluation Parameters.}
The main issue introduced by the \emph{overlay} deployment in the PURSUIT architecture is the traffic management. This is mainly due to the existing Internet applications and protocols, which are not completely compatible with the techniques implementing \gls{ICN} over \gls{TCP}/\gls{IP} or UDP~\cite{ccnxudp,Zhang,NDNLP,ccnx-1.0} for traffic transport. Thus, many applications and protocols, such as \gls{HTTP} based multimedia streaming protocols, might face false throughput estimations \cite{wowmom2018}. This is due to the \gls{TCP} aggressiveness in presence of variations in content source location (e.g., dynamic caching and interest aggregation) \cite{CONTI2018209}.

\vspace{0.4cm}
\subsection{\gls{NetInf}}
The \gls{NetInf} architecture \cite{Dannewitz:2013:NII:2459510.2459643} is the approach proposed by the European \gls{FP7} project SAIL~\cite{sail}, started in January 2010 and ended in February 2013. The key component of the \gls{NetInf} architecture  is the \gls{CL}, which is able to map the information, expressed through any protocol (e.g., \gls{HTTP}, \gls{TCP}, \gls{IP}, Ethernet), into specific messages compliant to a general communication paradigm. In particular, when two nodes communicate between each other, the functionality of a \gls{CL} is to provide framing and message integrity to  NetInf requests and responses.      

\begin{figure}[!htbp]
	\centering
	\includegraphics[width=0.3\textwidth]{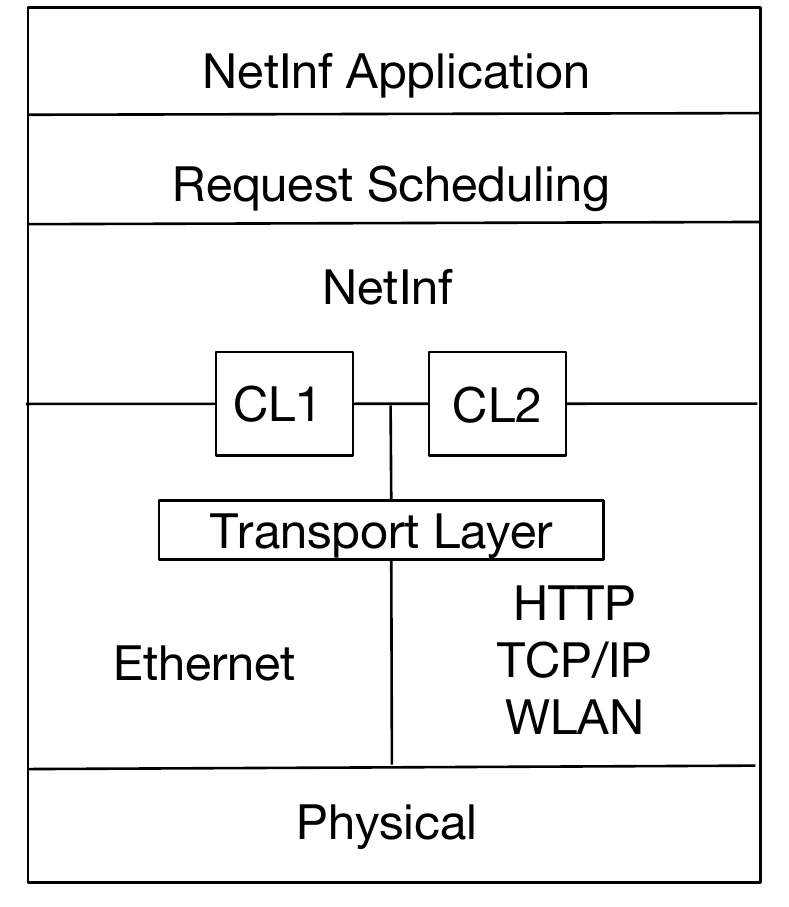}
	\caption{Internal architecture of a \gls{NetInf} node.}
	\label{fig:NetInf-node} 
\end{figure}

\review{Fig.}~\ref{fig:NetInf-node} depicts the different \gls{CL}s designed within the \gls{NetInf} stack. In particular, \gls{CL}s encompass an additional function (i.e., \emph{Request Scheduling}) between the \emph{NetInf Application} and the \emph{NetInf Protocol}. The \emph{CL1} functions over Ethernet, while \emph{CL2} makes \gls{NetInf} able to function over a variety of networks links and protocols such as HTTP, \gls{TCP}/\gls{IP}, \gls{WLAN}. The \gls{CL}s also provide transport layer functions across different nodes such as flow control, congestion control and reliability.   

\textbf{Deployment Approach.}
\gls{NetInf} adheres to the \emph{overlay} deployment approach, as it is confirmed by its first prototypes, deployed as an \emph{overlay} strategy over \gls{TCP}/\gls{UDP}. 

\textbf{Deployment Scenarios.}
The \gls{NetInf} architecture supports the \emph{ICN-ICN communication in IP ``ocean''} scenario. 

\textbf{Addressed Coexistence Requirements.}
The coexistence requirements provided by \gls{NetInf} are as follows: 
\begin{itemize}
    \item Forwarding - \gls{NetInf} guarantees both name-based forwarding and name resolution; \gls{NetInf} message forwarding protocol relies on the lower-layer networking technology (e.g., \gls{TCP} connection between two Internet hosts) and this communication is provided by the \gls{CL}s.
    \item Storage - \gls{NetInf} nodes support both on-path and off-path caching.
    \item Security - the \gls{CL}s are responsible for the integrity of the messages exchanged in the architecture.
    %\item Interoperability: the CLs are able to interconnect different types of networks.
\end{itemize}

%\question{Why was storage previously indicated?} 
%\textcolor{black}{Storage: Generic NetInf nodes supports both on-path caching (on the request/data path) as well as off-path caching. In addition, the protocol  is able to integrate any available copy when retrieving data: the original server, redundant copies, as well as replicas stored on user devices (if so permitted). Thereby, NetInf can access the best available copy}

\textbf{\review{Additional architecture or Technology Used.}}
Besides the standard \gls{TCP}/\gls{UDP}/\gls{IP} tunneling, which is part of the \emph{overlay} approach, \gls{NetInf} does not rely on additional architectures. 

\textbf{Evaluation Parameters.}
The deployment of the \gls{NetInf} architecture in a coexistence scenario introduces the following challenges: traffic management, due to the absence of interaction among the \gls{CL}s transport functions and the \gls{NetInf} transport functions, and access control. The first issue refers to the \gls{CL}s, which are responsible for the interconnection of different types of networks into a single \gls{ICN} network. For example, the interaction among the underlying protocols that provide really different communication services creates new challenges (e.g., from uni-directional, opportunistic message forwarding to flow- and congestion-controlled higher layer communication services; from delay-challenged to high-speed optical backbone networks). Concerning the access control limitation, in \gls{NetInf}, it is not possible to apply controls over the accessibility levels of the information. Thus, anyone can access the published data without any restriction.  

\vspace{0.4cm}
\subsection{\gls{NDN} and \gls{CCN}}
Among the the existing implementations of the \gls{CCN} paradigm \cite{Jacobson}, funded by the \gls{NSF}~\cite{NSF} as part of the Future Internet Architectures program, there is the \gls{NDN} research project~\cite{NDNProject}. From its first design late in 2010, the \gls{NDN} main idea is to shift the existing \gls{IP} host-to-host communication into a data oriented one by leveraging on an increased responsibility of the routers. Upon receiving a request for a content, the routers first check whether the content is already present in their cache (i.e., Content Store). If this is the case, they immediately return the content back, otherwise, they check the \gls{PIT}, searching for a pending request issued for the same content. If the PIT already contains an entry for the specific content, routers just collapse the current request into the PIT. If none of the previous cases verifies, routers forward the request to the next node in the network using the FIB, and keep waiting for the associated data to  return back. Once the data packet arrives, all the pending interests for that content are satisfied just by sending the copy of data back to all the hosts which have requested it.

\begin{figure}[H]
	\centering
	\includegraphics[width=0.33\textwidth]{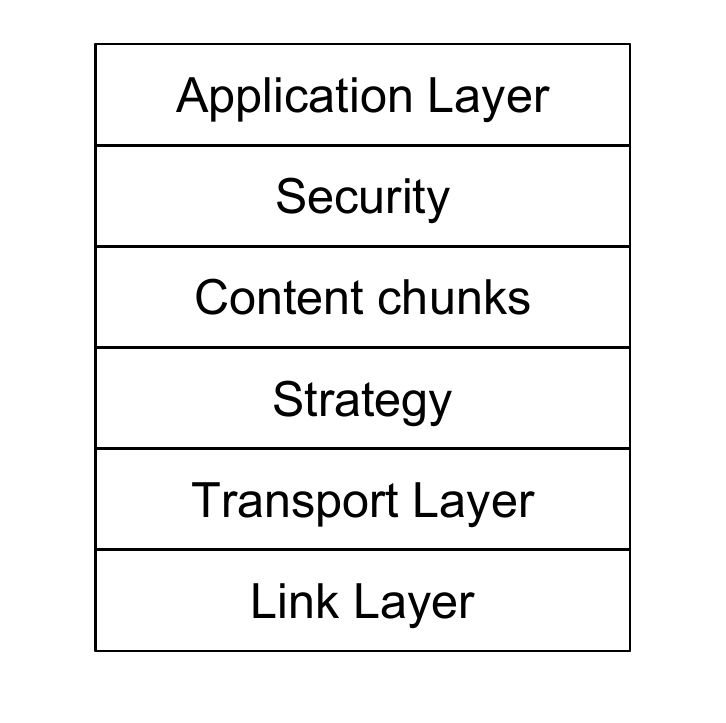}
	\caption{\gls{NDN} network stack~\cite{Zhang}.}
	\label{fig:NDN-node}
\end{figure}

As shown in \review{Fig.}~\ref{fig:NDN-node}, \gls{NDN} introduces some changes into the \gls{IP} stack by adding the \emph{Security} and \emph{Strategy} novel layers: the first refers to the \gls{NDN} design addressing the security of the content instead of the security of the communication channel between two nodes (which is how \gls{IP} works); the second substitutes the network layer and provides the forwarding plane to forward \emph{Content chunks} by giving the best choices to maintain multiple connectivities under varying conditions. In addition, the \emph{Strategy} layer also supports security, scalability, efficiency and resiliency. Finally, \gls{NDN} modifies the \emph{Transport Layer} making it consumer-driven instead of producer-driven~\cite{6193510, CAROFIGLIO2016104}, importing it into the \gls{NDN} forwarding plane.

\textbf{Deployment Approach.}
The common implementation of \gls{NDN} and \gls{CCN} includes \emph{overlay} protocols, such as CCNx~\cite{ccnx-1.0} and NDNLP~\cite{NDNLP}, which are deployed over existing \gls{IP} infrastructure.  For instance, CCNx~\cite{ccnxudp} showcases the explicit example of overlay by implementing \gls{CCN}-over-\gls{UDP}. In particular, it provides a method to transport CCNx messages between two nodes over \gls{UDP}. Moreover, a concrete example of \gls{NDN} overlay architecture is provided by the ndn-testbed\footnote{https://named-data.net/ndn-testbed/}, which connects multiple \gls{NDN} nodes located in several continents over  existing \gls{TCP}/\gls{IP}. The services provided in the trials of \gls{CCN}/\gls{NDN} include various projects, such as  real-time video-conferencing \cite{Gusev2015NDNRTCRV}, adaptive bit-rate streaming (not limited to end-to-end) \cite{dashoverccn,dashovericn,ndnavs} and  ndnSIM (\gls{NDN} simulator module on NS-3) \cite{399}. 

\textbf{Deployment Scenarios.}
\gls{NDN} supports the \emph{ICN-ICN communication in IP ``ocean''} scenario, as it is confirmed by the ndn-testbed.

\textbf{Addressed Coexistence Requirements.}
\gls{NDN} guarantees the following three coexistence requirements: 
\begin{itemize}
    \item Forwarding - the router's FIB is responsible for forwarding interests towards the content provider via one or more network interfaces based on the routes to the origin node(s). The requested data packet is then forwarded towards the requester by simply traversing, in reverse, the path of the preceding interest~\cite{Zhang}. \gls{NDN} supports also the multicast data routing, which improves receiver-driven multimedia delivery.
    \item Storage - \gls{NDN} routers are enabled to cache contents.
    \item Security - \gls{NDN} provides a data-centric security model where each data unit is uniquely signed by the data producer \cite{8539023}. 
\end{itemize}

\textbf{\review{Additional architecture or Technology Used.}}
Besides the standard TCP/UDP/\gls{IP} tunneling, which is part of the \emph{overlay} approach, the \gls{NDN} project does not rely on additional architectures. 

\textbf{Evaluation Parameters.}
The tunneling approach, where \gls{NDN}/\gls{CCN} endpoints communicate over \gls{IP} \cite{TCP/ICN,367}, disowns the fundamental advantages of the content oriented networking (i.e., in-network caching and multicast forwarding) and the architectures implementing hop-to-hop connection-less (/oriented) connectivity (i.e., over \gls{TCP}/\gls{UDP}) suffer from a lack of traffic management \cite{CONTI2018209}. 
In \gls{NDN}/\gls{CCN} networks, \gls{CA} is operated by the consumer rather than by the producer (server). This means that the Interests transmission rate is adapted in order to ensure that the delivery of a requested resource can make maximum fair use of the network. Existing \gls{NDN}/\gls{CCN} CA algorithms are largely based on the \gls{TCP} \gls{CA} algorithms, which assume that the bandwidth-delay product of the network fluctuates relatively slowly, as all the data packets traverse the same path from server to client. However, in \gls{NDN}/\gls{CCN} network content objects may be retrieved from various locations and may reach the consumer through different paths. Thus, the concept of a bandwidth-delay related to a single path and the use of \gls{TCP} \gls{CA} algorithms do not fit for \gls{NDN}/\gls{CCN} networks. In the \gls{NDN}/\gls{CCN} community, this is an active research area~\cite{Schneider:2016:PCC:2984356.2984369}.

\vspace{0.4cm}
\subsection{O-ICN}
\gls{O-ICN}~\cite{7084921} is a novel architecture, which leverages the \gls{SDN} technology for separating data plane activities (i.e., forwarding  and  storing/caching of \gls{ICN} contents) from control plane activities (i.e., naming, name resolution and routing). In particular, O-ICN introduces the \gls{ICN} Manager as an extended version of a \gls{DNS} server, which performs name resolution for both \gls{ICN} and non-\gls{ICN} requests. In case of an \gls{ICN} request, the \gls{ICN} Manager identifies the source of the content and sends to it the user's address, so that the source can route back the requested content to the user. In case of a non-\gls{ICN} request, the standard routing mechanism of \gls{TCP}/\gls{IP} is followed. The naming scheme adopted by O-ICN is hybrid, i.e., both human readable and self-certifying as in the SAIL architecture~\cite{sail}. Finally, the existing routers are modified to cache contents and communicate with the \gls{ICN} Manager. 

\begin{figure}[b]
\centering
\begin{subfigure}[t]{\columnwidth}
	\centering
        \includegraphics[width=0.70\textwidth]{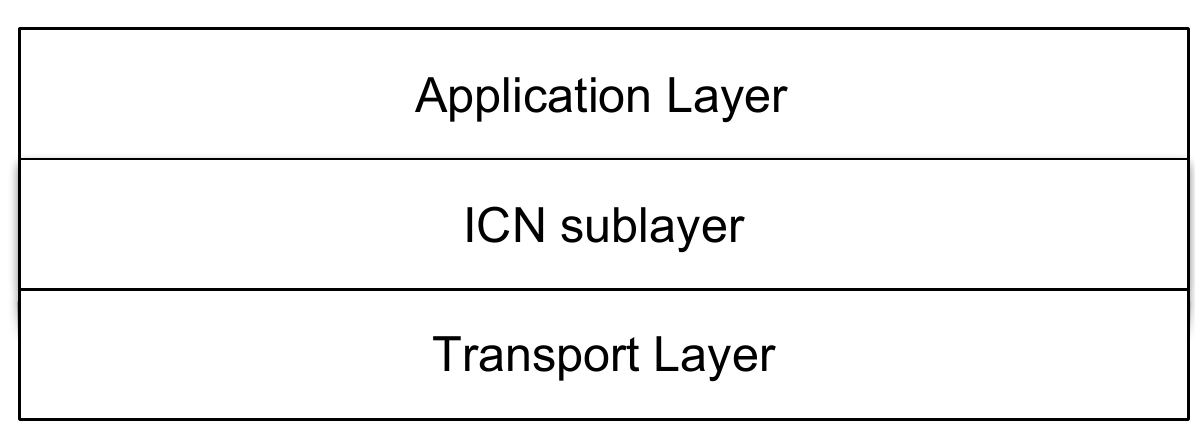}
        \caption{Position of the \gls{ICN} sublayer.}
    	\label{fig:oicn_1}
\end{subfigure}
~
\begin{subfigure}[t]{\columnwidth}
	\centering
        \includegraphics[width=0.70\textwidth]{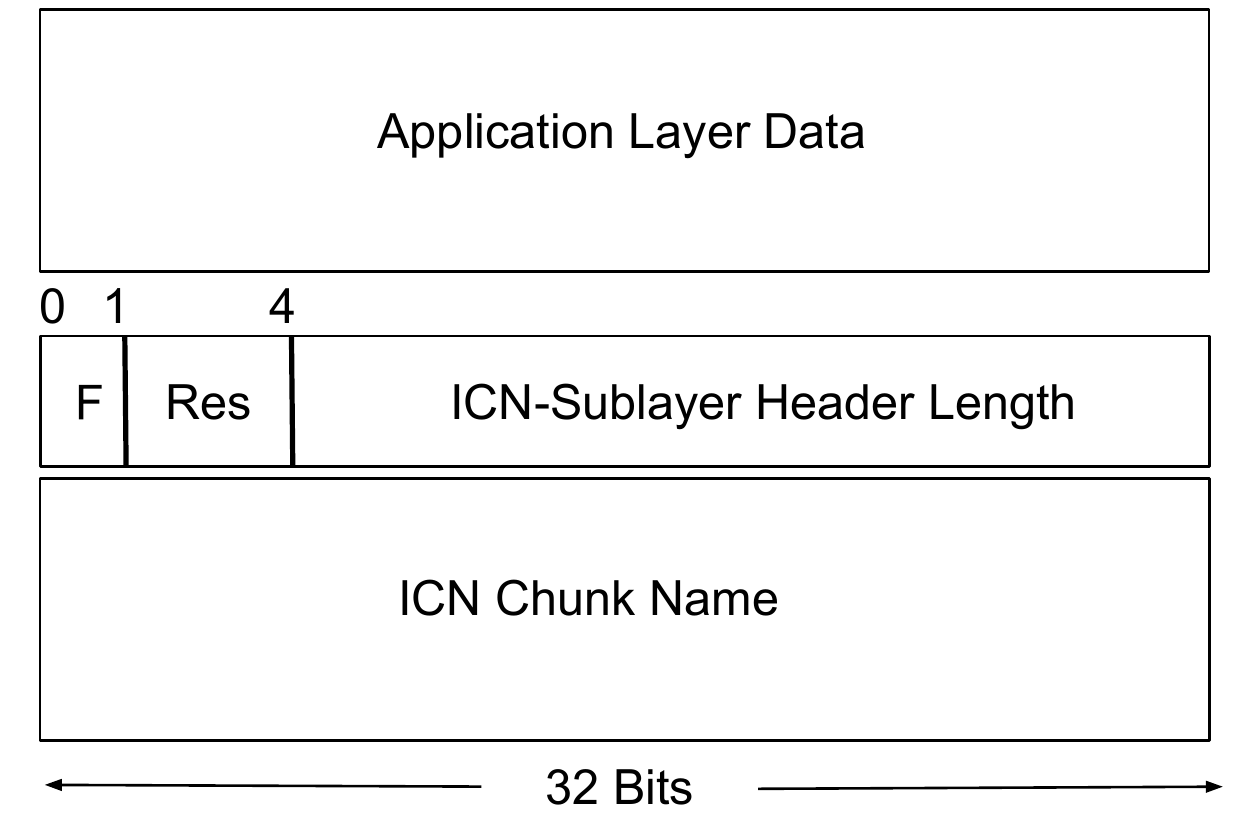}
        \caption{Detail of the \gls{ICN} sublayer header format.}
    	\label{fig:oicn_2}
\end{subfigure}
    \caption{Internal architecture of an O-ICN node.}
    \label{fig:OICN-node}
\end{figure}

\review{Fig}~\ref{fig:oicn_1} depicts the position of the novel \emph{ICN-sublayer} proposed by O-ICN, which lies between the \gls{TCP}/\gls{IP} \emph{Application Layer} and \emph{Transport Layer}. More specifically, \review{Fig.}~\ref{fig:oicn_2} describes the fields used by the new layer: the \gls{ICN} flag bit (\emph{F}), equal to 0 for an \gls{ICN} request or to 1 for an \gls{ICN} content; the three subsequent bits (1-4) reserved for additional purposes, and the remaining 28 bits for the total \gls{ICN} header~\cite{Agrawal2018}.
%\begin{figure}
%	\centering
%	\includegraphics[width=0.40\textwidth]{./images/OICN-stack.pdf}
%	\caption{Internal architecture of an O-ICN node.}
%	\label{fig:OICN-node}
%\end{figure}

\textbf{Deployment Approach.}
O-ICN relies on an \emph{overlay} deployment solution by leveraging on the \gls{ICN} Manager, which performs dual tasks: name resolution, along with routing functionalities for \gls{ICN} requests, and standard \gls{DNS} resolution for the existing Internet requests. \review{To evaluate the O-ICN architecture, authors in~\cite{Agrawal:2018:OSN:3265997.3266000} present the Overlay \gls{ICN} simulator (OICNSIM)\footnote{https://www.nsnam.org/wiki/Contributed\_Code}, an ns-3 based simulator where each O-ICN component is provided with helper classes and it is able to satisfy a wide variety of deployment scenarios. As an example, in~\cite{Agrawal:2018:OSN:3265997.3266000}, the authors studied the performance of OICNSIM for different \gls{ICN} caching policies.}  

\textbf{Deployment Scenarios.}
O-ICN supports the \emph{ICN-ICN communication in IP ``ocean''} scenario. Moreover, thanks to the \gls{ICN} manager capability of manipulating both \gls{ICN} and not-\gls{ICN} requests, O-ICN can support also the \emph{Border Island} deployment scenario. 

\textbf{Addressed Coexistence Requirements.}
The coexistence requirements addressed by O-ICN are as follows: 
\begin{itemize}
    \item Forwarding - the \gls{ICN} Manager is responsible for the forwarding strategy.
    \item Storage - the data plane activities involve tactical storing/caching of \gls{ICN} contents at different locations/routers/gateways and are managed by \gls{ICN} routers.
\end{itemize}

\textbf{\review{Additional architecture or Technology Used.}}
O-ICN exploits the SAIL solution for the naming scheme and the \gls{SDN} technology for a separate management of data plane and control plane activities. 

\textbf{Evaluation Parameters.}
As for the previous \emph{overlay} approaches, O-ICN is affected from a lack of traffic management. In addition, the overall solution suffers from scalability problems and the \gls{ICN} manager is not able to guarantee its \gls{DNS} functionalities in case of dynamic network conditions. 

\vspace{0.4cm}
\subsection{CONET}
\begin{figure}[H]
	\centering
	\includegraphics[width=0.48\textwidth]{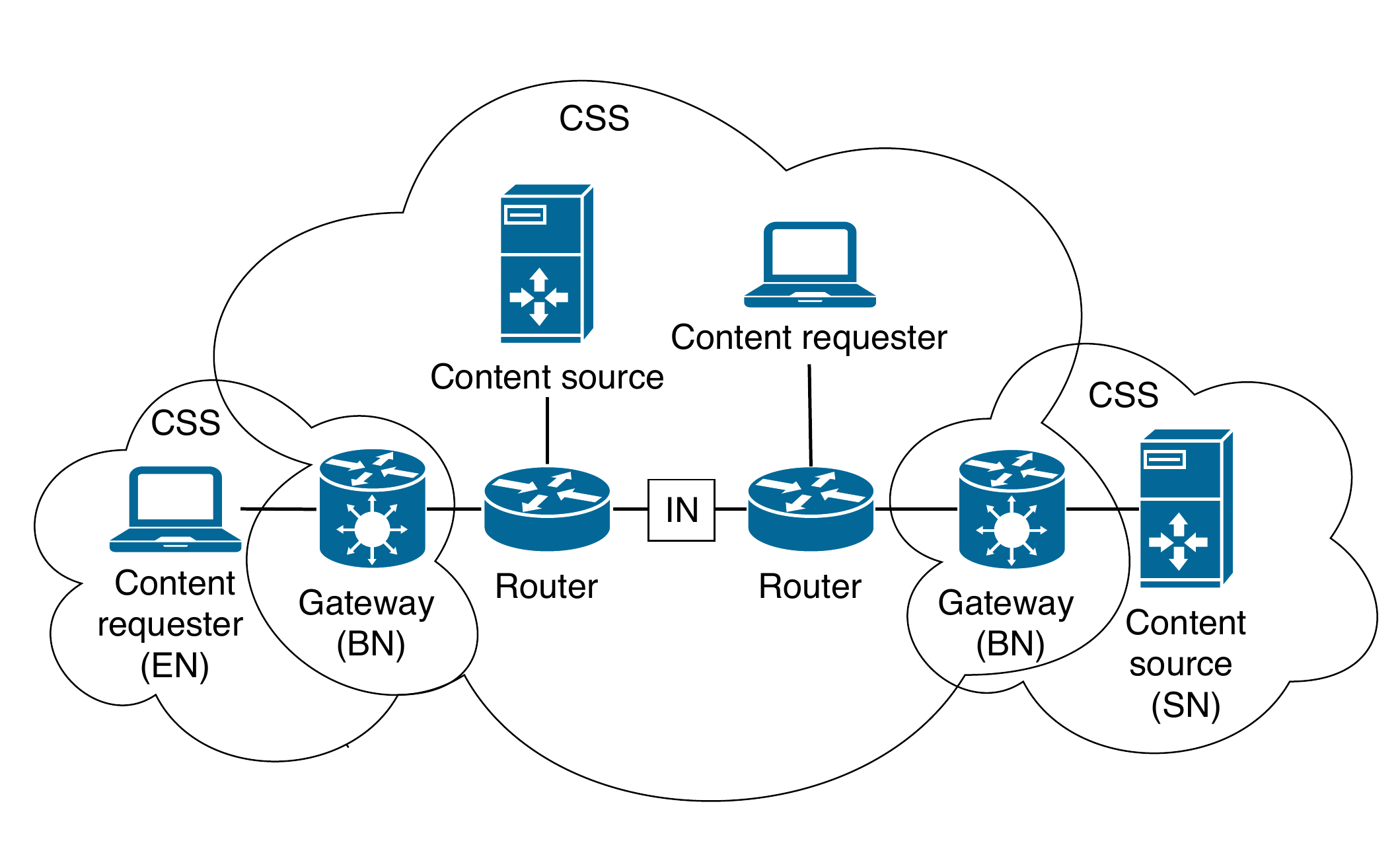}
	\caption{Simplified view of the CONET architecture.}
	\label{fig:conet}
\end{figure}

CONET~\cite{detti2011conet} is an architecture designed for connecting several \emph{\gls{CSS}}, which could be the whole Internet network, an \gls{IP} autonomous system or a couple of network connected components. The main components of the CONET design, shown in \review{Fig.}~\ref{fig:conet}, are as follows: \emph{\gls{EN}}, \emph{\gls{SN}}, \emph{\gls{BN}}, \emph{\gls{IN}}, and \emph{\gls{NSN}}. An \emph{EN} requests some named-data by issuing an interest routed by the \emph{\gls{BN}s}, which are located at the border of \emph{\gls{CSS}s}. The route-by-name process identifies the \emph{\gls{CSS}} address of the next \emph{\gls{BN}}, which is closest to the \emph{\gls{SN}} as soon as the appropriate \emph{\gls{CSS}} is reached. Then, the \emph{\gls{IN}s} forward the packet using the under-CONET routing engine. The \emph{\gls{CSS}} address of \emph{\gls{EN}} and the \emph{\gls{CSS}} addresses of the traversed nodes are appended to the packet. As soon as a CONET node is found to be able to provide the requested named-data, this is sent back on the reverse path to serve the requesting \emph{\gls{EN}}. All \emph{\gls{BN}s} and \emph{\gls{IN}s} along the traversed path may cache the content.

\textbf{Deployment Approach.}
The CONET architecture can follow either an \emph{overlay} or a \emph{hybrid} deployment approach. In the first case, CONET works on top of the \gls{IP} layer and the \emph{\gls{CSS}s} are nodes connected by overlay links (e.g, \gls{UDP}/\gls{IP} tunnels). In the second approach, the purpose is to make \gls{IP} content-aware by introducing a novel IPv4 option or an IPv6 extension header. The network components will have then hybrid routing tables with both \gls{IP} network addresses and names. 

\textbf{Deployment Scenarios.}
Considering the \emph{overlay} solution, CONET supports the \emph{ICN-ICN communication in IP ``ocean''} scenario. On the contrary, the \emph{hybrid} approach allows it to be deployed in the \emph{Border Island} scenario as well. 

\textbf{Addressed Coexistence Requirements.}
CONET guarantees the following three coexistence requirements. 
\begin{itemize} 
    \item Forwarding and Management - these are guaranteed by \emph{\gls{BN}s} and \emph{\gls{NSN}s}. In addition, \emph{\gls{EN}s} provide transport-level functionalities such as reliability and flow control. Since the logic for requesting a content involves sending separate interests, containing a small part of the named-data, the control of interest sending rate can be used as a \gls{TCP}-like flow control mechanism.
    \item Storage - \emph{\gls{BN}s} are able to store contents.
\end{itemize}

\textbf{\review{Additional architecture or Technology Used.}}
Besides the standard \gls{TCP}/\gls{UDP}/\gls{IP} tunneling, which is part of the \emph{overlay} approach, the CONET project does not rely on additional architectures. 

\textbf{Evaluation Parameters.}
The \emph{hybrid} deployment solution is hard to be introduced since it requires a new \gls{IP} option. However, with respect to the \emph{clean-slate} approach, the \emph{hybrid} one is less disruptive, and it allows the architecture deployment in different scenarios. 

%%%%%%%%%%%%%%%%%%%%%%%%%%%%%%%%%%%%%%%%%%%%%%%%%%%%%%%%%%%%%%%%%%%%%%%%%%%%%%%%%%%%%%%%%%%%%%%%%%%%%%%%%%%%%%%%%%%%%%%%
\vspace{0.4cm}
\subsection{GreenICN}
\begin{figure}[H]
	\centering
	\includegraphics[width=0.45\textwidth]{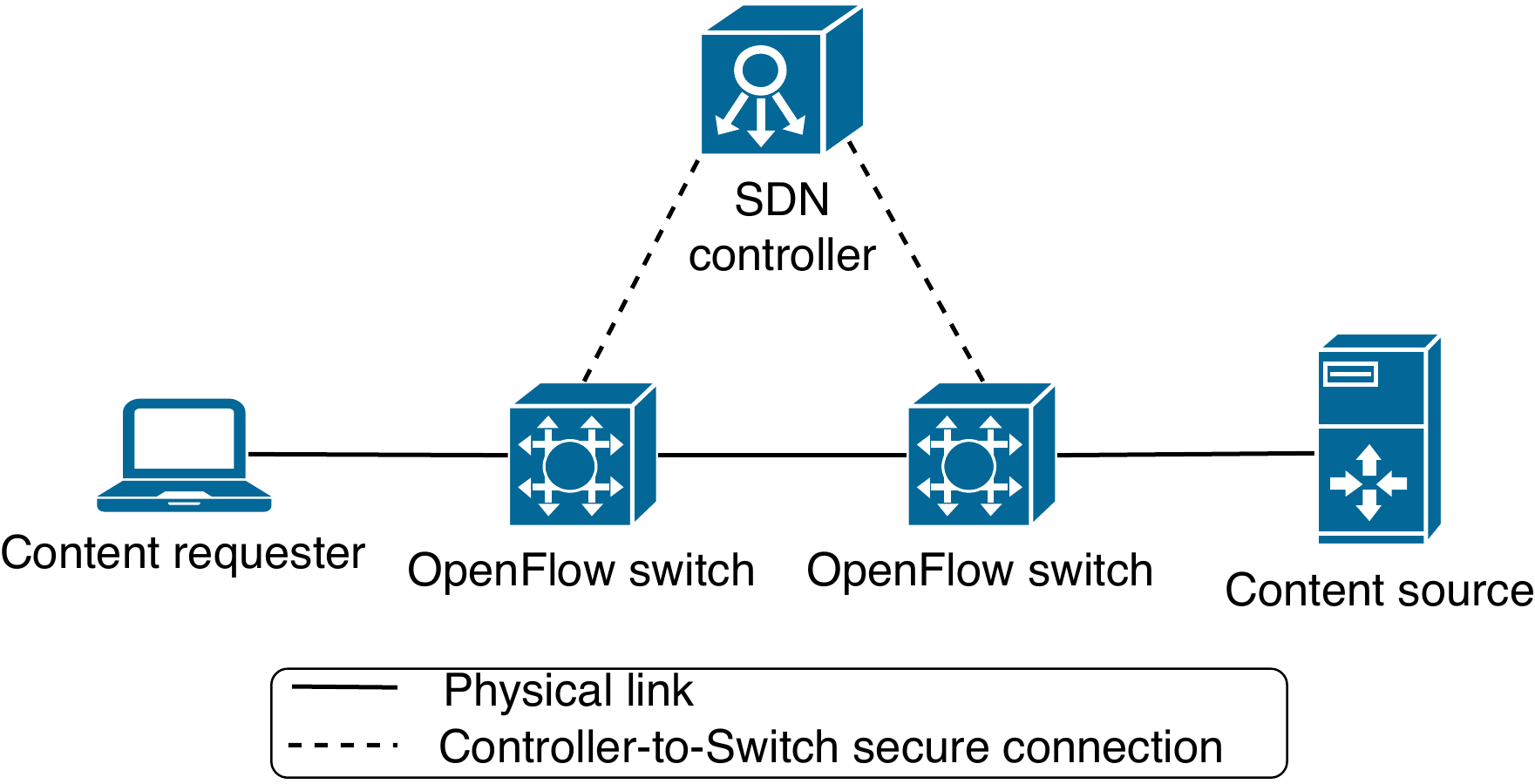}
	\caption{Simplified view of the GreenICN architecture.}
	\label{fig:Vahlenkamp13}
\end{figure}

The \gls{SDN} technology decouples control plane from data plane, and it provides a programmable, centrally managed network control that improves network performance and monitoring. \gls{SDN}-based implementations of \gls{ICN} exploit the centralized view available to \gls{SDN} controller, which enables the \gls{SDN} controller to install appropriate forwarding rules for \gls{ICN} requests/responses in such a manner that the network elements only have to support \gls{IP} forwarding. Vahlenkamp~et~al. in~\cite{vahlenkamp2013enabling} proposed an implementation of \gls{ICN} using \gls{SDN} under their GreenICN project. The proposal leverages \gls{ICN} protocol's Message IDs and features of \gls{SDN} instantiations such as OpenFlow to rewrite packet header information. \review{Fig.}~\ref{fig:Vahlenkamp13} presents a simplified view of this solution. Here, both the \emph{Content requester} and the \emph{Content source} are connected to \emph{OpenFlow-enabled switches} that are managed by the \emph{\gls{SDN} controller}. Routing information for the content requests and responses, upon arriving on OpenFlow switches, is handled/rewritten by the instructions from the controller. %When a request from a content requester arrives at the switch, the switch asks for the instruction (forwarding rules) from the SDN controller - that has the global view of the network. The controller installs all appropriate rules along the path till the content source. And, 

\textbf{Deployment Approach.} The proposed solution is an \emph{overlay} \gls{ICN} implementation as \gls{ICN} data is sent over the \gls{SDN}-managed \gls{IP} packets.

\textbf{Deployment Scenarios.}
Essentially, the authors in~\cite{vahlenkamp2013enabling} propose \gls{ICN} deployment over \gls{IP} network, where an \gls{ICN}-aware content source delivers the content to an \gls{ICN}-aware requester over \gls{IP} network. Hence, this solution supports both the \emph{ICN-ICN communication in IP ``ocean''} and the \emph{ICN-IP communication in IP ``ocean''} scenarios.

\textbf{Addressed Coexistence Requirements.} 
The architecture addresses the following coexistence requirements: 
\begin{itemize}
    \item Forwarding - network programmability offered by \gls{SDN} enables forwarding and routing for \gls{ICN}.
    \item Management - \gls{SDN} centrally managed network control supports load-balancing, traffic engineering, and explicit path steering (e.g., through \gls{ICN} caches).
\end{itemize}

\textbf{\review{Additional architecture or Technology Used.}} The authors argue that an ideal or native deployment of \gls{ICN}, in which user devices, content sources, and intermediary network elements are \gls{ICN} aware, may not be viable. Hence, the authors proposed to implement \gls{ICN}-awareness in the \gls{SDN}-enabled switches, where \gls{ICN} packets are carried over the \gls{IP} transport protocol. By using \gls{SDN}, the authors target all the services/applications of the \gls{TCP}/\gls{IP} protocol stack.

\textbf{Evaluation Parameters.} In the proposed \gls{ICN} implementation, \gls{SDN} controller must manage every \gls{ICN} request and rewrite several headers fields for every response packet, which might not scale with increased network size. Given that this solution is based on the widely accepted \gls{SDN} technology - that supports agile deployment and rapid alternation in networking - the hardware modifications required for its deployment are low in those scenarios where \gls{SDN} infrastructure already exists. Consequently, the time required for its deployment is also low. Nevertheless, the time and the hardware modifications required for its deployment would be higher if the \gls{SDN} infrastructure does not already exists.

%%%%%%%%%%%%%%%%%%%%%%%%%%%%%%%%%%%%%%%%%%%%%%%%%%%%%%%%%%%%%%%%%%%%%%%%%%%%%%%%%%%%%%%%%%%%%%%%%%%%%%%%%%%%%%%%%%%%%%%%
\vspace{0.4cm}
\subsection{coCONET}
\begin{figure}[h]
	\centering
	\includegraphics[width=0.45\textwidth]{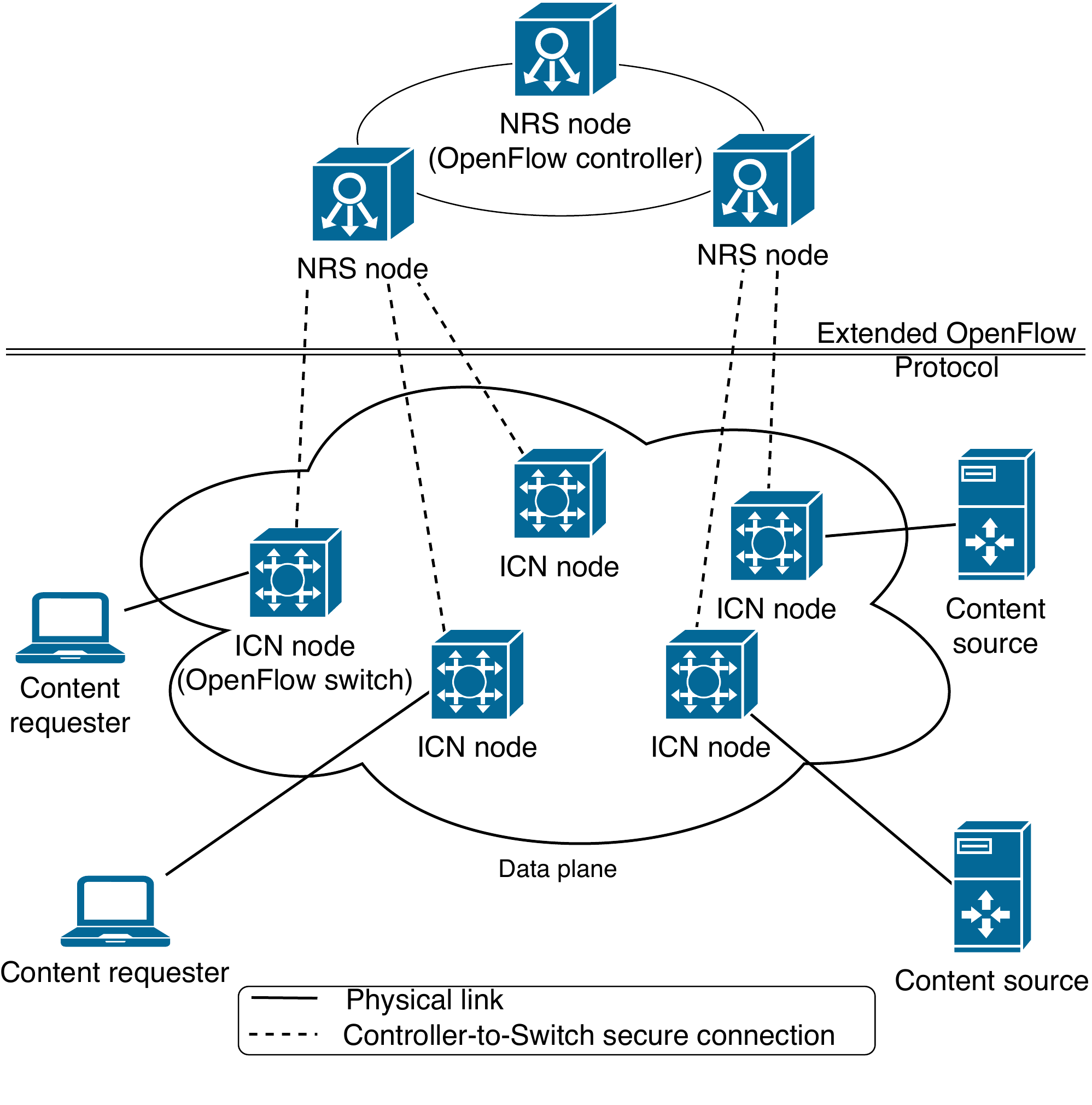}
	\caption{Simplified view of the coCONET architecture.}
	\label{fig:Veltri12}
\end{figure}

Similar to the work~\cite{vahlenkamp2013enabling}, Veltri~et~al.~\cite{veltri2012supporting} proposed a CONET~\cite{detti2011conet} inspired \gls{SDN}-based implementation of \gls{ICN}, called coCONET. \review{Fig.}~\ref{fig:Veltri12} presents a simplified view of this solution. In this architecture, \gls{ICN} nodes and user-terminals form the data plane and \emph{Name Resolution Service (NRS)} nodes are placed in the control plane. Moreover, \emph{\gls{ICN} node} works as an OpenFlow switch, while \emph{NRS node} works as an OpenFlow controller. To this end, the authors proposed to extend the OpenFlow protocol~\cite{mckeown2008openflow}.

\textbf{Deployment Approach.} Similar to the work~\cite{vahlenkamp2013enabling},  the proposed solution is an \emph{overlay} \gls{ICN} implementation as \gls{ICN} data is encapsulated inside the \gls{SDN}-based \gls{IP} packets.

\textbf{Deployment Scenarios.}
The proposed solution enables the \emph{ICN-ICN communication in IP ``ocean''} and the \emph{ICN-IP communication in IP ``ocean''} scenarios, where the underlying \gls{IP} network is managed by OpenFlow-based \gls{SDN} network.

\textbf{Addressed Coexistence Requirements.}
The present architecture provides the following coexistence requirements: 
\begin{itemize}
    \item Forwarding and Management - \gls{SDN}-based operations of the proposed approach support both forwarding and management of \gls{ICN} traffic.
    \item Storage - \gls{ICN} capable nodes cache the contents.
    \item Security - contents are cryptographically protected in order to assure content (and content generator) authentication and data integrity. This security service is provided through digital signature and can be verified through the public key associated to the private key of the content (or of the content generator). The proposed system enforces every \gls{ICN} node to verify such signature before forwarding the content toward the interested end-nodes, to protect the network against \gls{DoS} or other attacks.
\end{itemize}

%In addition to the forwarding and management issues, the authors discuss the possible solutions for storage and security issues of ICN's implementation via SDN's programability.

\textbf{\review{Additional architecture or Technology Used.}} Here, the authors focus specifically on OpenFlow-based \gls{SDN} implementations and target all the services/applications of the \gls{TCP}/\gls{IP} protocol stack. OpenFlow is a flavor of \gls{SDN}.

\textbf{Evaluation Parameters.}  The proposed solution requires \gls{ICN} capable OpenFlow network devices for \gls{ICN} operations. Due to such specific requirements, the hardware modifications and the time required for its deployment are high.
%%%%%%%%%%%%%%%%%%%%%%%%%%%%%%%%%%%%%%%%%%%%%%%%%%%%%%%%%%%%%%%%%%%%%%%%%%%%%%%%%%%%%%%%%%%%%%%%%%%%%%%%%%%%%%%%%%%%%%%%
\subsection{DOCTOR}
\gls{DOCTOR}~\cite{doctor} is an ongoing project funded by French Nation Research Agency. The project provides support towards the adoption of new standards by developing a secure use of virtualized network equipment. This leads to ease the deployment of novel networking architectures, thus enabling the coexistence of \gls{IP} and emerging stacks, such as \gls{NDN}, as well as the progressive migration of traffic from one stack to the other. DOCTOR proposes the use of \gls{NFV} infrastructure to achieve the incremental deployment of \gls{NDN} at a low cost. The project proposes an HTTP/\gls{NDN} gateway to interconnect \gls{ICN} ``islands'' to the \gls{IP} world, and an experimental architecture able to process the web traffic passing through a virtualized \gls{NDN} network. 
\par
\begin{figure}[ht!]
\centering
  \includegraphics[width=.9\columnwidth]{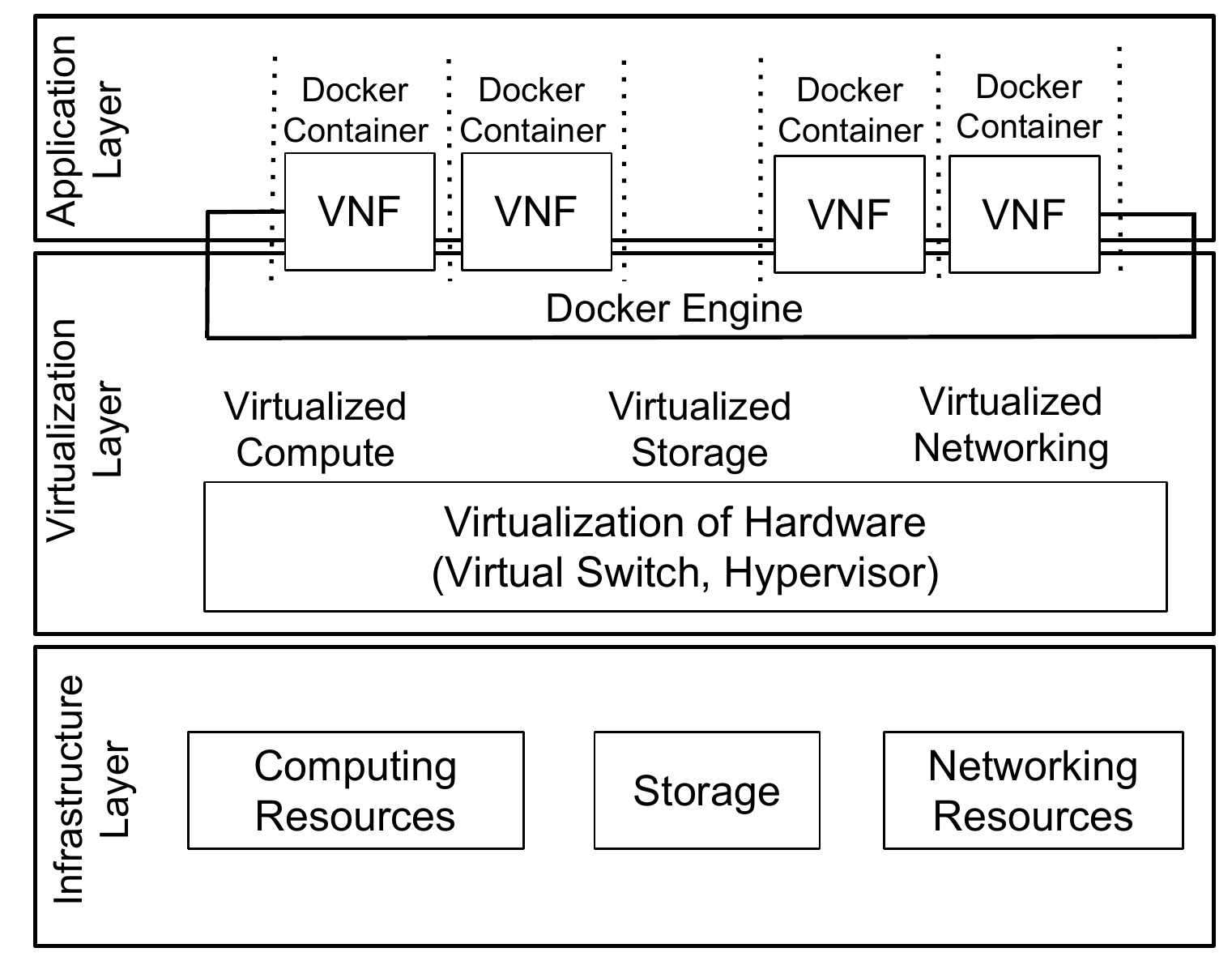}
\caption{Internal architecture of a DOCTOR virtualized node.}
  \label{Fig:doctor}
\end{figure}

 In particular, DOCTOR first deploys a virtual network based on OpenvSwitch to provide an end-to-end network connectivity between the virtualized network services and to enable a software control of the networking infrastructure. Then, it selects \gls{NDN} as an \gls{ICN} protocol stack. More specifically, the NDNx software is \textit{dockerized} to become a \gls{VNF}, deployable in DOCTOR architecture. In DOCTOR, \gls{NDN} is used both over \gls{IP} and over Ethernet since most \gls{NFV} tools are still \gls{IP}-dependent. To test the functionality of the coexistence, the web is considered as an application layer service due to its high popularity and predominance in the global network shares. However, since the current web clients and servers do not yet implement \gls{NDN}, dedicated gateways are used to perform an HTTP/\gls{NDN} conversion. Since these gateways are conceived as \gls{VNF}s, they can be deployed where and when required. In particular, two types of gateways are defined: (1) an \gls{iGW}, aimed at converting HTTP requests into \gls{NDN} Interest messages and \gls{NDN} Data messages into HTTP replies; (2) an \gls{eGW}, aimed at converting \gls{NDN} messages into HTTP requests, if the content is not available in the \gls{ICN} network, and HTTP replies into \gls{NDN} Data messages. \review{Fig.}~\ref{Fig:doctor} shows the high level architecture of a virtualized node in DOCTOR. The virtualized node is implemented on a single Linux server and it provides the required hardware resources for the \gls{VNF}s, which can act as various components (e.g., \gls{NDN} stack, \gls{IP} stack, and HTTP/\gls{NDN} gateway). 

\textbf{Deployment Approach.} DOCTOR uses an \textit{underlay} approach with the help of HTTP/\gls{NDN} gateways, that can map the HTTP protocol with \gls{NDN} messages and properly deliver the web content.

\textbf{Deployment Scenarios.} The \gls{iGW} and \gls{eGW} allow DOCTOR to support all the different deployment scenarios.

\textbf{Addressed Coexistence Requirements.} 
The DOCTOR architecture addresses the following coexistence requirements:
\begin{itemize}
    \item Forwarding - explicit name based routing of \gls{NDN} is performed at each router through the use of virtualized \gls{NDN} stack.
    \item Storage - content stores perform the content caching.
    \item Security - DOCTOR supports the same content oriented security as \gls{NDN}.
    \item Management - the control and management plane of \gls{VNF}s in DOCTOR has been designed with respect to the recommendations of the ETSI \gls{NFV} group, concerning the \gls{NFV} \gls{MANO}~\cite{docmang}.
\end{itemize}

\textbf{\review{Additional architecture or Technology Used.}} The architecture of DOCTOR is flexible, as it is based on \gls{NFV} and \gls{SDN} principles. Its main component is the \gls{NFV} infrastructure, which enables the resource virtualization to deploy the \gls{ICN} protocol stack over the data plane and the \gls{MANO} aspects over the control plane. As a computing virtualization framework, the architecture uses Docker, which relies on a lightweight virtualization principle

\textbf{Evaluation Parameters.} 
Among the key limitations of DOCTOR there is the latency, which occurs due to the repeated sending of requests to the \gls{ICN} servers, acting as gateways and attached to the content source. Since content names are different among each other, each new content name represents a new routing identifier to be given to the gateways. This results in a continuous interaction between content publisher and gateways for each HTTP request. 

%%%%%%%%%%%%%%%%%%%%%%%%%%%%%%%%%%%%%%%%%%%%%%%%%%%%%%%%%%%%%%%%%%%%%%%%%%%%%%%%%%%%%%%%%%%%%%%%%%%%%%%%%%%%%%%%%%%%%%%%
%\vspace{0.4cm}
\subsection{POINT}
The H2020 project \gls{POINT}~\cite{point} started in January 2015 and ended in December 2017. Its main purpose is to evaluate both quantitatively and qualitatively the improvements introduced by running \gls{ICN} over an \gls{IP} network. To achieve this aim, POINT designs an evolution of the PURSUIT architecture, which both leverages on the \gls{SDN} technology and on additional network components that enable \gls{IP}-based applications to run in the new setup without any modification. Those new elements are the \gls{NAP} and the \gls{ICN BGW}. The former directly interacts with the end user devices and is responsible for the translation of all the \gls{IP} protocol abstraction layers (e.g., HTTP, TCP and \gls{IP}) into the \gls{ICN} paradigm, while the latter controls the communication between \gls{ICN} and \gls{IP} networks. Furthermore, the \gls{NAP} provides standard gateway functions such as \gls{NAT}, firewall, and dynamic \gls{IP} address assignment. The core \gls{ICN} functionalities are provided by the PURSUIT components (i.e., \gls{TM}, \gls{FN}, and \gls{RP}). Usually, content items are assigned a \gls{RID} and are stored on the publisher, which advertises the contents availability in the network. Then, a user device sends a request for a content item and the \gls{NAP} transforms the interest into a subscription for a specific \gls{RID}. The subscription is then sent to the RP, which triggers the TM towards the identification of a path between publisher and subscriber. The TM identifies all the nodes that need to be traversed and it calculates the associated \gls{FIs}, which are placed in the packet header. At this point, the \gls{SDN} switches are responsible for forwarding the packets by using only the \gls{FIs} and not the routing tables. The \gls{SDN} switches are not aware of the POINT architecture and are, instead, coordinated by an \gls{SDN} controller, which communicates directly with the TM. This communication is bidirectional since the \gls{SDN} controller informs the TM about any topology modification, and the TM notifies the \gls{SDN} controller about the configuration to be placed on the \gls{SDN} switches. 

\begin{figure}[!htbp]
	\centering
	\includegraphics[width=0.4\textwidth]{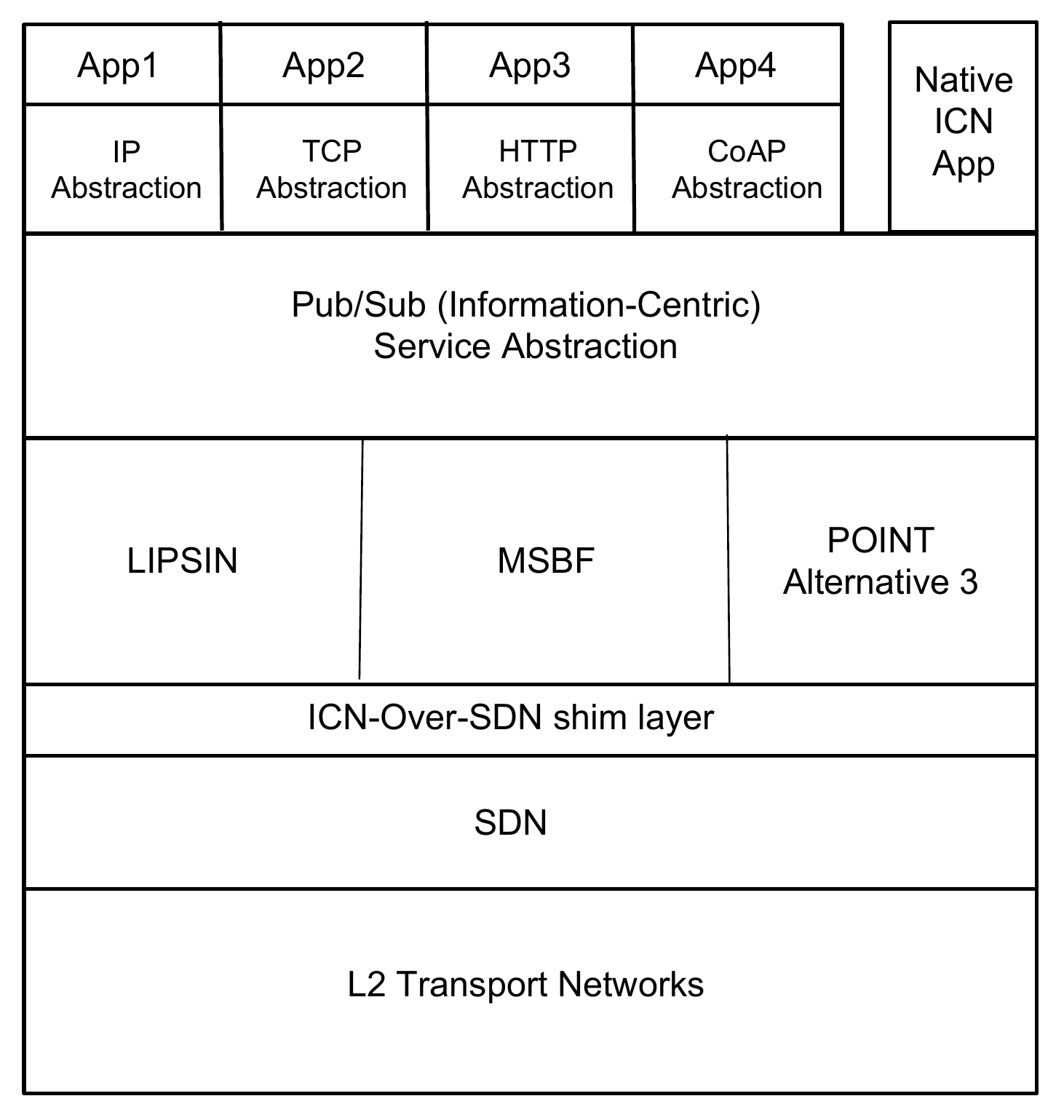}
	\caption{Internal architecture of a POINT node.}
	\label{fig:point}
\end{figure}

\review{Fig.}~\ref{fig:point} shows the internal architecture of a POINT node. In the upper layer of the node, there are generic applications (i.e., \emph{App1}, \emph{App2}, \emph{App3}, \emph{App4}) which interact with a set of abstractions provided by POINT (i.e., \emph{\gls{IP} Abstraction}, \emph{TCP Abstraction}, \emph{HTTP Abstraction}, \emph{CoAP Abstraction}). Those are aimed at enabling the communication between applications and \gls{ICN} networks without requiring any modification from the application interface side. Each abstraction, then, cooperates with the \emph{Pub/Sub (Information-centric) Service Abstraction} to adhere to a publish/subscribe paradigm, where information is delivered according to specific strategies (i.e., \emph{LIPSIN}, \emph{MSBF}, \emph{POINT Alternative3}). Finally, POINT exploits also the \gls{SDN} technology by introducing two new layers (i.e., \emph{ICN-over-\gls{SDN} shim layer} and \emph{\gls{SDN}}) just above the \emph{L2 Transport Network} layer. 

\textbf{Deployment Approach.}
The POINT project falls under the \emph{underlay} deployment approach due to the gateway components, which are responsible for the translation from the \gls{IP} semantics into the \gls{ICN} semantics. 

\textbf{Deployment Scenarios.}
The main purpose of the POINT architecture is to enable different subnetworks to communicate between each other. Thus, POINT supports the \emph{Border Island} scenario.  

\textbf{Addressed Coexistence Requirements.}
Given that POINT is an evolution of PURSUIT, they both share the same coexistence requirements, i.e., forwarding, storage, and security. 

\textbf{\review{Additional architecture or Technology Used.}}
The POINT solution relies on both the PURSUIT architecture and the \gls{SDN} technology.

\textbf{Evaluation Parameters.}
The challenges introduced by the POINT project involve scalability, dynamic network management and latency of data transmission. The first two challenges refer to the appropriate configuration of \gls{SDN} switches to face an automatic update of the network topology (e.g., a new host being attached). On the contrary, the third challenge might be due to the high frequency of interaction between \gls{NAP}s and \gls{RP}s. 

\subsection{RIFE}
The \gls{RIFE}~\cite{rife} architecture is a Horizon2020 funded project, which started in February 2015 and ended in January 2018. Its aim is to develop a new network infrastructure that brings connectivity to communities living in remote locations or unable to afford the communication network costs. To achieve the purpose, the RIFE project focuses on three different challenges regarding the current end-to-end communication paradigm: reduction of capacity, energy, and redundant contents available in the network. The first can be achieved through a time-shifted access to network services and applications. The energy consumed by connected devices can be reduced by introducing a tolerance delay in the communication, so that devices can stay in an idle mode during the absence of network activity. Finally, the third aim is achievable by serving the same content to all the clients that require it, instead of releasing each time a new copy. The architecture addressing those objectives is a combination of \gls{IP}, \gls{ICN}, and \gls{DTN} paradigms.  

\textbf{Deployment Approach.}
The RIFE architecture follows the \emph{underlay} approach because of the gateway components, which are responsible for the translation from the \gls{IP} semantics into the \gls{ICN} semantics. 

\textbf{Deployment Scenarios.}
RIFE supports the \emph{Border Island} scenario.

\textbf{Addressed Coexistence Requirements.}
RIFE is an evolution of the PURSUIT architecture. Thus, the coexistence requirements addressed are the same, i.e. forwarding, storage, and security. 

\textbf{\review{Additional architecture or Technology Used.}}
The architecture proposed in the RIFE project is a modification of the PURSUIT architecture and it relies on the coexistence of \gls{IP}, \gls{ICN} and \gls{DTN}. This last architecture is responsible for introducing the delay and disruption tolerance required to enable the time-shift requirement. 

\textbf{Evaluation Parameters.}
No challenges have been found for the RIFE project. 

\vspace{0.4cm}
\subsection{CableLabs}
Among the different \emph{underlay} approaches, there is a solution designed by CableLabs, which is a non-profit Innovation and R\&D lab focused on the introduction of fast and secure release of data, video, voice, and services to end users. CableLas proposes an incremental introduction of \gls{CCN}/\gls{NDN} in the existing \gls{CDN}s to improve the overall content distribution without modifying \gls{IP} routers~\cite{cableLabs}. The architecture designed by CableLabs requires first a migration of some services/applications to the \gls{ICN} paradigm, and then the introduction of proxies. Those are able to manage the translation between HTTP and \gls{CCN}. Once several \gls{ICN} ``islands'' are deployed in the network, the communication among them is provided through \gls{IP} tunneling. 

\textbf{Deployment Approach.}
The solution proposed by CableLabs adopts the \emph{underlay} approach because of the gateway components, which are responsible for the translation from the \gls{IP} semantics into the \gls{ICN} semantics. 

\textbf{Deployment Scenarios.}
Except for the \emph{Border Island}, the CableLabs architecture supports all the deployment scenarios. 

\textbf{Addressed Coexistence Requirements.}
The CableLabs architecture addresses the following coexistence requirements: 
\begin{itemize}
    \item Forwarding - the additional proxies introduced in the network to support the translations i.e., HTTP to \gls{CCN} and \gls{CCN} to HTTP, also work as \gls{CCN} forwarder.
    \item Storage - as the architecture is an evolution of a \gls{CDN}, by design the network nodes can cache contents.
\end{itemize}

\textbf{\review{Additional architecture or Technology Used.}}
Throughout this project, CableLabs investigates how the \gls{CCN} infrastructure is better in supporting a content-oriented network with respect to the current solutions, such as CDNs. Thus, CableLabs illustrates an incremental deployment of a \gls{CCN} network over a \gls{CDN} existing one. 

\textbf{Evaluation Parameters.}
The challenges identified by CableLabs with respect to their own architecture are as follows: traffic management, optimization of \gls{CCN} router implementation (e.g., FIB/PIT sizing and memory bandwidth), optimization of \gls{CCN} cache implementation, content object size and fragmentation (i.e., definition of the maximum content object size transmissible inside a network), \gls{CCN} to HTTP and HTTP to \gls{CCN} conversions (e.g, the computational complexity of the translation function).

%%%%%%%%%%%%%%%%%%%%%%%%%%%%%%%%%%%%%%%%%%%%%%%%%%%%%%%%%%%%%%%%%%%%%%%%%%%%%%%%%%%%%%%%%%%%%%%%%%%%%%%%%%%%%%%%%%%%%%%%
\vspace{0.4cm}
\subsection{NDN-LAN}
\begin{figure}[ht!]
\centering
  \includegraphics[width=.9\columnwidth]{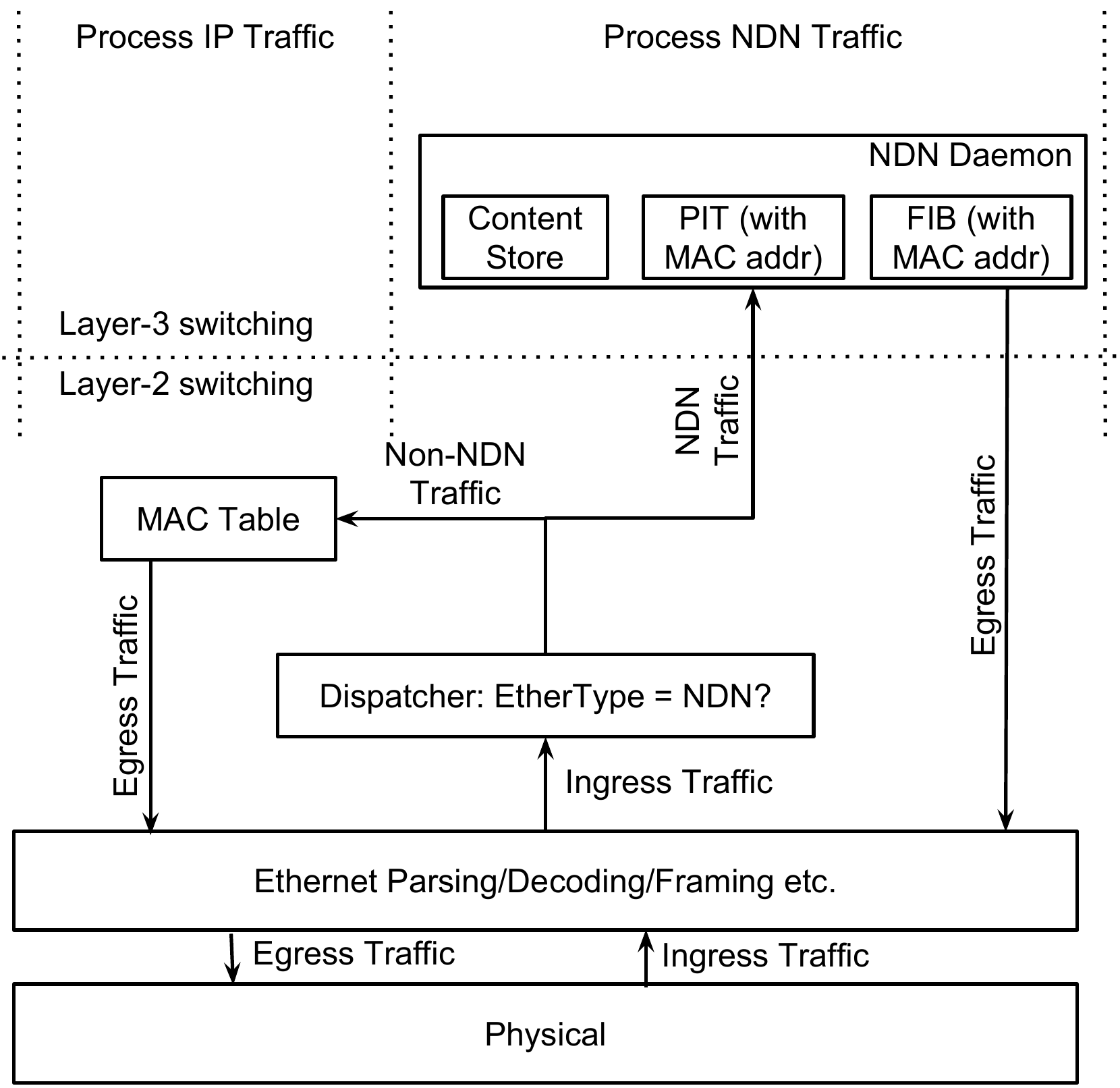}
\caption{Dual-stack switch internal architecture.}
  \label{Fig:D-switch}
\end{figure}

The authors in~\cite{NDNLAN} propose a \emph{hybrid} \gls{ICN} architecture in which content names are mapped to the MAC addresses. In particular, the authors present the design of a \gls{D-switch}, which provides name-based forwarding for \gls{NDN} traffic and address-based forwarding for conventional traffic such as \gls{IP}. It can be seen from \review{Fig.}~\ref{Fig:D-switch} that the key component of D-switch architecture is the \textit{Dispatcher}, which checks the \emph{EtherType} field in the header of a received frame. When an \gls{IP} frame is detected, the D-switch works like a traditional Ethernet switch and it forwards the frame using the MAC address. If an \gls{NDN} frame (i.e., Interest or Data packet) is detected, the D-switch processes/forwards the frame based on the content name carried in the \gls{NDN} header (i.e., Layer 3). In particular, the dispatcher either selects the \textit{Process \gls{IP} Traffic} or \textit{Process \gls{NDN} Traffic} module in the D-switch based on the value of \emph{EtherType} field. In the \textit{Process \gls{NDN} Traffic} module, the PIT and FIB tables are modified to store the mapping between the content names and MAC addresses. For instance, when an Interest packet is received, the D-switch will forward it by searching the content name and its corresponding MAC in the FIB, and then fill the destination MAC address field in Ethernet header with the recorded MAC address. 

\textbf{Deployment Approach.} This coexistence approach falls under the \emph{hybrid} approach because the D-switches are able to process both types of traffic (i.e., \gls{IP} and \gls{NDN}). In particular, a LAN consists (fully or partially) of D-switches that can process the data traffic received from \gls{NDN}-enabled hosts, as well as \gls{IP} hosts. However, a fully \emph{hybrid} scenario needs to be consistent with D-switches only, else other techniques or polices/rules are required to perform the data forwarding.

\textbf{Deployment Scenarios.}
Since the D-switches allow \gls{NDN} traffic to run within the \gls{IP} network, except for the \emph{Border Island}, NDN-LAN supports all the deployment scenarios. As a matter of fact, due to the use of MAC-layer encapsulation only, the inter-network communications are not possible and the \emph{Border island} scenario cannot be supported.

\textbf{Addressed Coexistence Requirements.}    
The present architecture provides the following coexistence requirements: 
\begin{itemize}
    \item Forwarding - full advantage of \gls{ICN} features, such as in-network caching and native multicast, is supported when the underlying LAN consists of D-switches only. However, when the LAN has both D-switch and conventional Ethernet switches, it has to be carefully designed to avoid conflict between name-based forwarding and address-based forwarding.
    \item Storage - in-network caching is only supported at D-switches, and it is responsibility of the network manager to prevent the conventional Ethernet switches from receiving \gls{ICN} packets.
    \item Management - management of such a deployment is challenging due to limitations of topology creation and forwarding rules installation.
\end{itemize}

\textbf{\review{Additional architecture or Technology Used.}}
NDN-LAN is mainly suitable for \gls{NDN} applications that run in small and private networks such as university campus and within an organization. However, the proposed coexistence solution aims to support a variety of applications which includes \gls{NDN} as well as \gls{IP} applications. This is achieved through the following design goals: (i) coexistence with \gls{IP} traffic, which ensures that the common mechanisms should run without any change or performance penalty, (ii) native \gls{NDN} support, by not relying on tunnels or overlays, and (iii) incremental deployment and general applicability. The proposed solution does not make use of any specific technology to implement the D-switch logic. Minor hardware and software changes in the D-switches allow them to process the \gls{IP} and \gls{NDN} traffic in a controlled environment (i.e., LAN).

\textbf{Evaluation Parameters.}
To implement the required logic and functionalities at D-switches so that it can support \gls{NDN}-enabled traffic processing, some changes are required in the switch hardware, as well as software. Additional forwarding polices need to be installed in scenarios where D-switches coexist with conventional Ethernet switches. Without any standardization of these new software/hardware components, the applicability of the proposed solution in real-world coexistence applications is limited. Designing mechanisms that support the name-based forwarding, meanwhile coexisting with address-based forwarding within the same \gls{LAN}, is a challenging task. Additionally, the process for D-switches to learn the forwarding table at Layer-2 and build name-based FIB at Layer-3 is an open problem that needs to be addressed. In \gls{LAN}, the implementation of the proposed solution is simple and straightforward. However, as the LAN size increases and communication between different \gls{LAN}s is needed, the deployment cost will increase significantly, and the current solution needs to be extended to deal with new issues such as interoperability and scalability. 

%%%%%%%%%%%%%%%%%%%%%%%%%%%%%%%%%%%%%%%%%%%%%%%%%%%%%%%%%%%%%%%%%%%%%%%%%%%%%%%%%%%%%%%%%%%%%%%%%%%%%%%%%%%%%%%%%%%%%%%%
\vspace{0.4cm}
\subsection{hICN} 
\begin{figure}[h!]
\centering
  \includegraphics[width=0.45\textwidth]{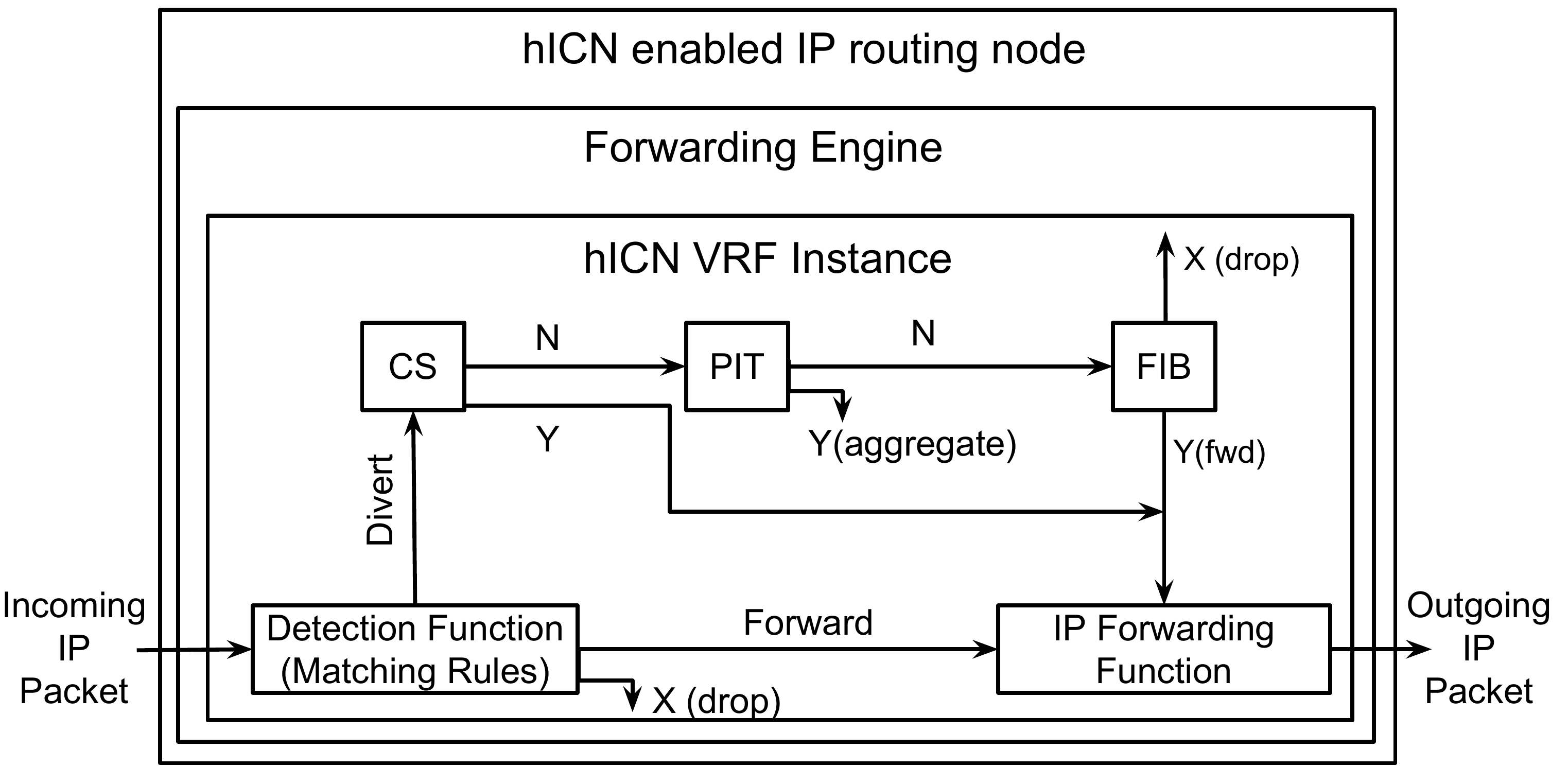}
\caption{Internal architecture of an hICN node.}
  \label{Fig:hICN_node}
\end{figure}

Authors in~\cite{hICN} propose methods and systems to facilitate the integration of \gls{ICN} into \gls{IP} networks. The \gls{hICN} communication system claims to have the ability to preserve \gls{ICN} features and advantages, while, at the same time, benefiting from exploiting an existing \gls{IP} infrastructure. The major components of hICN communication system are as follows: (i) hICN-enabled \gls{IP} router(s), capable of processing and forwarding both regular \gls{IP} packets and \gls{IP} packets enhanced with \gls{ICN} semantics, (ii) \gls{IP} router(s), capable of handling \gls{IP} packets, and (iii) hICN router(s), being provisioned with a consumer or producer application. The traditional \gls{IP} packet headers have been modified to add the \gls{ICN} semantics. As it is shown in \review{Fig.}~\ref{Fig:hICN_node}, when a router receives an \gls{IP} packet, then according to the \gls{IP} header content, it can identify how to process it, i.e., using \gls{ICN} or \gls{IP} stack. The authors suggest two possible name mapping schemes for hICN content names to \gls{IP}: (i) pure \gls{IP} mapping, in which content name components can be directly encoded in the \gls{IP} header, and (ii) optimized mapping, in which a subset of the content name component is encoded in the network header, while the remainder is encoded in the transport header.

\textbf{Deployment Approach.}
As the hICN-enabled \gls{IP} routers are able to process the \gls{IP}, as well as the \gls{ICN} traffic, hICN falls under the \emph{hybrid} deployment approach. However, unlike NDN-LAN, in which MAC-to-content name mapping and conversely is performed, in hICN, the \gls{IP}-to-content name and conversely is done.

\textbf{Deployment Scenarios.}
Due to the presence of dual stack routers, the proposed architecture supports all the deployment scenarios.

\textbf{Addressed Coexistence Requirements.}
hICN is among the best proposals supporting the coexistence because it retains most of the \gls{ICN} basic features (e.g., layer-3 name-based routing, partial symmetric routing, object-based security, anchorless mobility, and in-network reactive caching). This is because hICN exploits the IPv4 and IPv6 header fields content semantic to identify whether the received packet is an \gls{IP} Data packet or an \gls{IP} Interest packet. The use of IPv4 or IPv6 RFC compliant packet formats guarantees the communication between an IPv4/IPv6 router and a hICN router. More specifically, the hICN router processes and forwards both the regular \gls{IP} packets and the \gls{ICN}-semantic-based packets. Hence, it preserves pure \gls{ICN} behavior at Layer-3 and above by guaranteeing end-to-end service delivery between data producers and data consumers using \gls{ICN} communication principles. The present architecture provides the following coexistence requirements: 
\begin{itemize}
    \item Forwarding - the hICN-enabled \gls{IP} routers as well as \gls{IP} routers use the same forwarding module.
    \item Storage - the cache stores are available on hICN-enabled \gls{IP} routers, and the Interest packets could be satisfied by these routers if the requested content is available in the router cache. 
    \item Management - for large scale usage of this architecture, the consumer and producer applications must have the mapping of content-names with the corresponding \gls{IP} addresses, so that the \gls{ICN} packets can be processed seamlessly by the non-\gls{ICN} enabled routers as well. 
    \item Security - the architecture provides the same security features that are provided by \gls{ICN}. However, the \gls{IP}-only routers are not able to check the received data packets integrity and authentication, hence, at least one hICN-enabled \gls{IP} router must be available in the route between the consumer and producer.
    %\item Interoperability: The proposed protocol is interoperable as long as the consumer and producer applications are designed as per the design given in the hICN architecture.
\end{itemize}

\textbf{\review{Additional architecture or Technology Used.}}
The hICN proposal uses the \gls{IP} packet header semantics to differentiate the \gls{ICN} and \gls{IP} packets, and the mapping table at hICN-enabled router or \gls{DNS} is used for performing the mapping task. To support the interoperablity among different networks, the edge router could translate the incoming packets to hICN compliant packets using a proxy. Therefore, hICN does not use any specific architecture (e.g., \gls{SDN}) or technology (e.g., virtualization or tunnelling) to perform the coexistence.

\textbf{Evaluation Parameters.}
The major challenges of hICN are similar to the other \emph{hybrid} approaches and include a lack of support for heterogeneity, scalability, and standardization of the proposed changes in the traditional Internet protocols and network components. Moreover, the communication delay caused by the additional time used by hICN routers for the mapping could be an issue for delay sensitive applications. The hardware modifications are minimal because the hICN routers can be created by installing a software bundle in the existing \gls{IP} routers. However, the memory requirements will increase due to the need of storage cache. The deployment effort will be considerable due to the need of the modifications in the consumers and producers applications.

\subsection{OFELIA}
\begin{figure}[ht!]
	\centering
	\includegraphics[trim = 2mm 10mm 2mm 25mm, clip, width=0.4\textwidth]{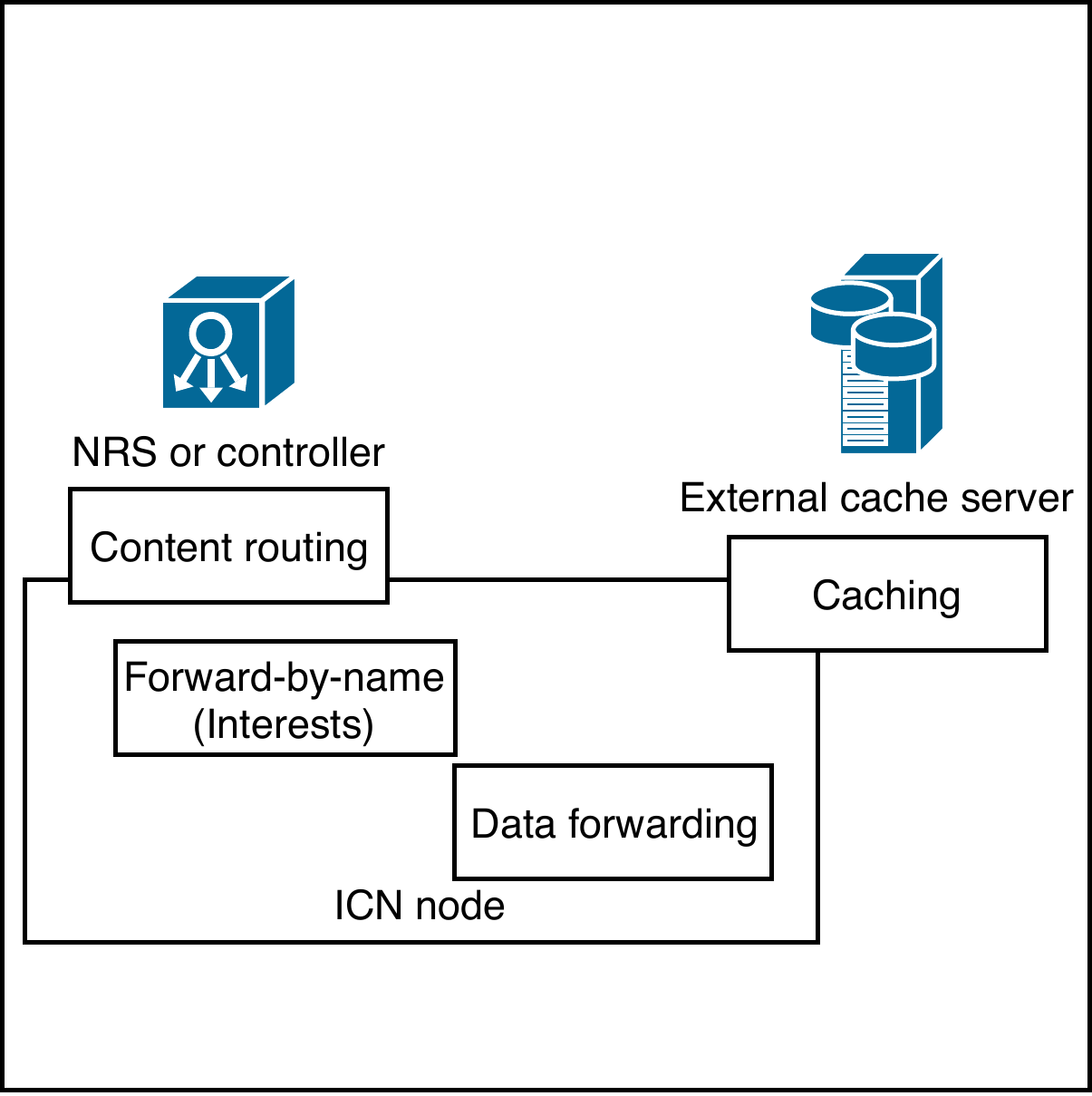}
	\caption{Simplified view of the solution proposed by OFELIA.}
	\label{fig:melazzi}
\end{figure}

Blefari Melazzi~et~al.~\cite{melazzi2012openflow} proposed an \gls{SDN}-based \emph{hybrid} implementation of \gls{ICN} under the OFELIA project. The proposed approach is an extension of the CONET architecture~\cite{detti2011conet} for OpenFlow networks, where dedicated \gls{BN}s perform name-to-location resolution, using an external system, for any requested \gls{NDO}. \review{Fig.}~\ref{fig:melazzi} presents a simplified view of this solution. The authors propose to include two different forwarding strategies in an \emph{\gls{ICN} node}: (1) to forward content requests; and (2) to deliver the data. \emph{Forward-by-name} feature of an \emph{\gls{ICN} node} applies to Interest packets, while \emph{Data Forwarding} is the mechanism that allows the content to be sent back to the device that issued a content request. \emph{Content routing} is used to disseminate information about location of contents, and \emph{Caching} is the ability of \gls{ICN} nodes to cache data and to directly reply to incoming content requests. \review{The OFELIA testbed was used in IRATI~\cite{rina} project for experimental activities.}

\textbf{Deployment Approach.} The proposed architecture adheres to a \emph{hybrid} approach.

\textbf{Deployment Scenarios.} 
The proposed implementation of \gls{ICN} is an extension of the CONET framework, in which \gls{BN}s interconnect different \gls{CSS}s. Hence, this solution supports the \emph{Border Island} scenario.

\textbf{Addressed Coexistence Requirements.}
The proposed system is based on CONET framework. Extending the primary goals of CONET framework, this architecture aims to support forwarding, storage, security and management for \gls{ICN} deployment.
%The present architecture provides the following coexistence requirements: 
%\begin{itemize}
%    \item Forwarding: \todo{Motivate}
%    \item Storage: \todo{Motivate}
%    \item Management: \todo{Motivate}
%    \item Security: \todo{Motivate}
%\end{itemize}

\textbf{\review{Additional architecture or Technology Used.}} The present solution strongly relies on the architecture proposed in the CONET project and, through \gls{SDN}/OpenFlow, it targets all the services/applications of the \gls{TCP}/\gls{IP} protocol stack. 

\textbf{Evaluation Parameters.} The architecture of the solution requires the networking elements to be OpenFlow compliant. Given that OpenFlow (\gls{SDN}) has been widely adopted in the networking domain, the hardware modifications and the time required for its deployment are low in scenarios where OpenFlow-based network is already present. On another side, the hardware modifications and the time required for its deployment would be higher if OpenFlow-based network is not already present.

%% file: discussion_conclusion.tex
\section{\review{Discussion}}
\label{discussion_conclusion}
%Here, we summarize the lessons learned (Section~\ref{lessons_learned}) and we provide possible future research directions (Section~\ref{future}).

\review{The purpose of this section is to summarize the findings achieved through our systematic analysis of all the existing coexistence architectures (Section~\ref{Summary}) and discuss the open challenges (Section~\ref{challenges}), along with some future directions concerning the coexistence between the current and the future Internet architectures (Section~\ref{future_directions}).}

%\subsection{Lessons Learned}
%\label{lessons_learned}

\subsection{\review{Summary of the survey}}
\label{Summary}
The main aim of this survey is to provide the necessary overview of the available solutions that already address the coexistence. We believe that it will help to move the research community towards the design of the most appropriate architecture for the future Internet. Thus, to guide the reader towards the interpretation of Table~\ref{table:comparison}, we add here two new tables, which are a summary of Table~\ref{table:comparison}. In particular, among all the features and evaluation parameters considered in this survey, the only ones that can be chosen by a network designer are the deployment approach and the possible \review{additional architecture or technology used} in the design of his solution. Thus, Table~\ref{LABEL1} and Table~\ref{LABEL2} are aimed at comparing each deployment approach and each \review{additional architecture or technology used} with respect to all the other features and evaluation parameters, respectively. As a matter of fact, the deployment scenarios, as well as the addressed coexistence requirements, directly depend on the deployment approach or on the \review{additional architecture or technology}, while the evaluation parameters are dynamic properties evaluated during the runtime deployment of an architecture. 
\par The content of the cells as well as their meaning is shared between Table~\ref{LABEL1} and Table~\ref{LABEL2}. More specifically, the content of each cell corresponds to the number of coexistence architectures addressing both the properties specified in the corresponding row and column (e.g., in the first cell of Table~\ref{LABEL1} the value equal to 7 means that there are 7 coexistence architectures adhering to the \emph{overlay} approach and supporting the \emph{forwarding} functionality). The meaning of the values in the cells is different throughout the table. \review{In the upper part (i.e., rows referring to addressed coexistence requirements and deployment scenarios), the value in the cell refers to the number of architectures that guarantee a specific addressed coexistence requirement or a deployment scenario by adopting a deployment approach (listed in the columns). On the contrary, in the lower part of the table (i.e., rows referring to the evaluation parameters), the value in the cells refers to the number of limitations an architecture is affected from.}

\par Table~\ref{LABEL1} shows on the columns the three different deployment approaches (i.e., \emph{overlay}, \emph{underlay} and \emph{hybrid}), while on the rows there are all the other features, except for the architectures or technologies used, considered in Table~\ref{LABEL2}. Considering the deployment approaches, we found six architectures adopting the \emph{overlay} solution, four the \emph{underlay}, three the \emph{hybrid} and one architecture (i.e., CONET) adhering to both \emph{overlay} and \emph{hybrid}. As it is shown in the table, a plausible reason for this greater adoption of the \emph{overlay} approach might be the higher number of addressed coexistence requirements provided by it. As a matter of fact, almost all the \emph{overlay} architectures guarantee the forwarding and storage features and the number of the architectures supporting security and management is higher than in the \emph{underlay} and \emph{hybrid} cases. While, adopting an \emph{overlay} approach prevents architectures from being deployed in all the deployment scenarios: none of the \emph{overlay} architectures covers either the \emph{ICN-IP communication in ICN ``ocean''} or the \emph{IP-IP communication in ICN ``ocean''} scenarios. Finally, considering the evaluation parameters, most \emph{overlay} architectures are not able to properly manage the network traffic, but the other limitations are comparable with the ones affecting the \emph{underlay} and \emph{hybrid} solutions. Moreover, even if the number of challenges under the last class (i.e., \emph{Other}) might be significant, we \review{note} that those limitations strongly depend on the design of each coexistence architecture. 

\begin{table}[ht!]
\centering
\resizebox{.95\columnwidth}{!}{
\begin{threeparttable}
\centering
\caption{Comparison of all the deployment approaches for coexistence architectures - The value of each cell refers to the number of coexistence architectures addressing both the properties specified in the corresponding row and column.}
\label{LABEL1}
\begin{tabular}{cc|c|c|c|}
\cline{3-5}
\textbf{} &  & \multicolumn{3}{c|}{\textbf{Deployment Approach}} \\ \cline{3-5} 
\textbf{} &  & Overlay & Underlay & Hybrid \\ \hline
\multicolumn{1}{|c|}{\multirow{4}{*}{\textbf{\begin{tabular}[c]{@{}c@{}}Addressed\\ coexistence\\ requirements\end{tabular}}}} & Forwarding & 7 & 4 & 4 \\ \cline{2-5} 
\multicolumn{1}{|c|}{} & Storage & 6 & 4 & 4 \\ \cline{2-5} 
\multicolumn{1}{|c|}{} & Security & 4 & 3 & 2 \\ \cline{2-5} 
\multicolumn{1}{|c|}{} & Management & 3 & 1 & 3 \\ \hline
\multicolumn{1}{|c|}{\multirow{5}{*}{\textbf{\begin{tabular}[c]{@{}c@{}}Deployment\\ scenarios\end{tabular}}}} & \begin{tabular}[c]{@{}c@{}}ICN-ICN\\ communication\\ in IP ``ocean''\end{tabular} & 7 & 2 & 3 \\ \cline{2-5} 
\multicolumn{1}{|c|}{} & \begin{tabular}[c]{@{}c@{}}ICN-IP\\ communication\\ in IP ``ocean''\end{tabular} & 2 & 2 & 2 \\ \cline{2-5} 
\multicolumn{1}{|c|}{} & \begin{tabular}[c]{@{}c@{}}ICN-IP\\ communication\\ in ICN ``ocean''\end{tabular} & 0 & 2 & 2 \\ \cline{2-5} 
\multicolumn{1}{|c|}{} & \begin{tabular}[c]{@{}c@{}}IP-IP\\ communication\\ in ICN ``ocean''\end{tabular} & 0 & 2 & 2 \\ \cline{2-5} 
\multicolumn{1}{|c|}{} & Border Island & 2 & 3 & 3 \\ \hline
\multicolumn{1}{|c|}{\multirow{6}{*}{\textbf{\begin{tabular}[c]{@{}c@{}}Evaluation\\ parameter\end{tabular}}}} & Traffic management & 4 & 1 & 1 \\ \cline{2-5} 
\multicolumn{1}{|c|}{} & Access control & 1 & 0 & 0 \\ \cline{2-5} 
\multicolumn{1}{|c|}{} & Scalability & 2 & 1 & 2 \\ \cline{2-5} 
\multicolumn{1}{|c|}{} & \begin{tabular}[c]{@{}c@{}}Dynamic network\\ management\end{tabular} & 1 & 1 & 1 \\ \cline{2-5} 
\multicolumn{1}{|c|}{} & Latency & 0 & 2 & 2 \\ \cline{2-5} 
\multicolumn{1}{|c|}{} & Other & 4 & 4 & 2 \\ \hline
\end{tabular}
\begin{tablenotes}
\end{tablenotes}
\end{threeparttable}
}
\end{table}

Table~\ref{LABEL2} contains the same rows as Table~\ref{LABEL1}, while on the columns it shows all \review{the additional architectures or technologies used} in the analyzed coexistence solutions. Throughout this survey, we found the following results: one coexistence solution relying on the \gls{PSIRP} architecture, two on LAN, one on SAIL, six on \gls{SDN}, two on \gls{PURSUIT}, one on \gls{CDN}, one on \gls{DTN}, one on CONET, and one on \gls{DNS}. As it is clearly visible from the table, the reason for adopting the \gls{SDN} technology in a coexistence scenario is given by its numerous benefits in terms of both features and evaluation parameters with respect to the other possible solutions.

\begin{table*}[!htbp]
\centering
\scalebox{1}{
\begin{threeparttable}
\centering
\caption{Comparison of all \review{the additional architectures or technologies used} in coexistence architectures - The value of each cell refers to the number of coexistence architectures addressing both the properties specified in the corresponding row and column.}
\label{LABEL2}
\begin{tabular}{cc|c|c|c|c|c|c|c|c|c|}
\cline{3-11}
 &  & \multicolumn{9}{c|}{\textbf{\review{Additional architecture or technology used}}} \\ \cline{3-11} 
 &  & PSIRP & LAN & SAIL & SDN & PURSUIT & CDN & DTN & CONET & DNS \\ \hline
\multicolumn{1}{|c|}{\multirow{4}{*}{\textbf{\begin{tabular}[c]{@{}c@{}}Addressed\\ coexistence\\ requirements\end{tabular}}}} & Forwarding & 1 & 2 & 1 & 6 & 2 & 1 & 1 & 1 & 1 \\ \cline{2-11} 
\multicolumn{1}{|c|}{} & Storage & 1 & 2 & 1 & 5 & 2 & 1 & 1 & 1 & 1 \\ \cline{2-11} 
\multicolumn{1}{|c|}{} & Security & 1 & 1 & 0 & 4 & 2 & 0 & 1 & 1 & 1 \\ \cline{2-11} 
\multicolumn{1}{|c|}{} & Management & 0 & 0 & 0 & 4 & 0 & 0 & 0 & 1 & 1 \\ \hline
\multicolumn{1}{|c|}{\multirow{5}{*}{\textbf{\begin{tabular}[c]{@{}c@{}}Deployment\\ scenarios\end{tabular}}}} & \begin{tabular}[c]{@{}c@{}}ICN-ICN\\ communication\\ in IP ``ocean''\end{tabular} & 1 & 2 & 1 & 4 & 0 & 1 & 0 & 0 & 1 \\ \cline{2-11} 
\multicolumn{1}{|c|}{} & \begin{tabular}[c]{@{}c@{}}ICN-IP\\ communication\\ in IP ``ocean''\end{tabular} & 0 & 1 & 0 & 3 & 0 & 1 & 0 & 0 & 1 \\ \cline{2-11} 
\multicolumn{1}{|c|}{} & \begin{tabular}[c]{@{}c@{}}ICN-IP\\ communication\\ in ICN ``ocean''\end{tabular} & 0 & 1 & 0 & 1 & 0 & 1 & 0 & 0 & 1 \\ \cline{2-11} 
\multicolumn{1}{|c|}{} & \begin{tabular}[c]{@{}c@{}}IP-IP\\ communication\\ in ICN ``ocean''\end{tabular} & 0 & 1 & 0 & 1 & 0 & 1 & 0 & 0 & 1 \\ \cline{2-11} 
\multicolumn{1}{|c|}{} & Border Island & 0 & 0 & 1 & 4 & 2 & 0 & 1 & 1 & 1 \\ \hline
\multicolumn{1}{|c|}{\multirow{6}{*}{\textbf{\begin{tabular}[c]{@{}c@{}}Evaluation\\ parameter\end{tabular}}}} & Traffic management & 1 & 2 & 1 & 1 & 0 & 1 & 0 & 0 & 0 \\ \cline{2-11} 
\multicolumn{1}{|c|}{} & Access control & 0 & 0 & 0 & 0 & 0 & 0 & 0 & 0 & 0 \\ \cline{2-11} 
\multicolumn{1}{|c|}{} & Scalability & 0 & 1 & 1 & 3 & 1 & 0 & 0 & 0 & 1 \\ \cline{2-11} 
\multicolumn{1}{|c|}{} & \begin{tabular}[c]{@{}c@{}}Dynamic network\\ management\end{tabular} & 0 & 1 & 1 & 2 & 1 & 0 & 0 & 0 & 0 \\ \cline{2-11} 
\multicolumn{1}{|c|}{} & Latency & 0 & 1 & 0 & 2 & 1 & 0 & 0 & 0 & 1 \\ \cline{2-11} 
\multicolumn{1}{|c|}{} & Other & 0 & 0 & 0 & 3 & 0 & 4 & 0 & 1 & 0 \\ \hline
\end{tabular}
\begin{tablenotes}
\end{tablenotes}
\end{threeparttable}
}
\end{table*}

\subsection{\review{Open Challenges}}
\label{challenges}
\review{According to our findings, the following challenges need to be addressed while designing an efficient and secure coexistence architecture.} 
\begin{itemize}
     
\item \textbf{\review{Traffic management:}} \review{the existing Internet applications are not completely compatible with architectures implementing the \emph{overlay} approach~\cite{ccnxudp,Zhang,NDNLP,ccnx-1.0} due to the issues that these applications introduce on the transport layer. Changing the addressing scheme from host-based to content-based, as well as changing network models from push to pull, are indeed the two obstacles in adapting the existing transport layer protocols to the \gls{NDN} and \gls{CCN} architectures. A vast number of existing applications and protocols, such as the HTTP based multimedia streaming protocols, might face false throughput estimations due to the aggressiveness of the underlying \gls{TCP} in case of content source location variations~\cite{wowmom2018,CONTI2018209}.}
\item \textbf{\review{Latency:}} \review{one fundamental issue introduced by the solutions supporting the translation of \gls{IP} and HTTP-level semantics into \gls{ICN}~\cite{point,rife} is latency. This occurs due to the frequent requests sent to the \gls{NAP}, that is attached to the source (also referred to as sNAP). Assuming a meaningful interaction between consumer and producer, the URIs are likely different for each content and for each new published content at sNAP, a new \gls{RID} has to be added to the \gls{cNAP} through the \gls{RF}. Thus, for each HTTP get request, sNAP and RF have to interact, causing an increasing network latency.}
\item \textbf{\review{Topological limitations:}} \review{in \emph{underlay} approaches, there might be several publishers for the same content that belong to the same network. In this case, whenever a consumer asks for a content released by different publishers, the RF should identify the best publisher and suggest the best content route. However, in the current architectures, the RF only announces which is the most appropriate publisher, leaving the other ones in a \textit{silent} phase. This might lead to the generation of multi-point forwarding identifiers, which create unnecessarily long routing tables.}
\item \textbf{\review{Routing and scalability:}} \review{the number of content objects, and its continuous growing in the current Internet, introduce a limitation in \gls{ICN} solutions, which have to handle content names of a possibly indefinite length. Thus, the existing networking devices might not support the content-based routing and might have to face special requirements and optimizations.}
\item \textbf{\review{Security issues in coexistence architectures:}} 
\review{below, we illustrate the security risks affecting the coexistence architectures.}
\begin{itemize}
\item \review{\textbf{Attacks against NAP nodes:} in \emph{underlay} approaches, an attack performed against a NAP node can cause much more damage than one performed against the rendezvous system. This is because a NAP is a node in an \gls{ICN} network, which can be used by an attacker to launch prefix hijacking, replay attacks and many more attacks against the \gls{ICN} core network.}
\item \review{\textbf{\gls{DoS} attacks:} an external user sending a new \gls{IP} address causes the introduction of a state into a NAP. The same action can cause the introduction of states in centralized functions, such as the TF or the RF. Thus, if arbitrary users have a direct access to the centralized TF/RF, as it was the case in pure \gls{PURSUIT}/\gls{PSIRP} architectures~\cite{6231280}, they could also easily generate a \gls{DoS} attack.} 
\item \review{\textbf{Lack of authorization and access control:} for every new node added to a network, the entire topology needs to be updated to guarantee the proper link among the new and the old network nodes. Thus, an enhanced access control policy is required in \gls{ICN} networks.}
\item \review{\textbf{Attacks against the SDN controller:} there have been increasing concerns about the security of \gls{SDN}-based networks. Many of these concerns are related to the fact that \gls{SDN} controller may parse an arbitrary part of a packet's content, and use this information to set up states in the flow tables (and possibly in the controller). Moreover, systems that parse user generated packet input (e.g. Wireshark packet analyzer and Snort intrusion detection system) have been the frequent cause of security vulnerabilities due to the large permutation of potential cases. Since numerous \gls{ICN} coexistence solutions propose to use \gls{SDN}, they are potentially open to the inherent vulnerabilities of an \gls{SDN} controller. Moreover, considering that an \gls{SDN} controller is the logically centralized entity that affects the entire network, the risk is even higher.}
\end{itemize}

\end{itemize}

\subsection{Future Research Directions}
\label{future_directions}

As confirmed by the large number of coexistence projects (e.g., POINT, DOCTOR, and hICN) that we surveyed in this paper, Industry and Government are pushing towards the definition of a new Internet architecture (i.e., \gls{ICN}) and its coexistence with the current one (i.e., \gls{IP}). Over the years, the research community has significantly grown around \gls{ICN}, following different coexistence design approaches. As mentioned before, a clean slate deployment of \gls{ICN} requires overhauling the entire Internet infrastructure and changing all the host and producer applications, thus, it is difficult to simply transit from research testbeds to operational networks. Based on the experience received from the initial \gls{ICN} architecture efforts (e.g., \gls{NDN}), researchers have realized that it is difficult, as well as infeasible, to replace a greatly successful imperative architecture with a clean slate approach. A plausible reason for this is that \gls{ICN} remains unproven due to the lack of large scale testbeds, and the consequently limited number of users in a trial, and that it has been tested on a limited number of applications so far.

\par In the past few years, a significant effort put by Governments, Industry, and Academia to assess the feasibility and effectiveness of \gls{ICN} indicates that \gls{ICN} paradigm is being considered as a possible replacement for the current \gls{IP}-based host-centric Internet infrastructure. Hence, we now present few research directions that need to be explored in this research field.

\begin{itemize}
     
\item \textbf{Secure transition phase:} from its start, \gls{ICN} was purposefully designed to have certain inherent security properties such as authentication of delivered content and (optional) encryption of the content. Additionally, relevant advances in the \gls{ICN} research community have occurred, promising to address each of the identified security gaps~\cite{8725179}~\cite{7009958}. However, due to the lack of real deployments, an array of security features in \gls{ICN} networks are still under-investigated, including access control~\cite{7447763}, security of in-network caches, protection against various network attacks (e.g., DDoS), and consumer privacy~\cite{8027034}. For instance, due to the distributed nature of content availability in \gls{ICN}, securing the content itself is much more important than securing the infrastructure or the end points. This lack of addressing security goals in the final \gls{ICN} paradigm is even more critical when considering the coexistence of \gls{TCP}/\gls{IP} and \gls{ICN}, which could lead to the introduction of new attacks and security issues. One of the main limitations of existing projects is that all of them address only the existence of a transition phase without investigating the impact of coexistence on the security and privacy of the system. We believe that not only passing through this intermediate step is unavoidable, but also that it is important to assess the security and privacy vulnerabilities that might come up under the coexistence of both architectures.

\item \textbf{Selection of an efficient coexistence approach:} in the literature, three main approaches (i.e., \emph{underlay}~\cite{8442635}, \emph{overlay}~\cite{detti2011conet}, and \emph{hybrid}~\cite{hICN}) have been used to deploy coexistence architectures. The \emph{underlay} approach introduces communication latency due to the required mapping between \gls{IP} and name addresses, which limits its usability for real-time and delay-sensitive applications. On the contrary, the \emph{underlay} approach maintains an unaltered quality of service under both normal and exceptional conditions, such as failure, server and link congestion, which are common in operator networks. Considering the \emph{overlay} approach, a major drawback is that it requires the definition and standardization of a new packet format, together with protocols that manage the mapping between \gls{ICN} faces and \gls{IP} addresses in the \gls{ICN} routers FIB. Thus, \emph{overlay} poses a significant challenge to network operators and developers. Additionally, upon new deployment, the tunnel configurations in \emph{overlay} needs to be manually changed to include the newly deployed \gls{ICN} nodes, and these point-to-point tunnels limit the \gls{ICN} capability in utilizing the underlying broadcast media. Finally, the \emph{hybrid} approach offers an interesting alternative as it allows \gls{ICN} semantics to be embedded in standard IPv4 and IPv6 packets so that the packets can be routed through either \gls{IP} routers or hybrid \gls{ICN} routers. However, the detailed performance results for \emph{hybrid} solutions are still incomplete, which limits its usage in real deployment scenarios.

\item \textbf{Coexistence solutions that preserve inherent \gls{ICN} advantages:} due to its inherent features such as in-network caching, interest aggregation, and content oriented security, \gls{ICN} provides improved communication system and security by design. Therefore, these essential features of \gls{ICN} should be protected while designing a coexistence architecture.

%\item  \textbf{Incremental deployments for interoperablility and standardization:}

\item \textbf{Optimized \gls{ICN}-\gls{IP} name-space mapping:} an important issue in the state-of-the-art solutions, that provide translation of \gls{IP}/HTTP-level services into \gls{ICN} (or vice versa), is to ensure that the communication latency is comparable with the one in the current network. In most of the coexistence solutions, that use some sort of translation at any networking layer (e.g., transport or network), the main problem is the repeated sending of newly published content information towards the translation server, which generates delay in the response path of requester and congestion in the network. The problem lies in the fact that the URL is likely different for every request (assuming some form of meaningful service interaction between \gls{IP} client and \gls{ICN} producer). Additionally, the existing channel semantics cannot be applied directly because the corresponding routing identifier at the \gls{ICN} level is different for each publication, from the translation server to \gls{IP} client. Also, realizing the rendezvous function approach, which is responsible for the response of new publications, requires continue interaction between server and content publisher. This causes an additional latency for the client requests, waiting for a fresh mapping of \gls{ICN}-\gls{IP} at each published event.

\item \textbf{Data protection and confidentiality:} ensuring privacy for network entities (e.g., consumer and producer) in coexistence architecture is not a trivial task, mainly due to the poor privacy support provided in \gls{ICN}~\cite{BERNARDINI201913}. Hence, it is important to investigate how the privacy issues were dealt in the current coexistence architectures. Ideally, names should reveal no more than what is currently revealed by an \gls{IP} address and port. However, in \gls{ICN} the name prefix reveals some information about the content, and the in-network caching and data in PIT might expose the consumer identity~\cite{7874168}. Therefore, the researchers should focus on the specific issues concerning the privacy and data protection in the coexistence scenarios. For instance, in a coexistence architecture, \gls{IP} to name-prefix mapping is performed when an \gls{IP} packet travels from \gls{IP} to \gls{ICN} network. In this scenario, the \gls{IP} header does not reveal any information about the payload, but the prefix name does, thus, the data confidentiality is threatened when these data packets are traveling through the \gls{ICN} ``island''. In particular, since the use of name prefix for addressing the data in \gls{ICN} reveals sufficient information to the passive eavesdropper, ensuring privacy means that names and payloads cannot be correlated. However, such privacy requirement would need an upper-layer service similar to the one that would resolve non-topological identifiers (e.g., \gls{ICN} name prefix) to topological names (e.g., \gls{IP} network address).

\item \textbf{\gls{SDN}/\gls{NFV} for efficient coexistence:} as mentioned earlier, the \gls{SDN} technology separates the control plane from the data plane. The decoupled control plane is programmable and has a global view of the network that provides easier network management monitoring. \gls{SDN}-based implementations of \gls{ICN} exploit the centralized view available to the \gls{SDN} controller, which enables the \gls{SDN} controller to install appropriate rules in the data-plane to process \gls{ICN} requests/responses. In the state-of-the-art, both \textit{overlay} and \textit{hybrid} \gls{ICN} deployments have leveraged \gls{SDN} to address different coexistence requirements, e.g., forwarding, storage, management, security, and interoperability. \gls{SDN} has already been successfully adopted for network deployment; it makes \gls{SDN} an appropriate choice for quick deployment of \gls{ICN} with low hardware modifications. On the another side, \gls{NFV} can help to virtualize several network functions that were previously implemented via physical devices.
\end{itemize}

\section{Conclusion}
\label{conclude}

In this paper, we survey various efforts done by researchers and industries in recent years to propose a design of \gls{ICN}-\gls{IP} coexistence architecture. All these architectures differ from each other according to their specific design, but they all adhere to the \gls{ICN} paradigm, which means a content-oriented communication model in replacement of the current host-centric one. In our survey, we identify that all these architectures have important limitations: none of them has been designed through a comprehensive approach that considers all the new challenges introduced by a coexistence scenario. Instead, the main aim for most of them is to improve the current Internet by exploiting some of the core \gls{ICN} features (i.e., forwarding, storage, management, and security). Even though security also belongs to that list of features, none of the existing architectures has considered it as the main purpose. In future, we believe appropriate coexistence architecture designs are needed to build a secure path towards the future Internet. This can be done by considering the limitations and necessary improvements of the existing coexistence solutions we have analyzed in this survey. With the set of future research directions and open questions that we have raised, our work will motivate researchers towards designing a complete solution for \gls{ICN}-\gls{IP} coexistence while tackling the key security and privacy issues.

%% file: acks.tex
\section*{Acknowledgment}
The work of M. Conti was supported by a Marie Curie Fellowship funded by the European Commission under the agreement PCIG11-GA-2012-321980. Ankit Gangwal is pursuing his Ph.D. with a fellowship for international students funded by Fondazione Cassa di Risparmio di Padova e Rovigo~(CARIPARO). This work is partially supported by EU LOCARD Project under Grant H2020-SU-SEC-2018-832735.

%% file: bio.tex
\begin{IEEEbiography}[{\includegraphics[width=1in,height=1.25in,clip]{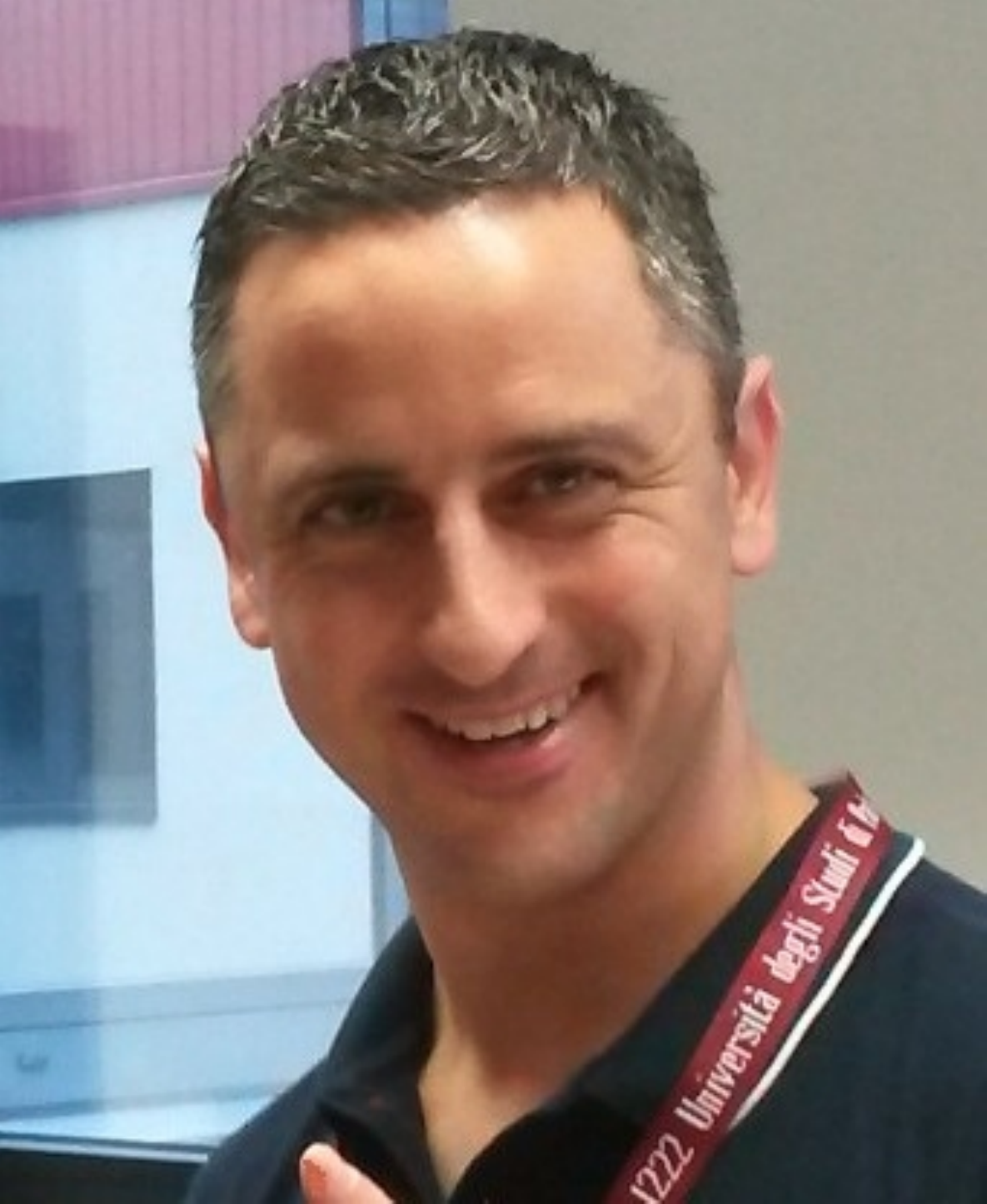}}]{Mauro Conti} is Full Professor at the University of Padua, Italy, and Affiliate Professor at the University of Washington, Seattle, USA. He obtained his Ph.D. from Sapienza University of Rome, Italy, in 2009. After his Ph.D., he was a Postdoc Researcher at Vrije Universiteit Amsterdam, The Netherlands. In 2011 he joined as Assistant Professor the University of Padua, where he became Associate Professor in 2015, and Full Professor in 2018. He has been Visiting Researcher at GMU (2008, 2016), UCLA (2010), UCI (2012, 2013, 2014, 2017), TU Darmstadt (2013), UF (2015), and FIU (2015, 2016, 2018). He has been awarded with a Marie Curie Fellowship (2012) by the European Commission, and with a Fellowship by the German DAAD (2013). His research is also funded by companies, including Cisco, Intel, and Huawei. His main research interest is in the area of security and privacy. In this area, he published more than 250 papers in topmost international peer-reviewed journals and conference. He is Area Editor-in-Chief for IEEE Communications Surveys \& Tutorials, and Associate Editor for several journals, including IEEE Communications Surveys \& Tutorials, IEEE Transactions on Information Forensics and Security, IEEE Transactions on Dependable and Secure Computing, and IEEE Transactions on Network and Service Management. He was Program Chair for TRUST 2015, ICISS 2016, WiSec 2017, and General Chair for SecureComm 2012 and ACM SACMAT 2013. He is Senior Member of the IEEE.
\end{IEEEbiography}
	
\begin{IEEEbiography}[{\includegraphics[width=1in,height=1.25in,clip]{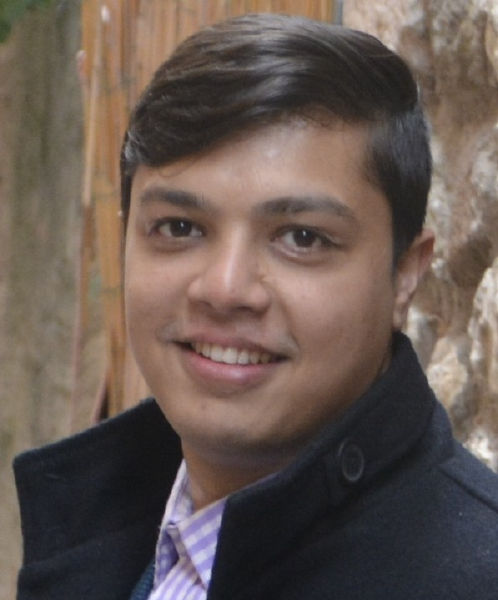}}]{Ankit Gangwal} received his BTech degree in Information Technology from RTU Kota, India in 2011 and his MTech degree in Computer Engineering from MNIT Jaipur, India in 2016. Currently, he is a Ph.D. student in the Department of Mathematics, University of Padua, Italy with a fellowship for international students funded by Fondazione Cassa di Risparmio di Padova e Rovigo (CARIPARO). His current research interest is in the area of security and privacy of the blockchain technology and novel network architectures.%, in particular, Software Defined Networks (SDN).
\end{IEEEbiography}

\begin{IEEEbiography}[{\includegraphics[width=1in,height=1.25in,clip]{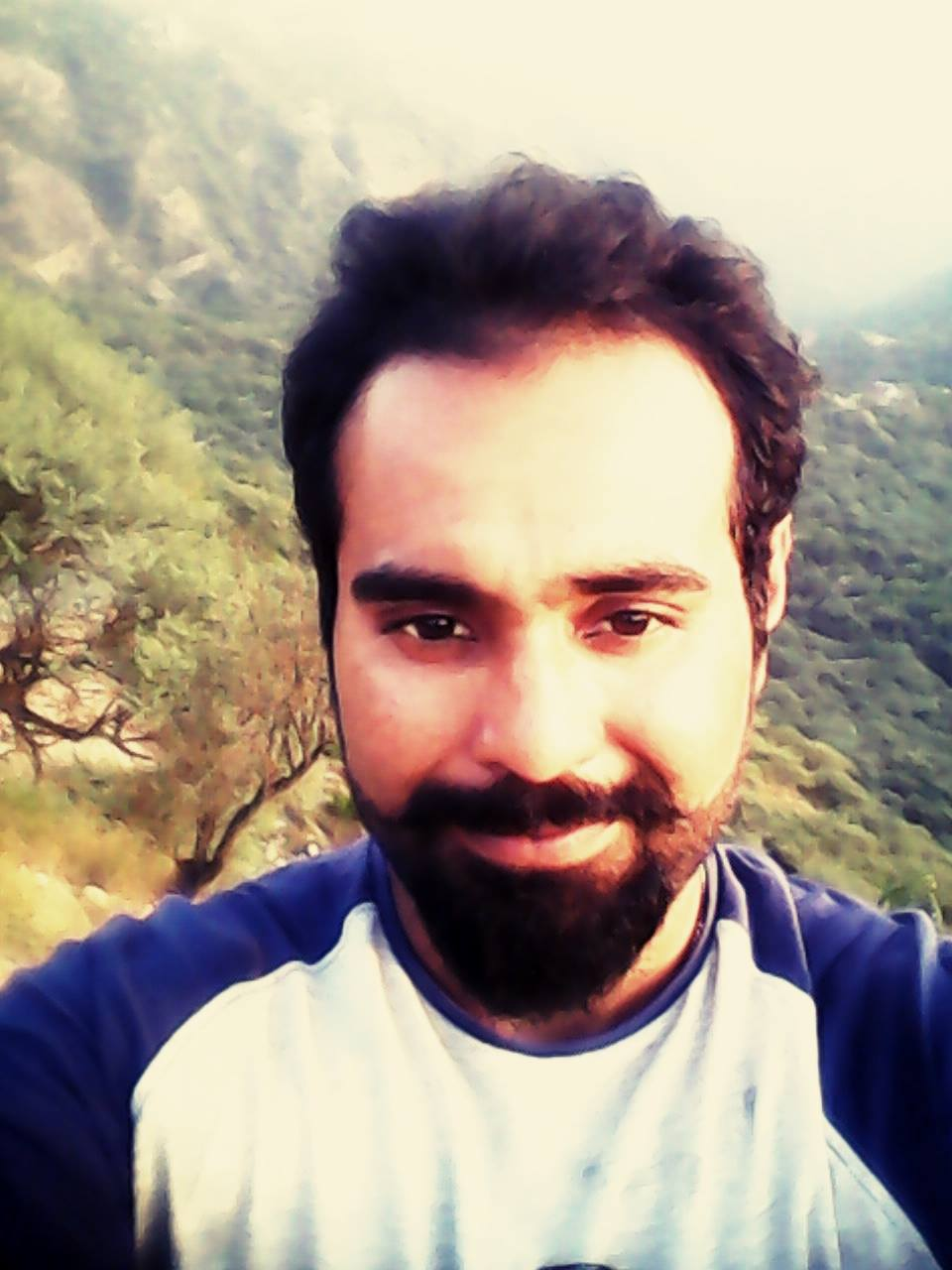}}]{Muhammad Hassan} completed the Bachelors in Electrical (Telecommunication) Engineering from COMSATS Institute of Information Technology, Pakistan in 2008 and the Masters in Computer Network Engineering from HALMSTAD University, Sweden in 2013. Currently, he is a PhD student in Brain, Mind and Computer Science at the University of Padua, Italy. He is also a part of the SPRITZ Security and Privacy research group, Padua. His research interests are in the area of securing ICN architecture and related studies such as secure integration of existing technologies in future Internet architecture. 
\end{IEEEbiography}

\begin{IEEEbiography}[{\includegraphics[width=1in,height=1.25in,clip]{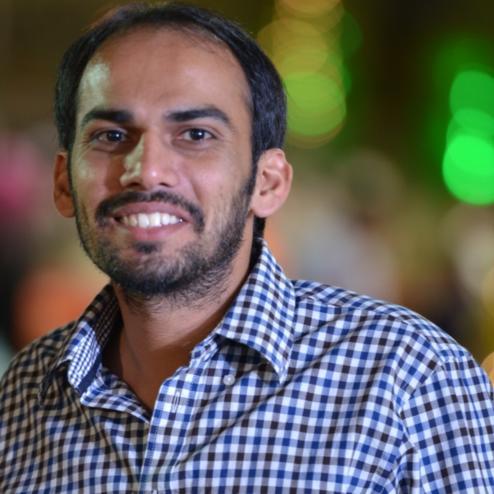}}]{Chhagan Lal}
is Postdoc fellow in Department of Mathematics, University of Padua, Italy. He obtained his Masters degree in Information Technology with specialization in Wireless communication from Indian Institute of Information Technology, Allahabad in 2009, and Ph.D. in Computer Science and Engineering from Malaviya National Institute of Technology, Jaipur, India in 2014. He have been awarded with Canadian Commonwealth scholarship in 2012 under Canadian Commonwealth Scholarship Program to work in University of Saskatchewan in Saskatoon, Canada. His current research areas include Information Centric Networking, Blockchain-based Application Design, and Security in Software-defined networking and Internet of Things (IoT) networks.
\end{IEEEbiography}

\begin{IEEEbiography}[{\includegraphics[width=1in,height=1.25in,clip]{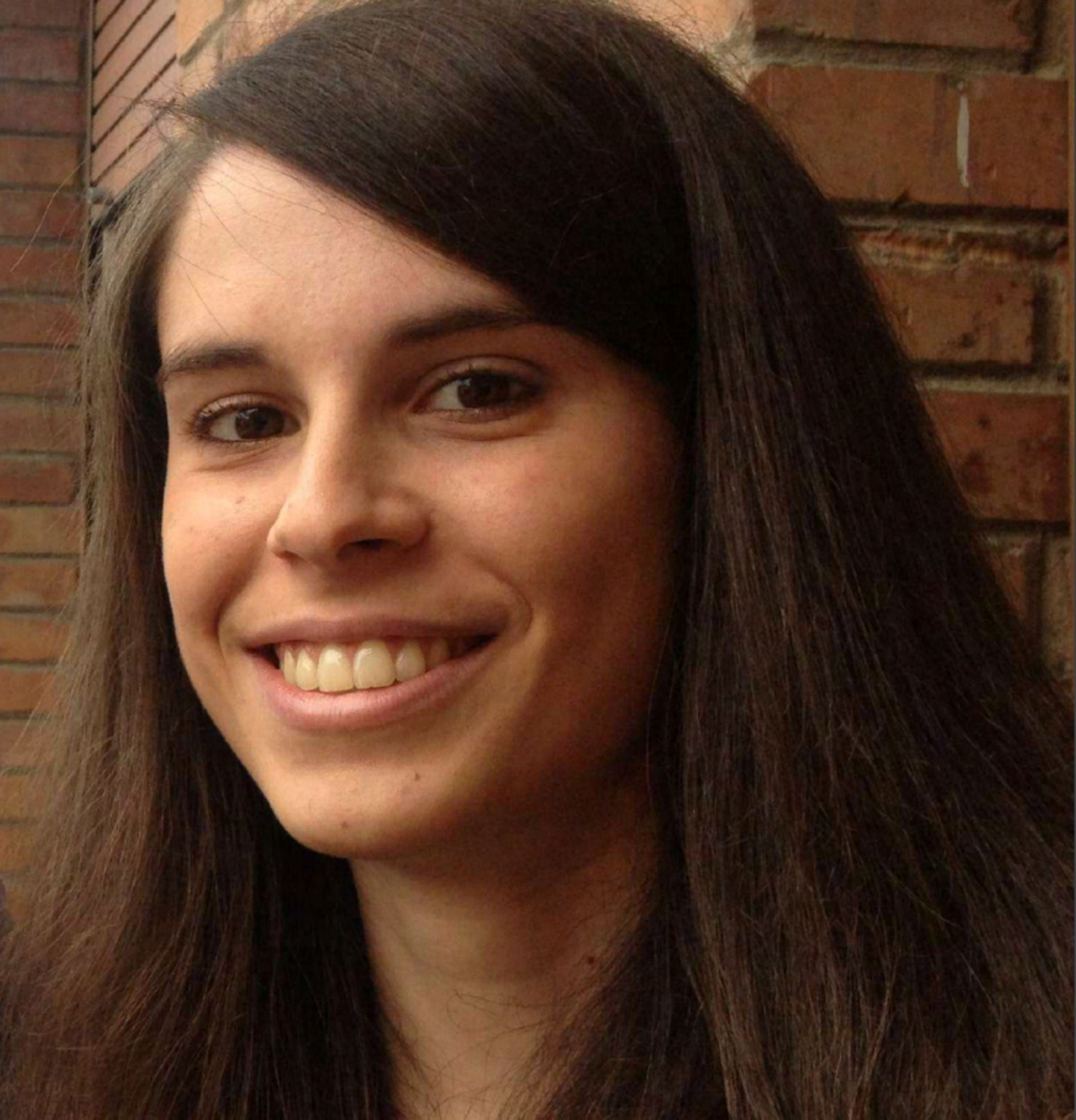}}]{Eleonora Losiouk}
Eleonora Losiouk is a Postdoc Fellow working in the SPRITZ Group of the University of Padova, Italy. In 2018, she obtained her Ph.D. in Bioengineering and Bioinformatics from the University of Pavia, Italy. She has been a Visiting Fellow at the Ecole Polytechnique Federale de Lausanne in 2017. Her main research interests regard the security and privacy evaluation of the Android Operating System and the Information-Centric Networking. During her Ph.D. she published several papers in peer-reviewed journals and IEEE conferences.
\end{IEEEbiography}

%% file: main.bbl
\begin{thebibliography}{100}
\providecommand{\url}[1]{#1}
\csname url@samestyle\endcsname
\providecommand{\newblock}{\relax}
\providecommand{\bibinfo}[2]{#2}
\providecommand{\BIBentrySTDinterwordspacing}{\spaceskip=0pt\relax}
\providecommand{\BIBentryALTinterwordstretchfactor}{4}
\providecommand{\BIBentryALTinterwordspacing}{\spaceskip=\fontdimen2\font plus
\BIBentryALTinterwordstretchfactor\fontdimen3\font minus
  \fontdimen4\font\relax}
\providecommand{\BIBforeignlanguage}[2]{{%
\expandafter\ifx\csname l@#1\endcsname\relax
\typeout{** WARNING: IEEEtran.bst: No hyphenation pattern has been}%
\typeout{** loaded for the language `#1'. Using the pattern for}%
\typeout{** the default language instead.}%
\else
\language=\csname l@#1\endcsname
\fi
#2}}
\providecommand{\BIBdecl}{\relax}
\BIBdecl

\bibitem{numberConnDevices}
\BIBentryALTinterwordspacing
``{Internet of Things (IoT) Connected Devices Istalled Base Worldwide from 2015
  to 2025 (in Billions)}.'' [Online]. Available:
  \url{https://tinyurl.com/yyfukyk2}
\BIBentrySTDinterwordspacing

\bibitem{RFC3022}
P.~Srisuresh and K.~Egevang, ``{RFC 3022: Traditional IP Network Address
  Translator (Traditional NAT)},'' RFC 3022, January 2001.

\bibitem{cisco2021}
\BIBentryALTinterwordspacing
\emph{{Cisco visual networking index: Forecast and methodology: 2016–2021}},
  Rep., Sep. 2017, accessed: July 28, 2019. [Online]. Available:
  \url{https://tinyurl.com/y2jptcbn}
\BIBentrySTDinterwordspacing

\bibitem{zetta}
\BIBentryALTinterwordspacing
\emph{{The Zettabyte Era: Trends and Analysis}}, Rep., Jun. 2017, accessed:
  July 28, 2019. [Online]. Available: \url{https://tinyurl.com/y32f4jwe}
\BIBentrySTDinterwordspacing

\bibitem{what2016}
\BIBentryALTinterwordspacing
\emph{{What Happens in an Internet Minute in 2016}}, Intel, Santa Clara, CA,
  USA, accessed: August 2, 2019. [Online]. Available:
  \url{https://tinyurl.com/y5wrz8p6}
\BIBentrySTDinterwordspacing

\bibitem{RFC4301}
K.~Seo and S.~Kent, ``{RFC 4301: Security Architecture for the Internet
  Protocol},'' BBN Technologies, RFC 4301, December 2005.

\bibitem{RFC8446}
E.~Rescorla, ``{RFC 8446: The Transport Layer Security (TLS) Protocol Version
  1.3},'' RFC 8446, August 2018.

\bibitem{rina}
\BIBentryALTinterwordspacing
E.~Grasa, L.~Bergesio, M.~Tarzan, E.~Trouva, B.~Gaston, F.~Salvestrini,
  G.~Maffione, V.~Carrozzo, D.~Staessens, S.~Vrijders, D.~Colle, A.~Chappell,
  J.~Day, and C.~L., \emph{{Recursive InterNetwork Architecture (RINA),
  Investigating RINA as an Alternative to TCP/IP (IRATI)}}, ser. Building the
  Future Internet through FIRE.\hskip 1em plus 0.5em minus 0.4em\relax Rivers
  Publishers, 2016. [Online]. Available: \url{https://tinyurl.com/y4kbyt5j}
\BIBentrySTDinterwordspacing

\bibitem{ccnx-1.0}
M.~Mosko, ``{CCNx 1.0 Protocol Specifications Roadmap}.''

\bibitem{Diallo2011LeveragingCF}
M.~Diallo, S.~Fdida, V.~Sourlas, P.~Flegkas, and L.~Tassiulas, ``{Leveraging
  Caching for Internet-Scale Content-Based Publish/Subscribe Networks},''
  \emph{IEEE International Conference on Communications (ICC)}, pp. 1--5, 2011.

\bibitem{TANG2019590}
\BIBentryALTinterwordspacing
Y.~Tang, K.~Guo, J.~Ma, Y.~Shen, and T.~Chi, ``{A Smart Caching Mechanism for
  Mobile Multimedia in information Centric Networking with Edge Computing},''
  \emph{Future Generation Computer Systems}, vol.~91, pp. 590 -- 600, 2019.
  [Online]. Available: \url{https://tinyurl.com/y5qnozwp}
\BIBentrySTDinterwordspacing

\bibitem{7467400}
A.~{Ioannou} and S.~{Weber}, ``{A Survey of Caching Policies and Forwarding
  Mechanisms in Information-Centric Networking},'' \emph{IEEE Communications
  Surveys Tutorials}, vol.~18, no.~4, pp. 2847--2886, Fourthquarter 2016.

\bibitem{8057300}
A.~{Seetharam}, ``{On Caching and Routing in Information-Centric Networks},''
  \emph{IEEE Communications Magazine}, vol.~56, no.~3, pp. 204--209, March
  2018.

\bibitem{8539022}
X.~{Fu}, D.~{Kutscher}, S.~{Misra}, and R.~{Li}, ``{Information-Centric
  Networking Security},'' \emph{IEEE Communications Magazine}, vol.~56, no.~11,
  pp. 60--61, 2018.

\bibitem{Anastasiades2014}
C.~Anastasiades, T.~Braun, and V.~A. Siris, \emph{{Information-Centric
  Networking in Mobile and Opportunistic Networks}}, 2014, pp. 14--30.

\bibitem{8303694}
C.~{Fang}, H.~{Yao}, Z.~{Wang}, W.~{Wu}, X.~{Jin}, and F.~R. {Yu}, ``A survey
  of mobile information-centric networking: Research issues and challenges,''
  \emph{IEEE Communications Surveys Tutorials}, vol.~20, no.~3, pp. 2353--2371,
  2018.

\bibitem{8478349}
S.~{Arshad}, M.~A. {Azam}, M.~H. {Rehmani}, and J.~{Loo}, ``Recent advances in
  information-centric networking-based internet of things (icn-iot),''
  \emph{IEEE Internet of Things Journal}, vol.~6, no.~2, pp. 2128--2158, April
  2019.

\bibitem{NOUR201995}
B.~Nour, K.~Sharif, F.~Li, S.~Biswas, H.~Moungla, M.~Guizani, and Y.~Wang, ``A
  survey of internet of things communication using icn: A use case
  perspective,'' \emph{Computer Communications}, vol. 142-143, pp. 95 -- 123,
  2019.

\bibitem{8263145}
C.~{Xu}, M.~{Wang}, X.~{Chen}, L.~{Zhong}, and L.~A. {Grieco}, ``Optimal
  information centric caching in 5g device-to-device communications,''
  \emph{IEEE Transactions on Mobile Computing}, vol.~17, no.~9, pp. 2114--2126,
  Sep. 2018.

\bibitem{Kumar2019}
A.~{Kumar}, J.~{Lu}, and K.~K. {Afridi}, ``Power density and efficiency
  enhancement in icn dc–dc converters using topology morphing control,''
  \emph{IEEE Transactions on Power Electronics}, vol.~34, no.~2, pp.
  1881--1900, Feb 2019.

\bibitem{8624408}
L.~{Bracciale}, P.~{Loreti}, A.~{Detti}, R.~{Paolillo}, and N.~B. {Melazzi},
  ``Lightweight named object: An icn-based abstraction for iot device
  programming and management,'' \emph{IEEE Internet of Things Journal}, vol.~6,
  no.~3, pp. 5029--5039, June 2019.

\bibitem{8027034}
R.~{Tourani}, S.~{Misra}, T.~{Mick}, and G.~{Panwar}, ``{Security, Privacy, and
  Access Control in Information-Centric Networking: A Survey},'' \emph{IEEE
  Communications Surveys Tutorials}, vol.~20, no.~1, pp. 566--600, 2018.

\bibitem{8240926}
I.~U. {Din}, S.~{Hassan}, M.~K. {Khan}, M.~{Guizani}, O.~{Ghazali}, and
  A.~{Habbal}, ``Caching in information-centric networking: Strategies,
  challenges, and future research directions,'' \emph{IEEE Communications
  Surveys Tutorials}, vol.~20, no.~2, pp. 1443--1474, 2018.

\bibitem{RFC}
\BIBentryALTinterwordspacing
A.~Rahman, D.~Trossen, D.~Kutscher, and R.~Ravindran, ``{Deployment
  Considerations for Information-Centric Networking (ICN)},'' \emph{ICNRG
  draft}, 2019. [Online]. Available: \url{https://tinyurl.com/y24d4oly}
\BIBentrySTDinterwordspacing

\bibitem{Cheriton00triad:a}
D.~Cheriton and M.~Gritter, ``{TRIAD: A New Next-Generation Internet
  Architecture},'' 2000.

\bibitem{Koponen:2007:DNA:1282380.1282402}
T.~Koponen, M.~Chawla, B.-G. Chun, A.~Ermolinskiy, K.~H. Kim, S.~Shenker, and
  I.~Stoica, ``{A Data-oriented (and Beyond) Network Architecture},'' in
  \emph{Conference on Applications, Technologies, Architectures, and Protocols
  for Computer Communications (SIGCOMM)}, 2007, pp. 181--192.

\bibitem{Jacobson}
V.~Jacobson, D.~K. Smetters, J.~D. Thornton, M.~F. Plass, N.~H. Briggs, and
  R.~L. Braynard, ``{Networking Named Content},'' in \emph{5th ACM
  International Conference on emerging Networking EXperiments and Technologies
  (CoNEXT)}, 2009, pp. 1--12.

\bibitem{Zhang}
L.~Zhang, A.~Afanasyev, J.~Burke, V.~Jacobson, P.~Crowley, C.~Papadopoulos,
  L.~Wang, B.~Zhang \emph{et~al.}, ``{Named Data Networking},'' \emph{ACM
  SIGCOMM Computer Communication Review}, vol.~44, no.~3, pp. 66--73, 2014.

\bibitem{ICNRG}
``{Information-Centric Networking Research Group (ICNRG)},''
  \url{https://irtf.org/icnrg}.

\bibitem{mcquillan1980new}
J.~McQuillan, I.~Richer, and E.~Rosen, ``{The New routing algorithm for the
  ARPANET},'' \emph{IEEE transactions on Communications}, vol.~28, no.~5, pp.
  711--719, 1980.

\bibitem{Dimitrov:2010:PPP:1839379.1839409}
V.~Dimitrov and V.~Koptchev, ``{PSIRP Project -- Publish-subscribe Internet
  Routing Paradigm: New Ideas for Future Internet},'' in \emph{11th ACM
  International Conference on Computer Systems and Technologies (CompSysTech)},
  2010, pp. 167--171.

\bibitem{Dannewitz:2013:NII:2459510.2459643}
C.~Dannewitz, D.~Kutscher, B.~Ohlman, S.~Farrell, B.~Ahlgren, and H.~Karl,
  ``{Network of Information (NetInf) - An Information-Centric Networking
  Architecture},'' \emph{Elsevier Computer Communication}, vol.~36, no.~7, pp.
  721--735, 2013.

\bibitem{zimmermann1980osi}
H.~Zimmermann, ``{OSI Reference Model-The {ISO} Model of Architecture for Open
  Systems Interconnection},'' \emph{IEEE Transactions on Communications},
  vol.~28, no.~4, pp. 425--432, 1980.

\bibitem{TCPstack}
``{Requirements for Internet Hosts - Communication Layers},'' Technical Report
  1122, 1989.

\bibitem{icnstack1}
C.~Safitri, Y.~Yamada, S.~Baharun, S.~Goudarzi, Q.~Nguyen, and T.~Sato, ``An
  intelligent quality of service architecture for information-centric vehicular
  networking,'' \emph{Internetworking Indonesia Journal}, vol.~10, pp. 15--20,
  01 2018.

\bibitem{icnstack2}
F.~Drijver, ``Assessment of benefits and drawbacks of {ICN} for {IoT}
  applications,'' 2018.

\bibitem{point}
\BIBentryALTinterwordspacing
``{POINT (iP Over IcN - the betTer IP)}.'' [Online]. Available:
  \url{https://www.point-h2020.eu/}
\BIBentrySTDinterwordspacing

\bibitem{cisco}
``{White paper: Cisco {V}isual {N}etworking {I}ndex ({V}{N}{I}): Forecast and
  methodology, 2017\---2022},''
  \url{https://www.cisco.com/c/en/us/solutions/collateral/service-provider/visual-networking-index-vni/white-paper-c11-741490.html}.

\bibitem{dashovericn}
S.~Lederer, C.~Mueller, C.~Timmerer, and H.~Hellwagner, ``{Adaptive Multimedia
  Streaming in Information-Centric Networks},'' \emph{IEEE Network}, vol.~28,
  no.~6, pp. 91--96, 2014.

\bibitem{6649319}
S.~Lederer, C.~Mueller, B.~Rainer, C.~Timmerer, and H.~Hellwagner, ``{Adaptive
  streaming over Content Centric Networks in mobile networks using multiple
  links},'' in \emph{2013 IEEE International Conference on Communications
  Workshops (ICC)}, June 2013, pp. 677--681.

\bibitem{dashoverccn}
Y.~Liu, J.~Geurts, J.~C. Point, S.~Lederer, B.~Rainer, C.~Müller, C.~Timmerer,
  and H.~Hellwagner, ``{Dynamic Adaptive Streaming over CCN: A Caching and
  Overhead Analysis},'' in \emph{IEEE International Conference on
  Communications (ICC)}, 2013, pp. 3629--3633.

\bibitem{7169859}
S.~Petrangeli, N.~Bouten, M.~Claeys, and F.~D. Turck, ``{Towards SVC-based
  Adaptive Streaming in information-Centric networks},'' in \emph{2015 IEEE
  International Conference on Multimedia Expo Workshops (ICMEW)}, June 2015,
  pp. 1--6.

\bibitem{ndnavs}
B.~Rainer, D.~Posch, and H.~Hellwagner, ``{Investigating the Performance of
  Pull-Based Dynamic Adaptive Streaming in NDN},'' \emph{IEEE Journal on
  Selected Areas in Communications}, vol.~34, no.~8, pp. 2130--2140, 2016.

\bibitem{Carofiglio}
J.~Samain, G.~Carofiglio, L.~Muscariello, M.~Papalini, M.~Sardara, M.~Tortelli,
  and D.~Rossi, ``{Dynamic Adaptive Video Streaming: Towards a Systematic
  Comparison of ICN and TCP/IP},'' \emph{IEEE Transactions on Multimedia},
  vol.~19, no.~10, pp. 2166--2181, Oct 2017.

\bibitem{7562050}
Y.~Zhang, A.~Afanasyev, J.~Burke, and L.~Zhang, ``{A Survey of Mobility Support
  in Named Data Networking},'' in \emph{{IEEE Conference on Computer
  Communications Workshops (INFOCOM WKSHPS)}}, April 2016, pp. 83--88.

\bibitem{Ott2004WhyST}
J.~Ott and D.~Kutscher, ``{Why Seamless? Towards Exploiting WLAN-Based
  Intermittent Connectivity on the Road},'' in \emph{TERENA Networking
  Conference}, 2004.

\bibitem{Fall:2003:DNA:863955.863960}
\BIBentryALTinterwordspacing
K.~Fall, ``{A Delay-tolerant Network Architecture for Challenged Internets},''
  in \emph{Proceedings of the 2003 Conference on Applications, Technologies,
  Architectures, and Protocols for Computer Communications}, ser. SIGCOMM
  '03.\hskip 1em plus 0.5em minus 0.4em\relax New York, NY, USA: ACM, 2003, pp.
  27--34. [Online]. Available: \url{http://doi.acm.org/10.1145/863955.863960}
\BIBentrySTDinterwordspacing

\bibitem{rfc4838}
\BIBentryALTinterwordspacing
L.~Torgerson, S.~C. Burleigh, H.~Weiss, A.~J. Hooke, K.~Fall, D.~V.~G. Cerf,
  K.~Scott, and R.~C. Durst, ``{Delay-Tolerant Networking Architecture},'' RFC
  4838, Apr. 2007. [Online]. Available:
  \url{https://rfc-editor.org/rfc/rfc4838.txt}
\BIBentrySTDinterwordspacing

\bibitem{Compagno2018}
A.~Compagno, M.~Conti, and M.~Hassan, \emph{{An ICN-Based Authentication
  Protocol for a Simplified LTE Architecture}}.\hskip 1em plus 0.5em minus
  0.4em\relax Cham: Springer International Publishing, 2018.

\bibitem{farhady2015software}
H.~Farhady, H.~Lee, and A.~Nakao, ``{Software Defined Networking: A Survey},''
  \emph{Elsevier Computer Networks}, vol.~81, pp. 79--95, 2015.

\bibitem{kreutz2015software}
D.~Kreutz, F.~M. Ramos, P.~E. Verissimo, C.~E. Rothenberg, S.~Azodolmolky, and
  S.~Uhlig, ``{Software Defined Networking: A Comprehensive Survey},''
  \emph{Proceedings of the IEEE}, vol. 103, no.~1, pp. 14--76, 2015.

\bibitem{mckeown2008openflow}
N.~McKeown, T.~Anderson, H.~Balakrishnan, G.~Parulkar, L.~Peterson, J.~Rexford,
  S.~Shenker, and J.~Turner, ``{OpenFlow: Enabling Innovation in Campus
  Networks},'' \emph{ACM SIGCOMM Computer Communication Review}, vol.~38,
  no.~2, pp. 69--74, 2008.

\bibitem{li2015software}
Y.~Li and M.~Chen, ``{Software Defined Network Function Virtualization: A
  Survey},'' \emph{IEEE Access}, vol.~3, pp. 2542--2553, 2015.

\bibitem{1250586}
A.~Vakali and G.~Pallis, ``{Content Delivery Networks: Status and Trends},''
  \emph{IEEE Internet Computing}, vol.~7, no.~6, pp. 68--74, 2003.

\bibitem{6674399}
A.~Binder and I.~Kotuliak, ``Content delivery network interconnect: Practical
  experience,'' in \emph{11th IEEE International Conference on Emerging
  eLearning Technologies and Applications (ICETA)}, 2013, pp. 29--33.

\bibitem{8046000}
M.~A. Salahuddin, J.~Sahoo, R.~Glitho, H.~Elbiaze, and W.~Ajib, ``{A Survey on
  Content Placement Algorithms for Cloud-Based Content Delivery Networks},''
  \emph{IEEE Access}, vol.~6, pp. 91--114, 2018.

\bibitem{6688724}
Y.~Nicolas, D.~Wolff, D.~Rossi, and A.~Finamore, ``{I Tube, YouTube, P2PTube:
  Assessing ISP benefits of peer-assisted caching of YouTube content},'' in
  \emph{IEEE P2P}, 2013, pp. 1--2.

\bibitem{7948965}
L.~Sun, M.~Ma, W.~Hu, H.~Pang, and Z.~Wang, ``{Beyond 1 Million Nodes -
  Crowdsourced Video CDN: Architecture, Technology, and Economy},'' \emph{IEEE
  MultiMedia}, pp. 1--1, 2018.

\bibitem{STOCKER20171003}
V.~Stocker, G.~Smaragdakis, W.~Lehr, and S.~Bauer, ``{The growing complexity of
  content delivery networks: Challenges and implications for the Internet
  ecosystem},'' \emph{Telecommunications Policy}, vol.~41, no.~10, pp.
  1003--1016, 2017.

\bibitem{Clark2005TheGO}
D.~Clark, Lehr, S.~Bauer, P.~Faratin, R.~Sami, and J.~Wroclawski, ``{The Growth
  of Internet Overlay Networks : Implications for Architecture , Industry
  Structure and Policy},'' 2005.

\bibitem{Medagliani2017OverlayRF}
P.~Medagliani, S.~Paris, J.~Leguay, L.~Maggi, C.~Xue, and H.~Zhou, ``{Overlay
  routing for fast video transfers in CDN},'' \emph{IFIP/IEEE Symposium on
  Integrated Network and Service Management (IM)}, pp. 531--536, 2017.

\bibitem{NivenJenkins2012ContentDN}
B.~Niven-Jenkins, F.~L. Faucheur, and N.~Bitar, ``{Content Distribution Network
  Interconnection (CDNI) Problem Statement},'' \emph{RFC}, vol. 6707, pp.
  1--32, 2012.

\bibitem{5770277}
M.~J. Khabbaz, C.~M. Assi, and W.~F. Fawaz, ``{Disruption-Tolerant Networking:
  A Comprehensive Survey on Recent Developments and Persisting Challenges},''
  \emph{IEEE Communications Surveys Tutorials}, vol.~14, no.~2, pp. 607--640,
  2012.

\bibitem{DTNstory}
A.~V. Vasilakos, Y.~Zhang, and T.~Spyropoulos, \emph{Delay Tolerant Networks:
  Protocols and Applications}, 1st~ed.\hskip 1em plus 0.5em minus 0.4em\relax
  Boca Raton, FL, USA: CRC Press, Inc., 2011.

\bibitem{storecarryforward}
J.~{Liu}, X.~{Jiang}, H.~{Nishiyama}, and N.~{Kato}, ``A general model for
  store-carry-forward routing schemes with multicast in delay tolerant
  networks,'' in \emph{2011 6th International ICST Conference on Communications
  and Networking in China (CHINACOM)}, Aug 2011, pp. 494--500.

\bibitem{ren2015deployment}
J.~Ren, L.~Li, H.~Chen, S.~Wang, S.~Xu, G.~Sun, J.~Wang, and S.~Liu, ``{On the
  Deployment of Information-centric Network: Programmability and
  Virtualization},'' in \emph{IEEE International Conference on Computing,
  Networking and Communications (ICNC)}, 2015, pp. 690--694.

\bibitem{6231280}
D.~Trossen and G.~Parisis, ``{Designing and Realizing an Information-Centric
  Internet},'' \emph{IEEE Communications Magazine}, vol.~50, no.~7, pp. 60--67,
  2012.

\bibitem{NDNProject}
\BIBentryALTinterwordspacing
L.~Zhang, D.~Estrin, J.~Burke, V.~Jacobson, J.~D. Thornton, D.~K. Smetters,
  B.~Zhang, G.~Tsudik, D.~Massey, C.~Papadopoulos \emph{et~al.}, ``{Named Data
  Networking (NDN) Project}.'' [Online]. Available:
  \url{https://tinyurl.com/y3o5auhw}
\BIBentrySTDinterwordspacing

\bibitem{7084921}
S.~{Shailendra}, B.~{Panigrahi}, H.~K. {Rath}, and A.~{Simha}, ``{A novel
  overlay architecture for Information Centric Networking},'' in \emph{21st
  National Conference on Communications (NCC)}, 2015, pp. 1--6.

\bibitem{detti2011conet}
A.~Detti, N.~{Blefari Melazzi}, S.~Salsano, and M.~Pomposini, ``{CONET: A
  Content Centric Inter-networking Architecture},'' in \emph{ACM SIGCOMM
  workshop on Information-Centric Networking}, 2011, pp. 50--55.

\bibitem{vahlenkamp2013enabling}
M.~Vahlenkamp, F.~Schneider, D.~Kutscher, and J.~Seedorf, ``{Enabling ICN in IP
  Networks using SDN},'' in \emph{IEEE International Conference on Network
  Protocols (ICNP)}, 2013, pp. 1--2.

\bibitem{veltri2012supporting}
L.~Veltri, G.~Morabito, S.~Salsano, N.~{Blefari Melazzi}, and A.~Detti,
  ``{Supporting Information Centric Functionality in Software Defined
  Networks},'' in \emph{IEEE International Conference on Communications (ICC)},
  2012, pp. 6645--6650.

\bibitem{doctor}
\BIBentryALTinterwordspacing
``{DeplOyment and seCurisaTion of new functiOnalities in virtualized networking
  enviRonments}.'' [Online]. Available: \url{http://www.doctor-project.org/}
\BIBentrySTDinterwordspacing

\bibitem{rife}
\BIBentryALTinterwordspacing
``{architectuRe for an Internet For Everybody (RIFE)}.'' [Online]. Available:
  \url{https://rife-project.eu/}
\BIBentrySTDinterwordspacing

\bibitem{cableLabs}
\BIBentryALTinterwordspacing
G.~White and G.~Rutz, ``{Content Delivery With Content-Centric Networking}.''
  [Online]. Available: \url{https://tinyurl.com/y5328v4s}
\BIBentrySTDinterwordspacing

\bibitem{NDNLAN}
H.~Wu, J.~Shi, Y.~Wang, Y.~Wang, G.~Zhang, Y.~Wang, B.~Liu, and B.~Zhang, ``{On
  Incremental Deployment of Named Data Networking in Local Area Networks},'' in
  \emph{ACM/IEEE Architectures for Networking and Communications Systems
  (ANCS)}, 2017, pp. 82--94.

\bibitem{hICN}
\BIBentryALTinterwordspacing
L.~Muscariello, G.~Carofiglio, and J.~Aug{\'e}, ``{System and Method to
  Facilitate Integration of Information-centric Networking into Internet
  Protocol Networks},'' in \emph{CISCO Technology, Inc.}, 2018. [Online].
  Available: \url{https://tinyurl.com/y46pc6w4}
\BIBentrySTDinterwordspacing

\bibitem{melazzi2012openflow}
N.~{Blefari Melazzi}, A.~Detti, G.~Mazza, G.~Morabito, S.~Salsano, and
  L.~Veltri, ``{An Openflow-based Testbed for Information Centric
  Networking},'' in \emph{IEEE Future Network \& Mobile Summit (FutureNetw)},
  2012, pp. 1--9.

\bibitem{Jokela2009LIPSINLS}
P.~Jokela, A.~Zahemszky, C.~E. Rothenberg, S.~Arianfar, and P.~Nikander,
  ``{LIPSIN: Line Speed Publish/Subscribe Inter-networking},'' in
  \emph{SIGCOMM}, 2009.

\bibitem{Kohler:2000:CMR:354871.354874}
E.~Kohler, R.~Morris, B.~Chen, J.~Jannotti, and M.~F. Kaashoek, ``{The Click
  Modular Router},'' \emph{ACM Transaction on Computer Systems}, vol.~18,
  no.~3, pp. 263--297, 2000.

\bibitem{ccnxudp}
PARC, ``{CCNx Over UDP},'' 2015.

\bibitem{NDNLP}
\BIBentryALTinterwordspacing
``{NDNLP: A Link Protocol for NDN},'' 2012. [Online]. Available:
  \url{https://tinyurl.com/yyx8ezmq}
\BIBentrySTDinterwordspacing

\bibitem{wowmom2018}
M.~Conti, R.~Droms, M.~Hassan, and S.~Valle, ``{QoE Degradation Attack in
  Dynamic Adaptive Streaming Over ICN},'' in \emph{19th IEEE International
  Symposium on "A World of Wireless, Mobile and Multimedia Networks" (WoWMoM)},
  2018, pp. 1--9.

\bibitem{CONTI2018209}
M.~Conti, R.~Droms, M.~Hassan, and C.~Lal, ``{Fair-RTT-DAS: A robust and
  efficient dynamic adaptive streaming over ICN},'' \emph{Computer
  Communications}, vol. 129, pp. 209 -- 225, 2018.

\bibitem{sail}
\BIBentryALTinterwordspacing
``{Scalable \& Adaptive Internet soLutions (SAIL) European Commission's 7th
  Framework Program}.'' [Online]. Available:
  \url{http://www.sail-project.eu/about-sail/index.html}
\BIBentrySTDinterwordspacing

\bibitem{NSF}
\BIBentryALTinterwordspacing
``{NSF Future Internet Architecture Project}.'' [Online]. Available:
  \url{http://www.nets-fia.net/}
\BIBentrySTDinterwordspacing

\bibitem{6193510}
G.~Carofiglio, M.~Gallo, and L.~Muscariello, ``{ICP: Design and Evaluation of
  an Interest Control Protocol for Content-Centric Networking},'' in \emph{IEEE
  INFOCOM Workshops}, March 2012, pp. 304--309.

\bibitem{CAROFIGLIO2016104}
G.~{Carofiglio}, M.~{Gallo}, L.~{Muscariello}, and M.~P. and, ``{Optimal
  multipath Congestion Control and Request Forwarding in Information-Centric
  Networks},'' in \emph{21st IEEE International Conference on Network Protocols
  (ICNP)}, 2013, pp. 1--10.

\bibitem{Gusev2015NDNRTCRV}
P.~Gusev and J.~Burke, ``{NDN-RTC: Real-Time Videoconferencing over Named Data
  Networking},'' in \emph{2nd ACM Conference on Information-Centric
  Networking}, 2015, pp. 117--126.

\bibitem{399}
S.~Mastorakis, A.~Afanasyev, I.~Moiseenko, and L.~Zhang, ``{NDNSIM 2: An
  updated NDN simulator for NS-3},'' NDN, Technical Report NDN-0028, Revision
  2, 2016.

\bibitem{8539023}
Z.~Zhang, Y.~Yu, H.~Zhang, E.~Newberry, S.~Mastorakis, Y.~Li, A.~Afanasyev, and
  L.~Zhang, ``{An Overview of Security Support in Named Data Networking},''
  \emph{IEEE Communications Magazine}, vol.~56, no.~11, pp. 62--68, 2018.

\bibitem{TCP/ICN}
I.~Moiseenko and D.~Oran, ``{TCP/ICN: Carrying TCP over Content Centric and
  Named Data Networks},'' in \emph{3rd ACM Conference on Information-Centric
  Networking}, 2016, pp. 112--121.

\bibitem{367}
\BIBentryALTinterwordspacing
A.~Afanasyev, I.~Moiseenko, and L.~Zhang, ``{NDNSIM: NDN Simulator for NS-3},''
  Technical Report NDN-0005, 2012. [Online]. Available:
  \url{https://tinyurl.com/y2ysqp8v}
\BIBentrySTDinterwordspacing

\bibitem{Schneider:2016:PCC:2984356.2984369}
K.~Schneider, C.~Yi, B.~Zhang, and L.~Zhang, ``{A Practical Congestion Control
  Scheme for Named Data Networking},'' in \emph{3rd ACM Conference on
  Information-Centric Networking}, 2016, pp. 21--30.

\bibitem{Agrawal2018}
S.~Agrawal, S.~Shailendra, B.~Panigrahi, H.~K. Rath, and A.~Simha, ``Oicnsim:
  An ns-3 based simulator for overlay information centric networking.''

\bibitem{Agrawal:2018:OSN:3265997.3266000}
A.~Suvrat, S.~Samar, P.~Bighnaraj, R.~Hemant, and S.~Anantha, ``{O-ICN
  Simulator (OICNSIM): An NS-3 Based Simulator for Overlay Information Centric
  Networking (O-ICN)},'' pp. 13--15, 2018.

\bibitem{docmang}
\BIBentryALTinterwordspacing
``{Network Functions Virtualisation (NFV); Management and Orchestration, ETSI
  GS NFV-MAN 001 V1.1.1 (2014-12)}.'' [Online]. Available:
  \url{https://tinyurl.com/y966jvg4}
\BIBentrySTDinterwordspacing

\bibitem{8725179}
F.~{Khan} and H.~{Li}, ``Ensuring trust and confidentiality for adaptive video
  streaming in icn,'' \emph{Journal of Communications and Networks}, vol.~PP,
  no.~99, pp. 1--9, May 2019.

\bibitem{7009958}
E.~G. {AbdAllah}, H.~S. {Hassanein}, and M.~{Zulkernine}, ``A survey of
  security attacks in information-centric networking,'' \emph{IEEE
  Communications Surveys Tutorials}, vol.~17, no.~3, pp. 1441--1454,
  thirdquarter 2015.

\bibitem{7447763}
B.~{Li}, D.~{Huang}, Z.~{Wang}, and Y.~{Zhu}, ``Attribute-based access control
  for icn naming scheme,'' \emph{IEEE Transactions on Dependable and Secure
  Computing}, vol.~15, no.~2, pp. 194--206, March 2018.

\bibitem{8442635}
G.~{Xylomenos}, Y.~{Thomas}, X.~{Vasilakos}, M.~{Georgiades},
  A.~{Phinikarides}, I.~{Doumanis}, S.~{Porter}, D.~{Trossen}, S.~{Robitzsch},
  M.~J. {Reed}, M.~{Al-Naday}, G.~{Petropoulos}, K.~{Katsaros}, M.~{Xezonaki},
  and J.~{Riihijarvi}, ``Ip over icn goes live,'' in \emph{2018 European
  Conference on Networks and Communications (EuCNC)}, June 2018, pp. 319--323.

\bibitem{BERNARDINI201913}
C.~Bernardini, S.~Marchal, M.~R. Asghar, and B.~Crispo, ``Privicn:
  Privacy-preserving content retrieval in information-centric networking,''
  \emph{Computer Networks}, vol. 149, pp. 13 -- 28, 2019.

\bibitem{7874168}
G.~{Acs}, M.~{Conti}, P.~{Gasti}, C.~{Ghali}, G.~{Tsudik}, and C.~A. {Wood},
  ``Privacy-aware caching in information-centric networking,'' \emph{IEEE
  Transactions on Dependable and Secure Computing}, vol.~16, no.~2, pp.
  313--328, March 2019.

\end{thebibliography}
